\documentclass[pra,twocolumn,showpacs,groupedaddress,superscriptaddress,nofootinbib,floatfix,preprintnumbers,longbibliography]{revtex4-1}
\usepackage{slashbox}
\usepackage{amssymb,amsmath,graphicx,color}
\usepackage{bm}
\usepackage[tight]{subfigure}
\usepackage[export]{adjustbox}
\usepackage{braket}
\usepackage[colorlinks=true,citecolor=blue,linkcolor=blue]{hyperref}
\usepackage[table]{xcolor}
\usepackage{cancel}
\usepackage{multirow}

\begin{document}
\title{
Search for Efficient Formulations for Hamiltonian Simulation
\\
of non-Abelian Lattice Gauge Theories
}

\author{Zohreh Davoudi}
\affiliation{
Maryland Center for Fundamental Physics and Department of Physics, 
University of Maryland, College Park, MD 20742, USA}
\affiliation{
RIKEN Center for Accelerator-based Sciences,
Wako 351-0198, Japan}

\author{Indrakshi Raychowdhury}
\affiliation{
Maryland Center for Fundamental Physics and Department of Physics, 
University of Maryland, College Park, MD 20742, USA}

\author{Andrew Shaw}
\affiliation{
Maryland Center for Fundamental Physics and Department of Physics, 
University of Maryland, College Park, MD 20742, USA}

\date{\today}

\preprint{UMD-PP-020-6}

\begin{abstract}
Hamiltonian formulation of lattice gauge theories (LGTs) is the most natural framework for the purpose of quantum simulation, an area of research that is growing with advances in quantum-computing algorithms and hardware. It, therefore, remains an important task to identify the most accurate, while computationally economic, Hamiltonian formulation(s) in such theories, considering the necessary truncation imposed on the Hilbert space of gauge bosons with any finite computing resources. This paper is a first step toward addressing this question in the case of non-Abelian LGTs, which further require the imposition of non-Abelian Gauss's laws on the Hilbert space, introducing additional computational complexity. Focusing on the case of SU(2) LGT in 1+1~dimensions coupled to one flavor of fermionic matter, a number of different formulations of the original Kogut-Susskind framework are analyzed with regard to the dependence of the dimension of the physical Hilbert space on boundary conditions, system's size, and the cutoff on the excitations of gauge bosons. The impact of such dependencies on the accuracy of the spectrum and dynamics obtained from a Hamiltonian computation is examined, and the (classical) computational-resource requirements given these considerations are studied. Besides the well-known angular-momentum formulation of the theory, the cases of purely fermionic and purely bosonic formulations (with open boundary conditions), and the Loop-String-Hadron formulation are analyzed, along with a brief discussion of a Quantum-Link-Model formulation of the same theory. Clear advantages are found in working with the Loop-String-Hadron framework which implements non-Abelian Gauss's laws \emph{a priori} using a complete set of gauge-invariant operators. Although small lattices are studied in the numerical analysis of this work, and only the simplest algorithms are considered, a range of conclusions will be applicable to larger systems and potentially to higher dimensions. Future studies will extend this investigation to the analysis of resource requirements for quantum simulations of non-Abelian LGT, with the goal of shedding light on the most efficient Hamiltonian formulation of gauge theories of relevance in nature.
\end{abstract}

\maketitle

\tableofcontents
\section{Introduction
\label{sec:intro}}
\noindent
Gauge field theories are at the core of our modern understanding of nature, from the descriptions of quantum Hall effect and superconductivity in condensed-matter physics~\cite{kleinert1989gauge, fradkin2013field}, to the mechanisms underlying the interactions of sub-atomic particles at the most fundamental level within the Standard Model of particle physics~\cite{Tanabashi:2018oca}, to emerging in the context of high-energy models proposed for new physics that are yet to be discovered~\cite{langacker1986standard}. Strongly interacting gauge theories coupled to matter, such as the quantum theory of strong interactions or quantum chromodynamics (QCD), are notoriously hard to simulate, and often demand applications of nonperturbative numerical strategies. Lattice-based methods, in which quantum fields are placed on a finite spacetime lattice, provide a natural regulation of ultraviolet modes and along with Renormalization Group (RG) methods, recover the continuum limit of the theory.\footnote{In asymptotically free theories like QCD.} Such a discretized theory also provides the framework for nonperturbative numerical studies, such as those based in path integral (Lagrangian) and Hamiltonian (canonical) formulations. While these two approaches are intrinsically equivalent, symmetries and constraints are manifested differently in each case~\cite{tong2018gauge}.

\noindent
\textbf{Path integral vs. Hamiltonian formulation of lattice gauge theories.} From a computational standpoint, the path-integral formulation has emerged as the primary tool in the LGT program given its parallels with the quantum statistical physics~\cite{Wilson:1974sk, Kogut:1979wt}, and its reliance on efficient state-of-the-art quantum Monte Carlo sampling techniques given this connection~\cite{Creutz:1983ev}. However, in order for such an analogy with statistical mechanics to be established, a Wick rotation to Euclidean spacetime is performed such that only imaginary-time correlation functions are directly accessed in this numerical program. This feature introduces two limitations: firstly the connection to real-time quantities is lost, and except in limited cases (e.g., Refs.~\cite{Alexandru:2016gsd,Ji:1996nm,Briceno:2019opb,Ji:2013dva,Radyushkin:2017cyf}), practical proposals are lacking for mapping a generic Euclidean correlation function that is obtained numerically on a spacetime lattice to dynamical amplitudes as measured in experiments. Second, a non-zero fermionic chemical potential introduces a sign problem by making the sampling weight in the Euclidean path integral imaginary~\cite{Aarts:2015tyj}, along with an inherently related signal-to-noise problem observed in correlation functions at zero chemical potential but with non-zero baryonic number~\cite{Wagman:2016bam}. On the other hand, the Hamiltonian formulation lacks the manifestly Lorentz covariance of the path-integral formulation, and further requires gauge fixing. In particular, in the most common gauge in which the temporal component of the gauge field is set to zero, the information about `constants of motion' is lost and the related constraint must be imposed \emph{a posteriori} on the Hilbert space. From a computational perspective, a Hamiltonian formulation enables both real-time and imaginary-time simulations. However, the most efficient numerical approach in Hamiltonian-based studies is no longer stochastic as in the path-integral formulation, and the cost of a typical numerical simulation scales with powers of the dimension of the Hilbert space (which itself grows exponentially with the size of the system). Extremely efficient Hamiltonian-simulation algorithms have been developed and implemented in recent years for low-dimensional LGTs using tensor-network methods~\cite{Banuls:2019rao}, but they rely on strict assumptions on the rate of entanglement growth in the physical system~\cite{Orus:2013kga}, assumptions that generally break down as system evolves arbitrarily in time.

Despite the drawbacks encountered in a Hamiltonian approach to LGTs, there has been revived interest in the Hamiltonian-simulation program given the improving prospects of quantum simulation and quantum computation. A plethora of ideas, proposals, and implementations have emerged for simulations of quantum many-body systems in general~\cite{abrams1997simulation, cirac2012goals, trabesinger2012quantum, schaetz2013focus, georgescu2014quantum}, and quantum field theories and LGTs in particular (see e.g., Refs.~\cite{Jordan:2011ne, Jordan:2011ci, Tagliacozzo:2012vg, Banerjee:2012pg, Zohar:2012xf, Jordan:2014tma, Zohar:2014qma, Mezzacapo:2015bra, Martinez:2016yna, Moosavian:2017tkv, Zache:2018jbt, Gorg:2018xyc, schweizer2019floquet, Klco:2018kyo, Lu:2018pjk, Bhattacharyya:2018bbv, Stryker:2018efp, Raychowdhury:2018osk, Mil:2019pbt, Klco:2019xro, Klco:2019evd, Bauer:2019qxa, Davoudi:2019bhy, Klco:2019yrb, Lamm:2019uyc, Mueller:2019qqj, Lamm:2019bik, Alexandru:2019nsa, Harmalkar:2020mpd, Yang:2020yer, Shaw:2020udc, Kharzeev:2020kgc, Chakraborty:2020uhf, Ciavarella:2020vqm, Liu:2020eoa, Kreshchuk:2020dla, Klco:2020aud, Haase:2020kaj, Paulson:2020zjd, Banuls:2019bmf}), in recent years, in light of advances in existing and upcoming digital and analog quantum-simulation technologies~\cite{preskill2018quantum, blatt2012quantum, gross2017quantum, schafer2020tools, schneider2012experimental, lanyon2011universal, monroe2019programmable, schmidt2013circuit, houck2012chip, paraoanu2014recent, lamata2018digital, Altman:2019vbv}. Generally speaking, a quantum-simulating/computing hardware will reduce the exponential cost of encoding the Hilbert space of a LGT onto the classical bits down to a polynomial cost, by storing information onto the quantum-mechanical wavefunctions of qubits. Nonetheless, quantum hardware will continue to exhibit small capacity and noise-limited capability for the foreseeable future. As a result, the search for an ultimate efficient formulation of LGTs for the quantum-simulation program is a crucial first step toward harnessing the power of quantum-simulating/computing platforms. In the meantime, as the classical Hamiltonian-simulation algorithms advance, such efficient formulation(s) can facilitate classical studies as well. In what follows, we elaborate on the meaning of an efficient formulation\footnote{A number of other proposals for general boson(gauge)-field digitizations exist as can be found in e.g., Refs.~\cite{somma2015quantum, Macridin:2018oli, Klco:2018zqz, Hackett:2018cel, Alexandru:2019nsa, Alexandru:2019ozf, Singh:2019uwd}. These will not be analyzed in the current work. Instead the focus is on the representation of the gauge theory itself in terms of the chosen basis states for fermions and bosons, and the gauge group will be kept exact despite the imposed truncation on the high excitations. One exception to this trend is the discussion of the Quantum Link Model~\cite{Chandrasekharan:1996ih, Brower:1997ha} of the SU(2) LGT that, given its popularity in the context of quantum simulation, is discussed in some length in Appendix~\ref{app:QLM}.} in the case of non-Abelian LGTs, and will make such efficiency considerations explicit by analyzing in depth the case of the SU(2) LGT in 1+1~Dimensions (D) coupled to matter. The focus is exclusively on the cost analysis of exact Hamiltonian-simulation algorithms using classical-computing hardware. Nonetheless, this study lays the groundwork for an analysis of resource requirements for simulating the same theory on quantum hardware, to be presented in future work.

\noindent
\textbf{A summary of the pros and cons of various representations of Kogut-Susskind theory.} In 1980s, Kogut and Susskind formulated a lattice Hamiltonian for Yang-Mills gauge-field theories coupled to matter~\cite{Kogut:1974ag} that recovered the continuum limit, and was shown to be equivalent to Wilson's path-integral formulation~\cite{Wilson:1974sk} of the same lattice theory. The Kogut-Susskind (KS) Hamiltonian further made it possible to perform quantum-mechanical perturbation theory around the strong-coupling vacuum, i.e., the ground state of the theory in the limit where the mass term and the electric-field term in the Hamiltonian dominate, see Sec.~\ref{sec:formalism}. Spectra and dynamics of both U(1) and SU(2) LGTs using the KS Hamiltonian in low dimensions were later studied toward the continuum limit, using high-order strong-coupling expansions, as well as non-perturbative numerical methods, see e.g., Refs.~\cite{Banks:1975gq, Hamer:1997dx, crewther1980eigenvalues, Hamer:1976bj, Hamer:1992ic, hamer19822}. Such investigations have continued to date using state-of-the-art algorithms such as those based on Matrix Product States~\cite{Banuls:2019rao, Banuls:2017ena, Sala:2018dui, Silvi:2016cas}. Given the requirement of generating and storing the Hamiltonian matrix whose size is determined by the size of the Hilbert space, and given the infinite dimensionality of the Hilbert space of a gauge theory, the truncation on excitations of the gauge degrees of freedom (DOF) is a necessity.\footnote{In the Quantum-Link-Model formulation of the SU(2) LGT, the gauge DOF are chosen to be fermionic, yielding a finite-dimensional Hilbert space~\cite{Chandrasekharan:1996ih, Brower:1997ha}. However, unless the continuum limit is taken through a dimensional-reduction procedure, the theory is not equivalent to the KS LGT.} Furthermore, only a small but still exponentially large (in the system's size) portion of the space of all possible basis states are those satisfying the gauge constraints, i.e., the local Gauss's laws. As will be demonstrated, exact and inexact Hamiltonian methods are not capable of simulating the full Hilbert space of a gauge theory even for small systems and small cutoff values on the gauge-field excitations. As a result, as a first step in the computation, a mechanism to construct the physical Hilbert space and its corresponding Hamiltonian must be implemented.

The representation chosen for fermionic and gauge DOE in a given LGT dictates the way the Hilbert-space truncation is performed and whether gauge invariance remains intact. It also determines the complexity of the construction of the physical Hilbert space and its associated Hamiltonian matrix for the sake of computation. For non-Abelian LGTs, in particular, the Gauss's law constraints are more complex to impose. In the case of the SU(2) LGT coupled to matter in the original proposal of Kogut and Susskind, the Hilbert space of the gauge DOF are expressed in terms of local angular-momentum basis states that mimic those associated with rigid-body rotations in fixed and body frames~\cite{Kogut:1974ag}. To construct the physical Hilbert space, these on-site basis states are combined with the fermionic DOF expressed in the fundamental representation of SU(2) in such a way that the net angular momentum is zero at each site. Furthermore, an Abelian constraint is satisfied such that the total angular momenta on the left and right of a link connecting two lattice sites are equal. As it will be demonstrated, even in 1+1~D, the computational complexity of imposing such constraints grows quickly with the size of the system, and with the cutoff imposed on the total angular momentum. The physical states, in general, become linear combinations of a large and growing number of terms in the basis chosen, adding to the complexity of Hamiltonian generation.

With open boundary conditions (OBC) in 1+1~D, it is possible to eliminate any dependence on the gauge DOE with the use of a gauge transformation, along with the application of Gauss's laws at the level of the Hamiltonian operator itself, as already proposed as a viable efficient basis for Hamiltonian simulation of the SU(2) LGT in Refs.~\cite{Banuls:2017ena, Sala:2018dui}. Such a trick significantly reduces the computational cost of imposing Gauss's laws on the Hilbert space and eliminates the need for a cutoff on the gauge DOF. However, as will be shown, there remain redundancies in this representation compared with the physical Hilbert space of the KS theory in the angular-momentum basis that grows slowly with the system's size. Furthermore, this formulation can not be extended to higher dimensions since there are not sufficient Gauss's laws present to eliminate all gauge DOF. So it will be beneficial to examine a formulation with better generalizability prospects. A recent bosonization proposal~\cite{Zohar:2019ygc, Zohar:2018cwb} for LGTs is explored as well, in which a subset of Gauss's laws corresponding to the Cartan sub-algebra of the SU(N) group coupled to one flavor of fundamental fermions are augmented with an additional U(1) Abelian Gauss's law to allow the elimination of fermionic DOF. While this procedure works in any dimension, it trades the finite-dimensional Hilbert space of the fermions with intrinsically infinite-dimensional Hilbert space of the gauge DOF, including an additional U(1) field. The Hilbert space of these bosonic fields must be cut off for practical purposes, which could lead to systematic uncertainties in computations with finite resources. Furthermore, constructing the Hamiltonian matrix in the physical Hilbert space remains computationally involved in the bosonic formulation.

The complexity associated with non-Abelian LGTs in the KS theory in its original formulation is the motivation behind the development of a recent framework called the Loop-String-Hadron (LSH) formulation for the SU(2) LGT coupled to matter, which is valid in any dimensions~\cite{Raychowdhury:2019iki, Raychowdhury:2018osk}. It is founded upon the prepotential formalism of pure LGTs~\cite{Mathur:2004kr, Mathur:2007nu,  Anishetty:2014tta, Raychowdhury:2014eta, Raychowdhury:2018tfj}, which is fundamentally a representation that re-expresses the angular-momentum basis in terms of the harmonic-oscillator basis of Schwinger bosons~\cite{schwinger1965quantum}.\footnote{Prepotential formulation for  SU(3)~\cite{Anishetty:2009nh} as well as SU(N)~\cite{Raychowdhury:2013rwa} LGTs have also been constructed in terms of irreducible Schwinger bosons~\cite{Anishetty:2009ai, Mathur:2010wc} in any dimension. These exhibit the same features as the SU(2) theory discussed in the present paper.} As a result, the SU(2) gauge-link and electric-field operators are expressed in terms of harmonic-oscillator creation and annihilation operators and allow gauge-invariant operators to be formed out of gauge and fermionic DOF at each site. These operators, therefore, excite only the states in the physical sector of the Hilbert space, as long as an Abelian Gauss's law is satisfied, which requires the number of oscillators at the left and right of the link to be equal. The LSH formulation constructs a complete set of properly normalized gauge-invariant operators and expresses the Hamiltonian in terms of this complete basis~\cite{Raychowdhury:2019iki}. As will be shown, the Hilbert space of the KS theory in the angular-momentum basis after imposing the Abelian and non-Abelian Gauss's laws, and for a given cutoff on the gauge DOF, is identical to that of the LSH Hamiltonian. Nonetheless, the computational cost of generating the associated Hamiltonian is far less in the LSH framework given its already gauge-invariant physical basis states. Consequently, the simplicity of the fermionic representation (with OBC) is enjoyed by the LSH formulation as well but without associated redundancies, and with the prospects of straightforward applications to higher dimensions.

\noindent
\textbf{Outline of the paper.} While all the different formulations studied here have been introduced, and to some extent implemented, in literature, the conclusions briefly stated above and those that will follow, are new and have resulted from a thorough comparative analysis that is conducted in this work. In particular, an analysis of the size of the full and physical Hilbert spaces as a function of the system's size and, when applicable, the cutoff on the gauge DOF is presented in Sec.~\ref{sec:Hilbert} for all the formulations of SU(2) LGT in 1+1~D enumerated above (and reviewed in Sec.~\ref{sec:formalism}). Here, empirical relations are obtained from a numerical study with small lattice sizes. These results lead to a discussion of the time complexity of exact classical Hamiltonian-simulation algorithms in Sec.~\ref{sec:cost}. Section~\ref{sec:specdyn} contains an analysis of the impact of the cutoff on the spectrum and dynamics of the theory. A detailed discussion of the global symmetries of SU(2) LGT in 1+1~D is presented in Sec.~\ref{sec:symmetry}, which allows the decomposition of the physical Hilbert space of the theory to even smaller decoupled sectors, hence simplifying the computation. While not a focus of this work, a brief comparative study of the KS SU(2) theory in 1+1~D with a QLM formulation~\cite{Brower:1997ha} is presented in Appendix~\ref{app:QLM}. Given the extent of discussions, and the spread of observations and conclusions made throughout this paper, Sec.~\ref{sec:conclusions} will summarize the main points of the study more crisply, along with presenting an outlook of this work.

In summary, the results presented here should offer a clear path to the practitioner of Hamiltonian-simulation techniques to evaluate the pros and cons of a given formulation of the SU(2) LGT in 1+1~D in connection to the simulation algorithm used. A similar study for the 2+1-dimensional theory can shed light on the validity of the conclusions made for higher-dimensional cases.\footnote{See a recent work on the efficient Hamiltonian simulation of the U(1) LGT in 2+1 D in Ref.~\cite{Haase:2020kaj}.} Moreover, the conclusions of this work will guide future studies of non-Abelian LGTs in the context of quantum simulation.

\section{An overview of the Kogut-Susskind SU(2) LGT and its various formulations
\label{sec:formalism}}
\begin{figure*}[t!]
\includegraphics[width=0.715\textwidth]{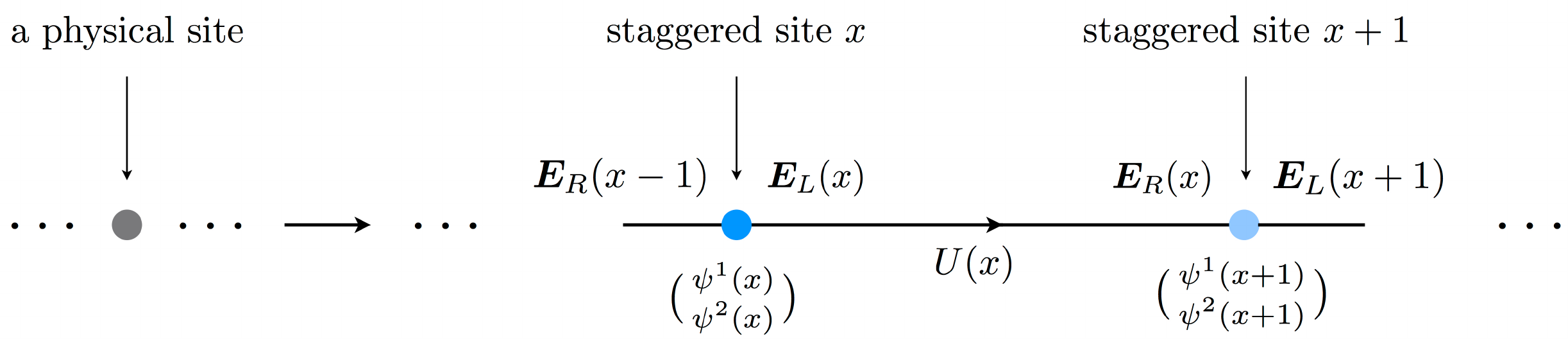}
\caption{A physical site along the spatial direction is split to two staggered sites in the KS Hamiltonian. These sites are connected by a gauge link. Corresponding to each staggered site, there is a two-component fermion field, a left electric field, and a right electric field, as indexed in the figure.
}
\label{fig:lattice}
\end{figure*}

\noindent
Within the Hamiltonian formulation of LGTs introduced by Kogut and Susskind, the temporal direction is continuous while the spatial direction is discretized. Each site along the spatial direction is split into two staggered sites, as shown in Fig.~\ref{fig:lattice}, such that matter and anti-matter fields occupy even and odd sites, respectively. The number of sites along this direction is denoted by $N$ and is called the lattice size throughout. The spacing between adjacent sites after staggering is denoted as $a$. For the SU(2) LGT in 1+1~D, the KS Hamiltonian can be written as:
\begin{eqnarray}
H^{({\rm KS})}&=&H^{({\rm KS})}_I+H^{({\rm KS})}_E+H^{({\rm KS})}_M.
\label{eq:HKS}
\end{eqnarray}
Here, $H^{({\rm KS})}_I$ denotes interactions among the fermionic and gauge DOF\footnote{Here and throughout, the position argument of the functions and the superscript of state vectors are assumed to be an index. A multiplication by the lattice spacing $a$ converts these to the absolute position.}
\begin{eqnarray}
H^{({\rm KS})}_I&=&\frac{1}{2a}\sum_{x=0}^{N_1}\left[\psi^{\dagger}(x) \hat{U}(x) \psi(x+1)+{\rm h.c.} \right]
,\label{eq:HIKS}
\end{eqnarray}
where $N_1=N-1$ for PBC and $N_1=N-2$ for OBC. The fermion field $\psi$ is in the fundamental representation of SU(2) and consists of two components, i.e., $\psi= \bigl( \begin{smallmatrix}\psi^1\\ \psi^2\end{smallmatrix}\bigr)$. The gauge link $\hat{U}(x)$ is a $2\times2$ unitary matrix operator which emanates from site $x$ along the spatial direction and ends at point $x+1$, as shown in Fig.~\ref{fig:lattice}. A temporal gauge is chosen which sets the gauge link along the temporal direction equal to unity.

$H^{({\rm KS})}_E$ corresponds to the energy stored in the electric field,
\begin{eqnarray}
H^{({\rm KS})}_E&=& \frac{g^2a}{2} \sum_{x=0}^{N_2} \hat{\bm{E}}(x)^2.
\label{eq:HEKS}
\end{eqnarray}
Here, $N_2=N-1$ for PBC, $N_2=N-2$ for OBC, and $g$ is a coupling. Further, $\hat{\bm{E}}^2 = (\hat{E}^1)^2+(\hat{E}^2)^2+(\hat{E}^3)^2 \equiv \hat{\bm{E}}_L^2=\hat{\bm{E}}_R^2$. $\hat{\bm{E}}_L$ and $\hat{\bm{E}}_R$ are the left and the right electric-field operators, respectively, as shown in Fig.~\ref{fig:lattice}. These satisfy the SU(2) Lie algebra at each site,
\begin{eqnarray}
[\hat{E}_L^a,\hat{E}_L^b]&=&-i\epsilon^{abc} \hat{E}_L^c,
\nonumber\\
{[\hat{E}_R^a,\hat{E}_R^b]}&=&i\epsilon^{abc} \hat{E}_R^c,
\nonumber\\
{[\hat{E}_L^a,\hat{E}_R^b]}&=&0,
\label{eq:ERELcomm}
\end{eqnarray}
where $\epsilon^{abc}$ is the Levi-Civita tensor and the spatial dependence of the fields is suppressed in these relations. Further, the electric fields on different sites commute. The electric fields and the gauge link satisfy the canonical commutation relations at each site,
\begin{eqnarray}
[\hat{E}_L^a,\hat{U}]&=&T^a\hat{U},
\nonumber\\
{[\hat{E}_R^a,\hat{U}]}&=& \hat{U}T^a,
\label{eq:EUcomm}
\end{eqnarray}
where $T^a=\frac{1}{2} \tau^a$, and $\tau^a$ is the $a^{\rm th}$ Pauli matrix. The corresponding commutation relations for fields with different site indices vanish.

Finally, $H^{({\rm KS})}_M$ in Eq.~(\ref{eq:HKS}) is a staggered mass term
\begin{eqnarray}
H^{({\rm KS})}_M&=&m\sum_{x=0}^{N_3}(-1)^x\psi^\dagger(x) \psi(x).
\label{eq:HMKS}
\end{eqnarray}
Here, $N_3=N-1$ for both PBC and OBC, and $m$ denotes the mass of each component of the fermions.

In this theory, a fermion SU(2)-charge-density operator defined at each site,
\begin{eqnarray}
\hat{\rho}^a(x) \equiv \psi^\dagger(x) T^a \psi(x),
\label{eq:rho}
\end{eqnarray}
which satisfies the SU(2) Lie algebra. It further satisfies the following commutation relation at each site,
\begin{eqnarray}
[\hat{\rho}^a,\psi]=-T^a \psi.
\end{eqnarray}
Such a commutation relation vanishes for fields at different sites. This SU(2)-charge-density operator also commutes with the electric fields and the gauge link. With these commutation relations, and those given in Eqs.~(\ref{eq:ERELcomm}) and (\ref{eq:EUcomm}), one can show that the Hamiltonian in Eq.~(\ref{eq:HKS}) commutes with the following operator,
\begin{equation}
\hat{G}^a(x)=-\hat{E}^a_L(x)+\hat{E}^a_R(x-1)+\hat{\rho}^a(x).
\label{eq:Ga}
\end{equation}
As a result, the Hilbert space of the theory is classified into sectors corresponding to each of the eigenvalues of the Gauss's law operators $G^a$, and these eigenvalues are the `constants of motion'. The physical sector of this Hilbert space is that corresponding to the zero eigenvalue of this operator.

\subsection{Angular-momentum formulation
\label{sec:KS}}
\noindent
The first step in forming the Hilbert space of a LGT for the sake of computation is to map the vacuum and the excitations of the fields to a state basis. In the absence of the magnetic Hamiltonian, which is the case in 1+1~D LGTs, the most efficient basis is formed out of eigenstates of the electric-field operator. The direct product of the fermionic eigenstates and the electric-field eigenstates forms the full Hilbert space. This is called the electric-field basis, or the strong-coupling basis, i.e., in the $g \to \infty$ limit, the interaction terms in Eq.~(\ref{eq:HIKS}) that involves transitions between different eigenvalues of the electric-field operator becomes insignificant compared with the electric-field term, Eq.~(\ref{eq:HEKS}), and the Hamiltonian becomes diagonal in the electric-field basis. Since the electric fields satisfy the SU(2) algebra, a familiar representation is the angular-momentum representation. In fact, as pointed out by Kogut and Susskind, the left and right electric field can be mapped to the body-frame ($\hat{\bm{J}}^b$) and space-frame ($\hat{\bm{J}}^s$) angular momenta of a rigid body. Explicitly, $\hat{\bm{E}}_L=-\hat{\bm{J}}^b(\equiv -\hat{\bm{J}}_L)$ and $\hat{\bm{E}}_R=\hat{\bm{J}}^s(\equiv \hat{\bm{J}}_R)$, satisfying $\hat{\bm{J}}_L^2=\hat{\bm{J}}_R^2$ on each link.

Given this correspondence, one may write the electric-field basis states for the KS formulation as
\begin{equation}
\ket{\Phi}_{({\rm KS})}^{(x)}=
\ket{J_R,m_R}^{(x-1)}\otimes
\ket{f_1,f_2}^{(x)}\otimes \ket{J_L,m_L}^{(x)},
\label{eq:KSbasis}
\end{equation}
for each site $x$. Here, $f_1$ and $f_2$ quantum numbers refer to the occupation number of the two components of the (anti)matter field, $\psi^1$ and $\psi^2$, each taking values 0 and 1, corresponding to the absence and presence of (anti)matter, respectively:
\begin{eqnarray}
\psi^1\ket{f_1,f_2}=(1-\delta_{f_1,0})\ket{f_1-1,f_2},
\\
{\psi^1}^\dagger\ket{f_1,f_2}=(1-\delta_{f_1,1})\ket{f_1+1,f_2},
\end{eqnarray}
at each site, and similarly for the other component of $\psi$. Here, $\delta$ denotes the Kronecker-delta symbol. Furthermore, the angular-momentum basis states satisfy
\begin{eqnarray}
\hat{\bm{J}}_R^2\ket{J_R,m_R}=J_R(J_R+1)\ket{J_R,m_R}
\\
\hat{\bm{J}}_L^2\ket{J_L,m_L}=J_L(J_L+1)\ket{J_L,m_L}
\end{eqnarray}
and
\begin{eqnarray}
\hat{J}_R^3\ket{J_R,m_R}=m_R\ket{J_R,m_R}
\\
\hat{J}_L^3\ket{J_L,m_L}=m_L\ket{J_L,m_L}
\end{eqnarray}
at each site $x$, where for brevity the site indices are suppressed. Here, $J_L,J_R=0,\frac{1}{2},1,\frac{3}{2},\cdots$, and $m_R$ and $m_L$ quantum numbers satisfy $ -J_R \leq m_R \leq J_R$ and $ -J_L \leq m_L \leq J_L$, as dictated by the angular-momentum group algebra. The action of the gauge-link operator on this basis can be written as:
\begin{eqnarray}
&&\hat{U}^{(\alpha,\beta)}(x)\left[ \cdots \ket{J_R,m_R}^{(x-1)}\otimes
\ket{f_1,f_2}^{(x)}\otimes \ket{J_L,m_L}^{(x)} \right .
\nonumber\\
&& \qquad \left . \otimes \ket{J_R,m_R}^{(x)}\otimes
\ket{f_1,f_2}^{(x+1)}\otimes \ket{J_L,m_L}^{(x+1)} \cdots \right] 
\nonumber\\
&& = \cdots \ket{J_R,m_R}^{(x-1)}\otimes
\ket{f_1,f_2}^{(x)} \otimes
\nonumber\\
&&\left [ \sum_{j=\{0,\frac{1}{2},1,...\}} \sqrt{\frac{2J+1}{2j+1}}
\braket{J,m_L;\frac{1}{2},\alpha | j,m_L+\alpha} \right .
\nonumber\\
&& \left . \braket{J,m_R;\frac{1}{2},\beta | j,m_R+\beta} \ket{j,m_L+\alpha}^{(x)}\otimes\ket{j,m_R+\beta}^{(x)} \right]
\nonumber\\
&&\qquad \qquad \qquad
\otimes \ket{f_1,f_2}^{(x+1)}\otimes \ket{J_L,m_L}^{(x+1)} \cdots,
\label{eq:UonState}
\end{eqnarray}
where $\alpha,\beta=\pm \frac{1}{2}$.\footnote{Note that: $U_{11}=U^{(\frac{1}{2},-\frac{1}{2})},~U_{12}=U^{(-\frac{1}{2},-\frac{1}{2})},~U_{21}=U^{(\frac{1}{2},\frac{1}{2})},~U_{22}=U^{(-\frac{1}{2},\frac{1}{2})}$.} Note that $J_L$ and $J_R$ on each link are equal, and as such we have defined $J \equiv J_L^{(x)}=J_R^{(x)}$ in this relation. Ellipses denote states that precede and follow those shown at site $x$ and $x+1$, respectively. 
 
The physical states $\ket{\phi}_{({\rm KS})}$ can be formed by identifying proper linear combinations of the basis states in Eq.~(\ref{eq:KSbasis}) such that the Gauss's laws are satisfied at each site, and by constructing the direct product of these combinations for adjacent sites along the lattice, following additional gauge and boundary conditions as is detailed below. First, given the Gauss's law operators defined in Eq.~(\ref{eq:Ga}), the physical states $\ket{\phi}_{({\rm KS})}$ are required to satisfy $G^a(x)\ket{\phi}_{({\rm KS})}=0$. Explicitly,
\begin{equation}
\left[\hat{J}^a_L(x)+\hat{J}^a_R(x-1)+\frac{1}{2}\psi^{\dagger}(x)\tau^a \psi(x)\right]\ket{\phi}_{({\rm KS})}=0,
\label{eq:GLawJ}
\end{equation}
for $a=1,2,3$, and for every $x$ where $x=0,1,\cdots,N-1$ along the one-dimensional lattice. So the Gauss's laws can be simply interpreted as the angular momenta $\hat{\bm{J}}^L$ (corresponding to $\hat{\bm{E}}_L$), $\hat{\bm{J}}_f$ with $\hat{\bm{J}}_f^2=\frac{3}{4}$ (corresponding to the presence of one and only one fermion), and $\hat{\bm{J}}^R$ (corresponding to $-\hat{\bm{E}}_L$) should add to zero at each site. When there is no fermion present or two fermions are present, $\hat{\bm{J}}_f=0$ and the left and right angular momenta are the same. Moreover, as mentioned before, $J_L$ and $J_R$ quantum numbers need to be equal on each link. These two requirements, in addition to the boundary conditions imposed on the $J_R$ value at site $x=0$ and the $J_L$ value at site $x=N-1$, constrain the Hilbert space to a physical gauge-invariant one, as analyzed in Sec.~\ref{sec:HilbertKS}

\subsection{Purely fermionic formulation
\label{sec:F}}
\noindent
The KS Hamiltonian in Eq.~(\ref{eq:HKS}) combined with the Gauss's law constraints on the Hilbert space, in essence, leaves no dynamical gauge DOF in 1+1~D beyond possible boundary modes. In particular, with OBC where the incoming flux of the (right) electric field is set to a fixed value, the value of electric-field excitations throughout the lattice is fixed. This, in fact, is a general feature of LGTs in 1+1~D, as is evident from the proof outlined below. As a result, the KS Hamiltonian acting on the physical Hilbert space can be brought to a purely fermionic form, in which the identification of (anti)fermion configurations is sufficient to construct the Hilbert space. This eliminates the need for adopting a state basis for the gauge DOF, and for solving the complex (non-diagonal) Gauss's laws locally which is the case in an angular-momentum basis. Such an elimination of gauge DOF in LGTs in 1+1~D was first discussed in Ref.~\cite{Hamer:1997dx} and is used in recent tensor-network simulations of the SU(2) LGT in Ref.~\cite{Sala:2018dui}. Here, we present a generic derivation of such a purely fermionic representation, before analyzing its Hilbert space in the following section.

Consider the following gauge transformation on the fermion fields in the KS Hamiltonian:
\begin{eqnarray}
\psi(x) &\to& \psi'(x) = \left[\prod_{y<x} U(y)\right] \psi(x),
\\
\psi^{\dagger}(x) &\to& \psi^{\dagger'}(x) = \psi^{\dagger}(x)\left[\prod_{y<x} U(y)\right]^{\dagger}.
\end{eqnarray}
Note that the products of gauge links are defined as $\prod_{y<x} U(y) = U(0)U(1) \cdots U(x-1)$. Consequently, the gauge links must transform as
\begin{equation}
U(x)\to U'(x) = \left[\prod_{y<x} U(y)\right] U(x)\left[\prod_{z < x+1} U(z)\right]^{\dagger},
\label{eq:Uprime}
\end{equation}
such that the gauge-matter interaction Hamiltonian, $H^{({\rm KS})}_{I}$, remains invariant:
\begin{eqnarray}
\psi^{\dagger}(x) U(x) \psi(x+1)+{\rm h.c.} \to \psi^{\dagger'}(x) U'(x) \psi'(x+1)+{\rm h.c.}
\nonumber\\
\end{eqnarray}
Now considering the unitarity condition on the gauge links, i.e., $U^\dagger U =\mathbb{I}$, reveals that Eq.~(\ref{eq:Uprime}) simplifies to
\begin{eqnarray}
U'=\mathbb{I},
\end{eqnarray}
where $\mathbb{I}$ is the unity matrix whose dimensionality is equal to that of the fundamental representation of the gauge group, e.g., two in the case of SU(2). Therefore, the interaction Hamiltonian becomes
\begin{eqnarray}
H^{({\rm F})}_{I}=\frac{1}{2a} \sum_{x=0}^{N_1}\left[\psi^{\dagger'}(x)\psi'(x+1)+{\rm h.c.}\right],
\label{eq:HIF}
\end{eqnarray}
where $N_1=N-2$ as noted after Eq.~(\ref{eq:HMKS}).

Now given the relation among the gauge link and the left and right electric fields belonging to the same link~\cite{Zohar:2015hwa},
\begin{eqnarray}
E_R(x) = U^\dagger (x) E_L(x) U(x),
\end{eqnarray}
one obtains the following relation in the new gauge:
\begin{eqnarray}
E_R(x) = E_L(x).
\end{eqnarray}
This relation, combined with the OBC set to $E^a_R(-1)=\epsilon^a_0$ for $a=1,2,3$, and the Gauss's laws $G^a \ket{\phi}_{(\rm KS)} =0$ with $G^a$ defined in Eq.~(\ref{eq:Ga}), fully fixes the values of $E_L$ and $E_R$ at all sites on the one-dimensional lattice in the physical Hilbert space:
\begin{eqnarray}
E_L^a(x) = \epsilon^a_0+\sum_{y=0}^x \rho^a(y) = E_R^a(x),
\end{eqnarray}
with $\rho^a$ defined in Eq.~(\ref{eq:rho}). Consequently, the electric-field Hamiltonian $H^{({\rm KS})}_{\rm E}$ becomes\footnote{Note that $\psi^\dagger(x) \psi(x) = \psi^{\dagger'}(x)\psi'(x)$.}
\begin{equation}
H^{({\rm F})}_E= \frac{g^2a}{2} \sum_{x=0}^{N_2} \sum_{a=1}^3 \left[ \epsilon_0^a+\sum_{y=0}^{x}\psi^{\dagger'}(y) T^a \psi'(y) \right]^2,
\label{eq:HEF}
\end{equation}
where $N_2=N-2$ for OBC as noted after Eq.~(\ref{eq:HEKS}). The consequence of applying Gauss's laws to arrive at Eq.~(\ref{eq:HEF}) is that the local electric-field Hamiltonian in the original formulation is replaced with arbitrary-range fermion-fermion interactions in the fermionic Hamiltonian. 

Finally, the mass term in the new gauge remains the same, as is expected from gauge invariance: 
\begin{eqnarray}
H^{({\rm F})}_M&=& m\sum_{x=0}^{N_3}(-1)^x\psi^{\dagger'}(x) \psi'(x),
\label{eq:HMF}
\end{eqnarray}
where $N_3=N-1$ as noted after Eq.~(\ref{eq:HMKS}). Note that upon expanding Eq.~(\ref{eq:HEF}), terms with a fermionic structure similar to the mass term arise, effectively modifying the mass in the new representation.

The procedure outlined above establishes that any explicit dependence on the gauge link and electric fields are eliminated in the KS Hamiltonian with OBC, giving rise to a purely fermionic Hamiltonian whose terms are specified in Eqs.~(\ref{eq:HIF}), (\ref{eq:HEF}), and (\ref{eq:HMF}), and which is identical to the original KS Hamiltonian only in the physical Hilbert space. As a result, any state in this formulation can be written in terms of a complete fermionic occupation-number basis,
\begin{equation}
\ket{\Phi}_{({\rm KS,F})}=\prod_{x=0}^{N-1}\ket{f_1,f_2}^{(x)},
\label{eq:Fbasis}
\end{equation}
where as before, $f_1$ and $f_2$ refer to the occupation number of the two components of the (anti)matter field, $\psi_1$ and $\psi_2$, respectively, each taking values 0 or 1.

\subsection{Purely bosonic formulation
\label{sec:B}}
\noindent
Gauge transformation, along with the imposition of the local Gauss's laws with OBC, led to the elimination of the gauge DOF in the previous section. Unfortunately, this procedure can obtain a purely fermionic theory only in 1+1~D, as in higher dimensions the number of constraints at each lattice site is not sufficient to eliminate the gauge DOF in all spatial directions. One could reversely consider eliminating the fermionic DOF with the use of the Gauss's laws, as proposed in Ref.~\cite{Zohar:2019ygc}, to obtain a fully bosonic theory. This protocol works in all dimensions, but in the case of $SU(2\mathcal{N})$ theories, requires enlarging the gauge group to $U(2\mathcal{N})$ to accommodate a sufficient number of constraints needed to eliminate the fermions.\footnote{As shown in Ref.~\cite{Zohar:2019ygc} for the case of the $SU(2\mathcal{N}+1)$ theory, the introduction of an auxiliary $Z_2$ gauge field on each link on the lattice is sufficient to eliminate the fermions, without the need to enlarge the group to $U(2\mathcal{N}+1)$. This enhancement also takes care of the fermionic statistics when fermions are replaced with the hardcore bosons and are subsequently eliminated. Since the focus of this work is the SU(2) theory, this case will not be analyzed here further.} One further needs to keep track of the fermionic statistics by encoding in the purely bosonic interactions, the nontrivial signs associated with the anti-commuting nature of the fermions~\cite{Zohar:2018cwb}. The extended theory can be shown to be equivalent to the original theory for all physical purposes, as long as the cutoff on the new gauge DOF of the extended symmetry is set sufficiently high, see Sec.~\ref{sec:HilbertB}. In the following, the bosonized form of the SU(2) LGT in 1+1~D is derived, following the procedure outlined in Ref.~\cite{Zohar:2019ygc} for general dimensions.

Consider the Gauss's laws in the KS formulation of the SU(2) LGT in 1+1~D, given in Eq.~(\ref{eq:GLawJ}). Although there exist three Gauss's laws at each site, only the Gauss's law corresponding to the $a=3$ component of Gauss's law operator in Eq.~(\ref{eq:Ga}) provides a diagonal relation in the angular momentum/fermionic basis. In other words, two of the Gauss's laws mix basis states with different quantum numbers, and only one of the Gauss's laws leads to an algebraic relation among the gauge and fermionic DOF. Explicitly, for a basis state $\ket{J_R,m_R}^{(x-1)}\otimes \ket{f_1,f_2}^{(x)}\otimes \ket{J_L,m_L}^{(x)}$ at site $x$, this relation reads
\begin{equation}
m_L(x)+m_R(x-1)=-\frac{1}{2}(f_1(x)-f_2(x)).
\label{eq:GLawDiag}
\end{equation}
However, in order to fully express the $\{f_1,f_2\}$ quantum numbers at each site in terms of the $\{J_R,m_R,J_L,m_L\}$ quantum numbers surrounding the site, at least one more independent relation is needed. Such a relation can be obtained by adding an extra U(1) symmetry to enlarge the gauge group, effectively modifying each link on the lattice by a U(1) link $U_0$, i.e., $U(x) \to U(x) \times U_0(x)$, where $U$ is the SU(2) link. This introduces a staggered U(1) charge,\footnote{The staggered term in the U(1) charge ensures that filled even and odd sites have opposite charges, corresponding to the presence of matter and antimatter at the site, respectively.}
\begin{equation}
\hat{\rho}^0(x) \equiv \frac{1}{2} \left[\psi^\dagger(x) \psi(x)-(1-(-1)^x)\right],
\label{eq:GLawDiag}
\end{equation}
along with a corresponding U(1) electric field $E(x)$ on each link emanating from site $x$.

The physical Hilbert space of the extended U(2) theory is the direct product of the physical Hilbert spaces of the KS SU(2) theory and the U(1) theory, i.e., $\ket{\phi}_{U(2)}=\ket{\phi}_{({\rm KS})} \otimes \ket{\phi}_{U(1)}$, where $\ket{\phi}_{({\rm KS})}$ was introduced in Sec.~\ref{sec:KS}, and
\begin{equation}
\ket{\phi}_{U(1)}=\ket{E}^{(0)} \otimes \ket{E}^{(1)} \cdots \otimes \ket{E}^{(N-1)},
\end{equation}
with
\begin{equation}
\hat{E}(x)\ket{E}^{(x)}=E(x)\ket{E}^{(x)},~~E(x) \in \mathbb{Z},
\end{equation}
for all values of the electric field that satisfy the U(1) Gauss's law $\hat{G}^0 \ket{\phi}_{U(2)}=0$ with
\begin{equation}
\hat{G}^0(x)=\frac{1}{2}\left[-\hat{E}(x)+\hat{E}(x-1)\right]+\hat{\rho}^0(x).
\label{eq:G0}
\end{equation}
Explicitly, the U(1) gauss's law acting on the new basis state $\ket{J_R,m_R}^{(x-1)}\otimes
\ket{f_1,f_2}^{(x)}\otimes \ket{J_L,m_L}^{(x)} \otimes \ket{E}^{(x)}$ at site $x$ gives
\begin{equation}
E(x)-E(x-1)=f_1(x)+f_2(x)-(1-(-1)^x).
\label{eq:GLawDiagU1}
\end{equation}
Here, OBC is considered with $E(-1)=\epsilon_0$, where $\epsilon_0$ is a constant integer. From Eqs.~(\ref{eq:GLawDiag}) and (\ref{eq:GLawDiagU1}), the $\{f_1,f_2\}$ quantum numbers at each site become redundant, as they can be written as
\begin{eqnarray}
f_i(x)&=&\frac{1}{2}\left[E(x)-E(x-1)+(1-(-1)^x)\right]+
\nonumber\\
&& s_i\left[m_L(x)+m_R(x-1)\right],
\label{eq:fi}
\end{eqnarray}
for $i=1,2$. Here, $s_1=-1$, and $s_2=1$.

As a consequence of Eq.~(\ref{eq:fi}), the action of the mass Hamiltonian $\propto \psi^\dagger(x)\psi(x)$ on $\ket{f_1,f_2}^{(x)}$ can be written as the action of the corresponding operators in the set $\{\hat{J}^3_R(x-1),\hat{J}^3_L(x),\hat{E}(x-1),\hat{E}(x)\}$ on the bosonic basis states in the physical Hilbert space, effectively rendering a purely bosonic term. However, the action of the matter-gauge interaction Hamiltonian $\propto \psi^\dagger(x)U(x)\psi(x+1)+{\rm h.c.}$ will be non-trivial due to the non-commuting nature of the fermions, and this feature must be built in the purely bosonic formulation explicitly. In other words, the gauge-matter interaction Hamiltonian must carry the information regarding fermionic signs in its purely bosonic form.

\begin{figure*}[t!]
\includegraphics[width=0.985\textwidth]{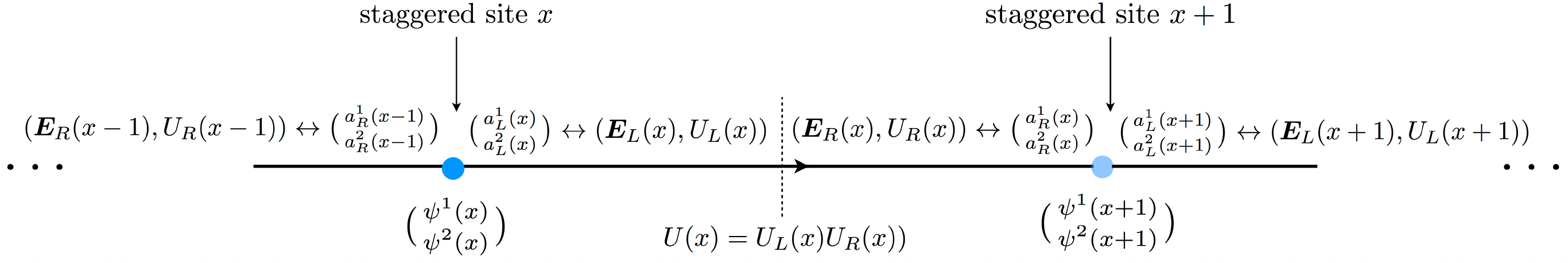}
\caption{The KS LGT is illustrated in terms of the DOF of the  LSH formulation. The left and right electric fields and the gauge link are replaced by a set of prepotential (Schwinger boson) doublets, $ \bigl( \begin{smallmatrix}a^1_{L(R)}\\ a^2_{L(R)}\end{smallmatrix}\bigr)$, at each end of a link. As a consequence of this construction, the gauge link explicitly breaks into a left part and a right part. The staggered matter remains the same as in original KS formalism, see Fig.~\ref{fig:lattice}. The LSH framework confines the gauge group at and around each site as indexed in the figure.
}
\label{fig:latticelsh}
\end{figure*}

While there are a number of protocols for transforming the fermions to hardcore bosons (spins) such that the fermionic anti-commutation relations are preserved, such as the familiar Jordan-Wigner transformation~\cite{jordan1993paulische}, an alternative protocol that preserves the locality of interactions is presented in Ref.~\cite{Zohar:2018cwb}. For a generic $SU(\mathcal{N})$ theory, this protocol amounts to augmenting the theory with an additional $Z_2$ gauge symmetry, where the new local Gauss's law corresponding to the auxiliary gauge keeps track of the fermionic signs, see also Refs.~\cite{Chen:2017fvr, Chen:2019wlx}. Back to the case of bosonized SU(2) theory, the extended U(2) theory (necessitated by the need for an extra diagonal Gauss's law) includes the $Z_2$ symmetry as a subgroup. Therefore, the protocol of Ref.~\cite{Zohar:2018cwb} does not require introducing an additional $Z_2$ gauge group. In other words, the transformation from fermions to hardcore bosons can already proceed by exploiting the local U(1) electric fields introduced above. Nonetheless, since the Hamiltonian terms are either local or nearest neighbor, and that only a 1+1~D theory is considered in the current work, all such transformation, local or non-local variants, are of comparable (low) complexity. Distinctions among different transformations become more relevant in the context of quantum simulation, in which the fermions need to be mapped to qubit DOF. Such considerations will be studied in future work.

Besides their coupling to the fermions in the modified matter-gauge interaction Hamiltonian, which guarantees the U(1) Gauss's law constraint, no further dynamics is introduced for the U(1) gauge DOF. As a result, apart from the issue of the fermionic statistics that needs to be dealt with via a separate transformation as discussed above, the Hamiltonian of the extended U(2) theory is a straightforward extension of the KS Hamiltonian presented in Sec.~\ref{sec:KS}, as shown in Ref.~\cite{Zohar:2019ygc}. The extended Hamiltonian involves nearest-link interactions, as a result of the replacement in Eq.~(\ref{eq:fi}) and the transformation to hardcore bosons, but is otherwise local. The physical Hilbert spaces of the SU(2) theory and the U(2) theory are isomorphic, meaning that in the limit where the U(1) gauge link approaches unity, the Hamiltonian matrix elements in the original theory is recovered from those in the extended theory. This is established straightforwardly for OBC, while for PBC, the isomorphism holds only in a given topological sector~\cite{Zohar:2019ygc}. In the remainder of this paper, we analyze the dimensionality of, and the resource requirement for constructing, the physical Hilbert space of the bosonized theory compared with the KS theory, along with the effect  of the U(1) cutoff on the spectrum.

\subsection{Loop-String-Hadron formulation
\label{sec:LSH}}
\noindent
An alternate reformulation of Kogut-Susskind Hamiltonian formalism in terms of Schwinger bosons, known as the prepotential formalism, has been developed over the past decade~\cite{Mathur:2004kr, Mathur:2007nu,  Anishetty:2014tta, Raychowdhury:2014eta, Raychowdhury:2018tfj, Anishetty:2009nh,Raychowdhury:2013rwa,Anishetty:2009ai, Mathur:2010wc}. In a recent work~\cite{Raychowdhury:2019iki}, the prepotential formalism of the SU(2) LGT has been made complete to include staggered fermions, explicit Hamiltonian, and the associated Hilbert space. In this section, the LSH formalism in 1+1~D will be introduced. Later on, we demonstrate the advantage of this formulation compared with the original KS theory within the angular-momentum basis in the physical sector, and with the purely fermionic and bosonic formulations.

Within the prepotential framework, the original canonical conjugate variables of the theory, i.e, electric-field and link operators are replaced by a set of harmonic-oscillator doublets, defined at each end of a link as shown in Fig.~\ref{fig:latticelsh}. Both the electric-field as well as the link operators can be re-expressed in terms of Schwinger bosons to satisfy all properties of these variables spelled out in section \ref{sec:KS}. However, the most important feature of the prepotential formalism is that the link operators $U$, originally defined over a link connecting neighboring sites $(x,x+1)$, are now split into a product of two parts:
\begin{eqnarray}
U(x)=U_L(x)U_R(x),
\end{eqnarray}
where $U_{L(R)}$ is  left (right) part of the link attached to site $x~(x+1)$.  As a result of this decomposition, the gauge group is now totally confined to each lattice site, which allows one to define gauge-invariant operators and states locally. For the pure gauge theory, these local gauge-invariant operators and states can be interpreted as local snapshots of Wilson-loop operators of the original gauge theory. One can now construct a local loop Hilbert space by the action of local loop operators on the strong-coupling vacuum defined locally at each site. At this point, it must be emphasized that mapping the local loop picture to the original loop description of the gauge theory requires one extra constraint on each link. This constraint demands that the states must satisfy
\begin{eqnarray}
\label{eq:AGL}
N_L(x)=N_R(x),
\end{eqnarray}
where $N_{L(R)}$ counts the total number of Schwinger bosons residing at the left (right) end of a link connecting sites $x$ and $x+1$.\footnote{In the notation of Ref.~\cite{Raychowdhury:2019iki}, these are indicated as $\mathcal{N}_{L(R)}$.} This constraint is a consequence of the relation $\bm{E}_L^2=\bm{E}_R^2$ on the link and is equivalent to the constraint $J_L=J_R$ in the angular-momentum basis. 

The inclusion of the staggered fermionic matter in the SU(2) LGT is straightforward, and combines smoothly with the local loop description obtained in the prepotential framework. The reason is that both the prepotential Schwinger bosons and the matter fields associated with a given site transform in the fundamental representation of the local SU(2). One can now combine matter and prepotential to construct local string operators, besides local loop operators. Acting on the strong-coupling vacuum, these build a larger local gauge-invariant Hilbert space, including string and `hadron' states. This complete description is named the LSH formalism in Ref.~\cite{Raychowdhury:2019iki}. The LSH formalism is briefly described in the following, focusing on necessary steps for working with this formalism in one spatial dimension.
\begin{figure*}[t!]
\includegraphics[width=0.815\textwidth]{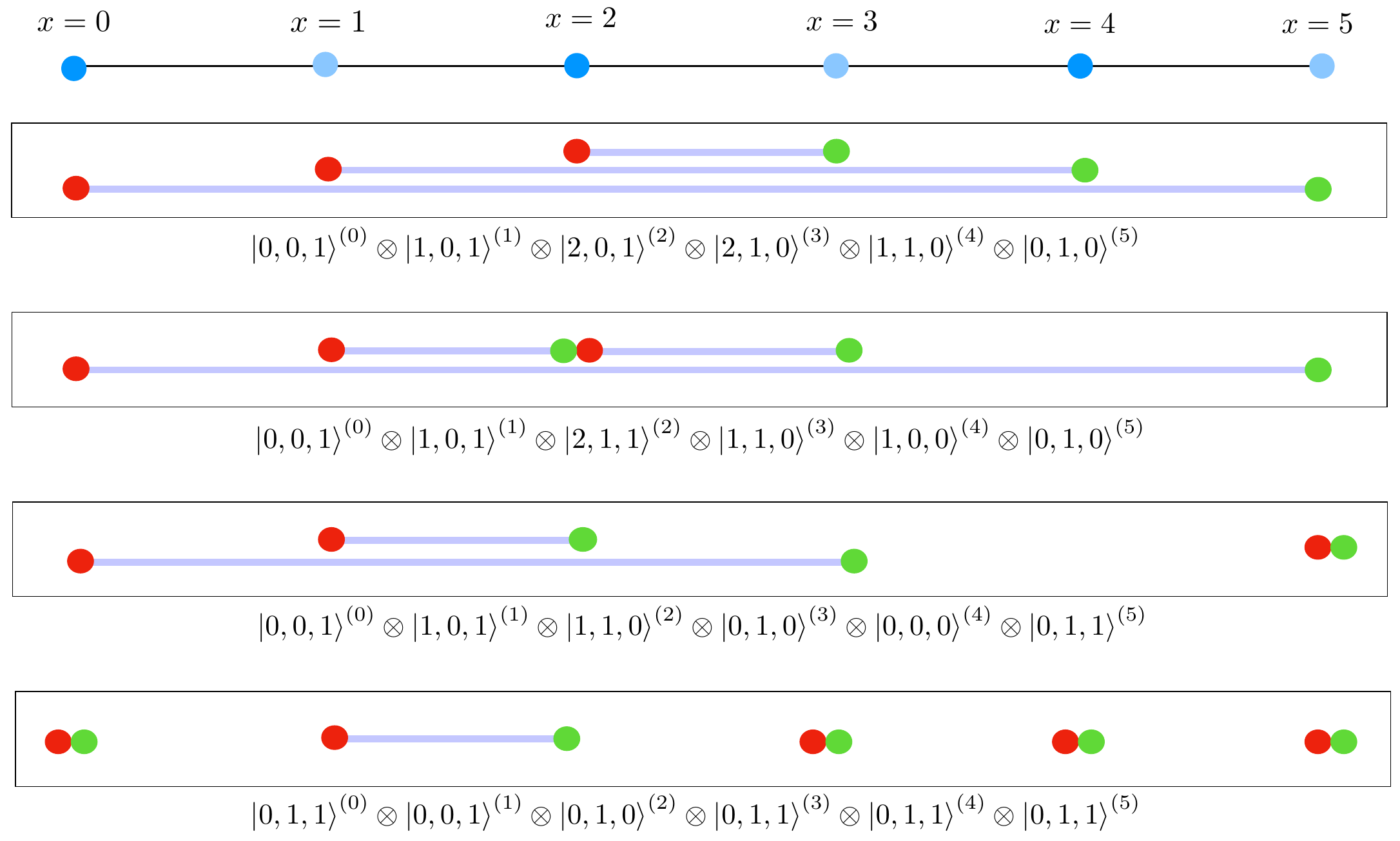}
\caption{The LSH states are illustrated on a one-dimensional lattice with $N=6$, where $N$ denotes the number of staggered sites. Here, a gauge-invariant state is characterized by $\ket{\phi}_{\rm (LSH)}=\prod_{x=0}^{N-1}\ket{n_l,n_i,n_o}^{(x)}$, with loop quantum numbers, $n_l$, and string quantum numbers: $n_i$ for an incoming string and $n_o$ for an outgoing string. Any state may consist of any number of open or closed strings (depending on the boundary conditions as well as the cutoff on the boson excitations), and any number of hadrons up to $N$. A sample of LSH states are shown on this lattice, denoted by a set of strings and/or hadrons. The number of solid thick lines passing through a site denotes the loop quantum number $n_l$. The number of solid disks (red or green) at a site from (to) which a thick line starts (ends) denotes outgoing (incoming) string quantum number $n_o$ $(n_i)$. If two strings start and end at the same site, it is equivalent to a hadron sitting on top of a loop, and is denoted as $|n_l=1,n_i=1,n_o=1\rangle$. A single hadron present at a site is denoted by $|n_l=0,n_i=1,n_o=1\rangle$. 
}
\label{fig:LSH_sample}
\end{figure*}

Within the LSH framework, a gauge-invariant and orthonormal basis is chosen, that is defined locally at each site and is characterized by a set of three integers:
\begin{equation}
n_l(x), n_i(x), n_o(x),
\end{equation}
for all $x$. The three quantum numbers signify loop, incoming string, and outgoing string at each site, respectively.\footnote{Note that the string quantum numbers were named `quark' quantum numbers in Ref.~\cite{Raychowdhury:2019iki} to remove the ambiguity associated with the absence of any string when a hadron is present at the site, see Fig.~\ref{fig:LSH_sample}. $n_i$ and $n_o$ will be called string quantum numbers throughout but a state with $n_i=n_o=1$ and $n_l=0$ should be understood as a state with no string starting and ending at the `quarks'.}  The allowed values of these integers are given by 
\begin{eqnarray}
&& 0\le n_l(x)\le \infty, \\
&& n_i(x) \in \{0,1\},\\
&& n_o(x) \in \{0,1\}.
\end{eqnarray}
It is clear from the range of the quantum numbers that $n_l$ is bosonic, whereas $n_i$ and $n_o$ are fermionic in nature. In terms of the LSH formalism, the operator building the local string Hilbert space consists of SU(2)-invariant bilinears of one bosonic prepotential operator and one fermionic matter field, yielding overall fermionic statistics, whereas the local loop Hilbert space is constructed by the action of SU(2)-invariant bilinears of two bosonic prepotential operators. Such operators will not be introduced in detail, instead the Hamiltonian will be written shortly in this operator basis. Characterization of gauge-invariant states on a one-dimensional lattice consisting of six staggered sites in terms of the three quantum numbers is illustrated in Fig.~\ref{fig:LSH_sample} via a few examples covering any situation that can occur within this theory.

Let us define a set of LSH operators consisting of diagonal and ladder operators locally at each site as following:\footnote{In the notation of Ref.~\cite{Raychowdhury:2019iki}, $\hat n_l$, $\hat n_i$, and $\hat n_o$ are indicated as $\hat{\mathcal{N}}_l$, $\hat{\mathcal{N}}_i$, and $\hat{\mathcal{N}}_o$, respectively. Further, in that reference $\hat \lambda^{\pm}$ is indicated as $\hat \Lambda^{\pm}$.}
\begin{eqnarray}
\label{nl}
\hat n_l|n_l, n_i, n_o\rangle &=& n_l|n_l, n_i, n_o\rangle, \\
\label{ni}
\hat n_i|n_l, n_i, n_o\rangle &=& n_i|n_l, n_i, n_o\rangle, \\
\label{no}
\hat n_o|n_l, n_i, n_o\rangle &=& n_o|n_l, n_i, n_o\rangle, 
\end{eqnarray}
\begin{eqnarray}
\label{nlpm}
\hat \lambda^{\pm}|n_l, n_i, n_o\rangle &=& |n_l\pm 1, n_i, n_o\rangle, \\
\label{nip}
\hat \chi_i^{+}|n_l, n_i, n_o\rangle &=& (1-\delta_{n_i,1}) |n_l, n_i+ 1, n_o\rangle, \\
\label{nim}
\hat \chi_i^{-}|n_l, n_i, n_o\rangle &=& (1-\delta_{n_i,0}) |n_l, n_i- 1, n_o\rangle, \\
\label{nop}
\hat \chi_o^{+}|n_l, n_i, n_o\rangle &=& (1-\delta_{n_0,1}) |n_l, n_i, n_o+ 1\rangle,\\
\label{nom}
\hat \chi_o^{-}|n_l, n_i, n_o\rangle &=& (1-\delta_{n_0,0}) |n_l, n_i, n_o- 1\rangle. 
\end{eqnarray}
Here, the site index $x$ is made implicit for brevity. Being an SU(2) gauge-invariant basis, one \textit{no longer needs to satisfy the SU(2) Gauss's laws} at each site. However, the neighboring sites still need to be glued together by the Abelian Gauss's law, i.e., Eq.~(\ref{eq:AGL}). In terms of the LSH operators, the Abelian Gauss's law reads as:
\begin{eqnarray}
\label{eq:AGL_LSH}
&&\hat n_l(x)+\hat n_o(x)(1-\hat n_i(x))\nonumber \\
&&\hspace{1 cm}= \hat n_l(x+1)+\hat n_i(x+1)(1-\hat n_o(x+1)).
\end{eqnarray}
Upon acting on the LSH basis states and comparing with Eq.~(\ref{eq:AGL}), one obtains:
\begin{eqnarray}
N_L(x)&=& n_l(x)+ n_o(x)(1- n_i(x)) \label{NL},\\
N_R(x)&=& n_l(x+1)+n_i(x+1)(1-n_o(x+1)),
\label{eq:NR}
\end{eqnarray}
where, $N_L(x)$ and $N_R(x)$ count bosonic occupation numbers at each end of the link connecting site $x$ and $x+1$. Pictorially, the left and right sides of Eq.~(\ref{eq:AGL}) are represented in Fig.~\ref{fig:LSH_sample} by the number of solid lines on the left and right side of each link, respectively. Another important relation is
\begin{eqnarray}
\hat{N}_{\psi}(x) \equiv \psi^\dagger(x) \psi(x) = \hat{n}_i(x)+\hat{n}_o(x),
\end{eqnarray}
which establishes the relation between the $\{f_1,f_2\}$ quantum numbers of the original formulation and the string quantum number of the LSH formulation.

The Hamiltonian of the SU(2) LGT coupled to matter in the LSH formulation can be written in terms of the LSH operators and is given by:
\begin{eqnarray}
H^{(\rm LSH)}=H^{(\rm LSH)}_I+H^{(\rm LSH)}_E+H^{(\rm LSH)}_M.
\end{eqnarray}
Here, $H^{(\rm LSH)}_I$ is the matter-gauge interaction term, $H^{(\rm LSH)}_E$ is the electric-energy term, and $H^{(\rm LSH)}_M$ is the mass term. Explicitly, in terms of the LSH operators defined in Eqs.~(\ref{nl})-(\ref{nom}), each part of the Hamiltonian can be written as~\cite{Raychowdhury:2019iki}:
\begin{eqnarray}
\label{HILSH}
H^{(\rm LSH)}_I &=&\frac{1}{2a}\sum_n \left\{ \frac{1}{\sqrt{\hat n_l(x)+\hat n_o(x)(1-\hat n_i(x))+1}}\right . \nonumber\\
&\times& \Big[
\hat S_o^{++}(x) \hat S_i^{+-}(x+1)
+\hat S_o^{+-}(x) \hat S_i^{--}(x+1)
\Big] \nonumber\\
&&\hspace{-1.5 cm} \left . \times 
\frac{1}{\sqrt{ \hat n_l(x+1)+\hat n_i(x+1)(1-\hat n_o(x+1))+1}} + {\rm h.c.} \right\},
\nonumber\\
\\
%
%
\label{HELSH}
H^{(\rm LSH)}_E&=&\frac{g^2a}{2}\sum_n\Bigg[ \frac{\hat n_l(x)+\hat n_o(x)(1-\hat n_i(x))}{2}\nonumber \\
&& \times \left( \frac{\hat n_l(x)+\hat n_o(x)(1-\hat n_i(x))}{2}+1 \right) \Bigg],
\\
%
%
\label{HMLSH}
H^{(\rm LSH)}_M &=& m\sum_n (-1)^x(\hat n_i(x)+\hat n_o(x)),
\end{eqnarray}
where (\ref{HILSH}) contains the LSH ladder operators in the following combinations (suppressing the site indices):
\begin{eqnarray}
\hat S_o^{++}&=& \hat \chi_o^+ (\lambda^+)^{\hat n_i}\sqrt{\hat n_l+2-\hat n_i}, \\
\hat S_o^{--}&=& \hat \chi_o^- (\lambda^-)^{\hat n_i}\sqrt{\hat n_l+2(1-\hat n_i)}, \\
\hat S_o^{+-}&=& \hat \chi_i^+ (\lambda^-)^{1-\hat n_o}\sqrt{\hat n_l+2\hat n_o}, \\
\hat S_o^{-+}&=& \hat \chi_i^- (\lambda^+)^{1-\hat n_o}\sqrt{\hat n_l+1+\hat n_o}, 
\end{eqnarray}
and
\begin{eqnarray}
\hat S_i^{+-}&=& \hat \chi_o^- (\lambda^+)^{1-\hat n_i}\sqrt{\hat n_l+1+\hat n_i}, \\
\hat S_i^{-+}&=& \hat \chi_o^+ (\lambda^-)^{1-\hat n_i}\sqrt{\hat n_l+2\hat n_i}, \\
\hat S_i^{--}&=& \hat \chi_i^- (\lambda^-)^{\hat n_o}\sqrt{\hat n_l+2(1-\hat n_o)}, \\
\hat S_i^{++}&=& \hat \chi_i^+ (\lambda^+)^{\hat n_o}\sqrt{\hat n_l+2-\hat n_o}.
\end{eqnarray}
The strong-coupling vacuum of the LSH Hamiltonian is given by
\begin{eqnarray}
n_l(x)&=& 0,~\text{for all $x$}, \nonumber \\
n_i(x)&=&0,~ n_o(x)=0,~\text{for $x$ even},\label{SCV_LSH} \\
n_i(x)&=&1,~ n_o(x)=1,~\text{for ~$x$ odd}.\nonumber
\end{eqnarray}
It is easy to verify that Eq.~(\ref{SCV_LSH}) satisfies the Abelian Gauss's law, Eq.~(\ref{eq:AGL_LSH}). 

This completes the introduction of the LSH formulation for the SU(2) LGT in 1+1~D. In later sections, the finite-dimensional Hilbert space of the theory will be constructed by imposing a cutoff on the $N_L$ and $N_R$ quantum numbers, and the associated cost of the classical simulation within this framework will be analyzed.

\section{Physical Hilbert-space analysis
\label{sec:Hilbert}}
\noindent
As introduced in the previous section, the naive basis states in the KS LGTs spans a Hilbert space that is predominantly unphysical. The physical sector corresponds to the zero eigenvalue of the Gauss's law operator in Eq.~(\ref{eq:Ga}). As mentioned before, in contrast to the U(1) LGT, in SU(2) LGT the Gauss's law is not a single algebraic constraint on the eigenvalues of the electric-field operator, but instead, it mixes states with different electric-field quantum numbers, and is therefore a non-diagonal constraint when expressed in the electric-field basis.\footnote{In $d>1+1~D$, another relevant basis is the magnetic-field basis, in which the magnetic Hamiltonian is diagonal. The Gauss's laws in such a basis remain non-diagonal conditions as well.} A major complexity in the Hamiltonian formulation of non-Abelian LGTs is to diagonalize the Gauss's law operator locally to form the physical Hilbert space, as otherwise the computation is prohibitively costly even in small systems.

A question worth addressing is how beneficial it is, from a computational perspective, to work with a formulation that solves the Gauss's law at the level of operators as opposed to states (e.g., the LSH formulation) compared with a formulation that sustains a simple mapping of the Hilbert space to operators in the Hamiltonian but requires solving the Gauss's laws for basis states subsequently (e.g., the KS formulation in the angular momentum representation). Such a cost analysis is presented in Sec.~\ref{sec:cost}, but it requires understanding and analyzing in more detail the steps involved in forming the physical Hilbert space in each formulation and the dimensionality of the Hilbert spaces involved. Another interesting question is how fast the dimensionality of the Hilbert space and its physical subsector grows as a function of the lattice size and the cutoff on the electric-field excitations in each of the formulations considered. Such questions are studied in various depth in this section for all the formulations introduced in Sec.~\ref{sec:formalism}, and briefly in Appendix~\ref{app:QLM} for the QLM.\footnote{In the following for the sake of brevity, the KS formulation in the angular momentum representation may be called KS formulation in short.}

\subsection{Gauge-invariant angular-momentum basis
\label{sec:HilbertKS}}
\begin{figure}[t!]
\includegraphics[width=0.480\textwidth]{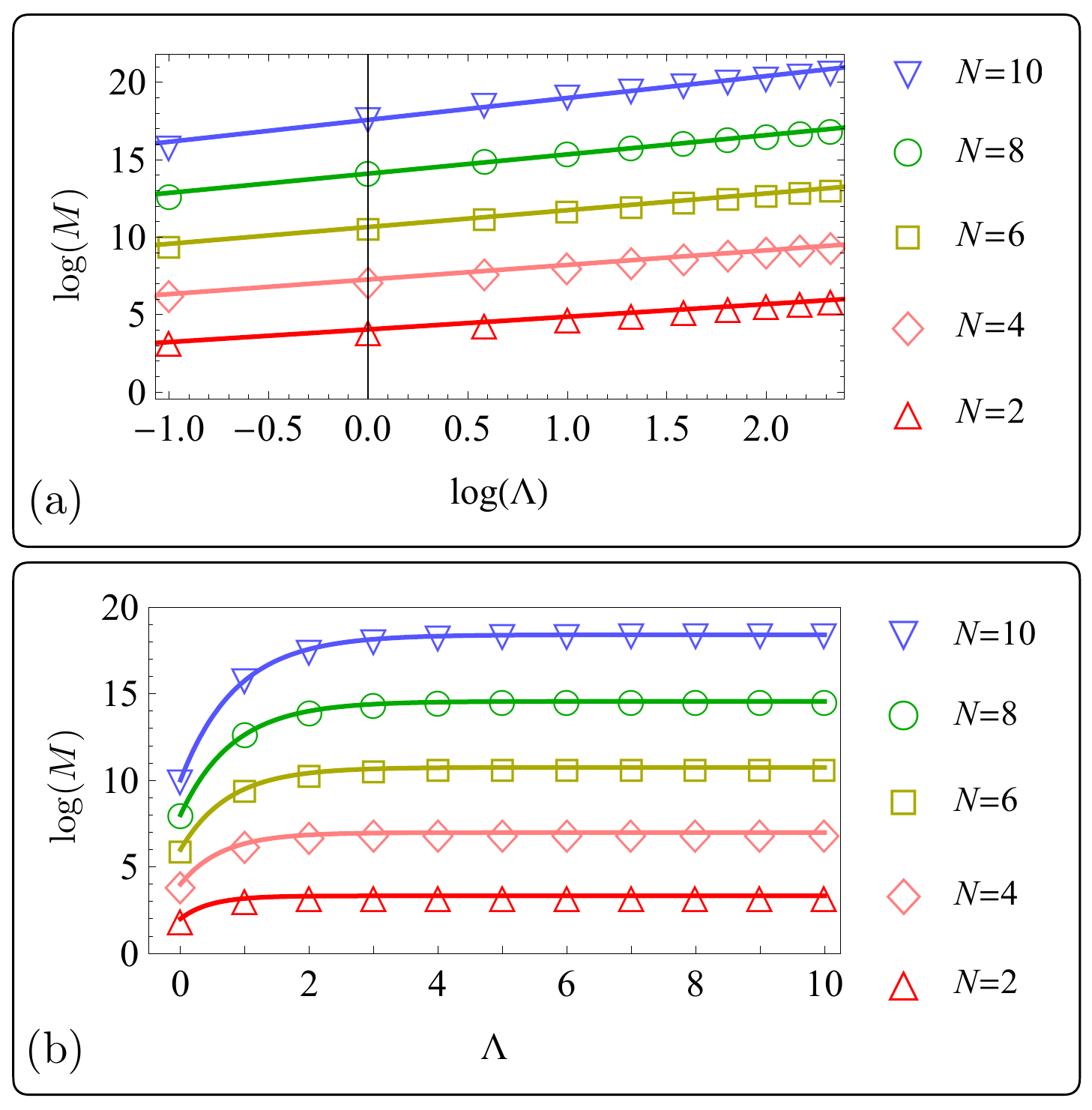}
\caption{
The dependence of the logarithm of the dimension of the physical Hilbert space, $M$, on (the logarithm of) the cutoff on the electric-field excitations, $\Lambda(=2J_{\rm max})$, within the KS (and LSH) formulation with PBC (a) and OBC (b), for various lattice sizes $N$. The lines are empirical fits to the points, with the fit values presented in Supplemental Material.
}
\label{fig:kspbcobclognloglambda}
\end{figure}
\begin{figure}[t!]
\includegraphics[width=0.480\textwidth]{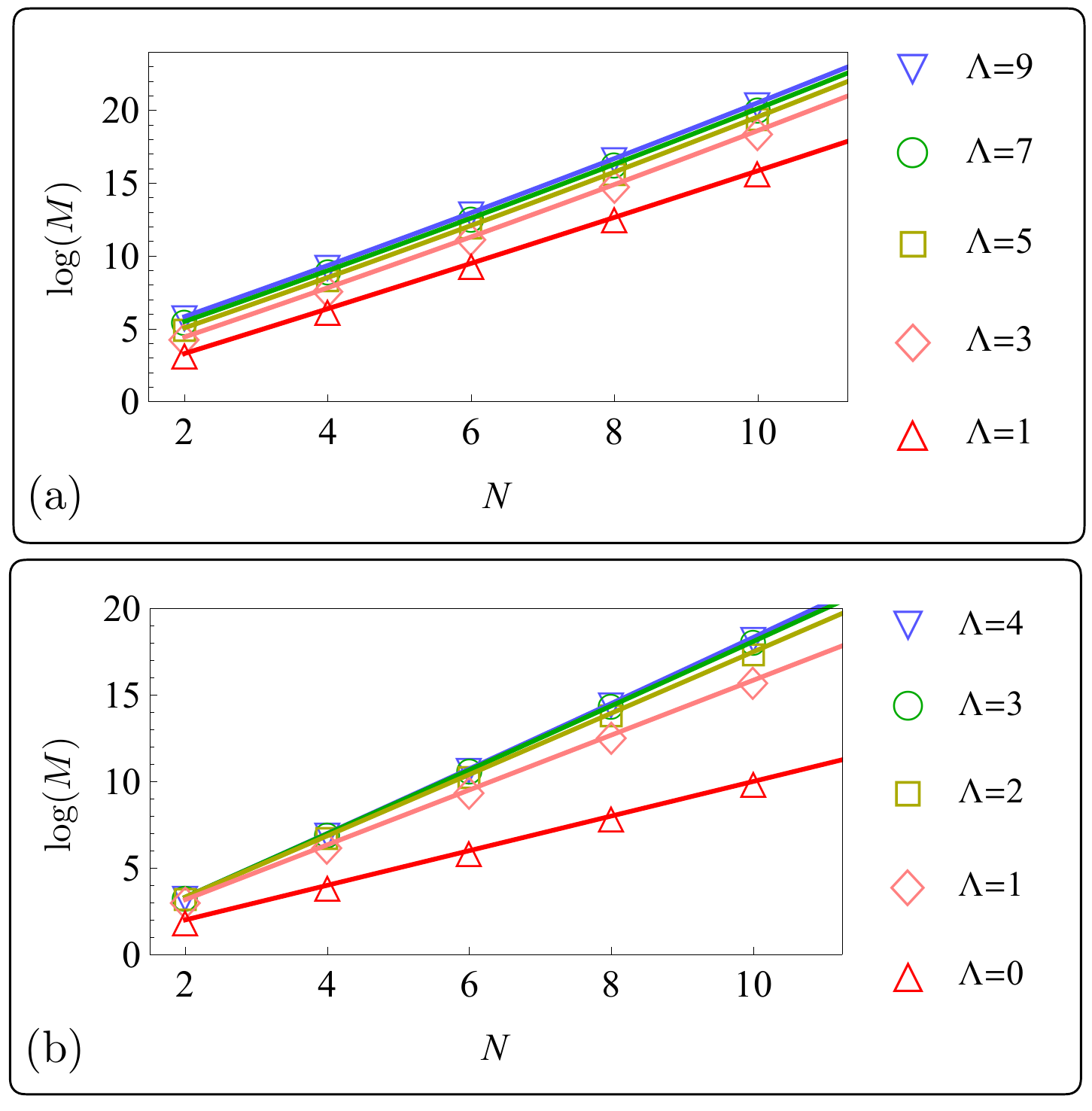}
\caption{
The dependence of the logarithm of the dimension of in the physical Hilbert space, $M$, on the lattice size $N$, within the KS (and LSH) formulation with PBC (a) and OBC (b), for various cutoffs on the electric-field excitations, $\Lambda(=2J_{\rm max})$. The lines are empirical fits to the points, with the fit values presented in Supplemental Material.
}
\label{fig:kspbcobclognn}
\end{figure}
\begin{figure}[!h]
\includegraphics[width=0.480\textwidth]{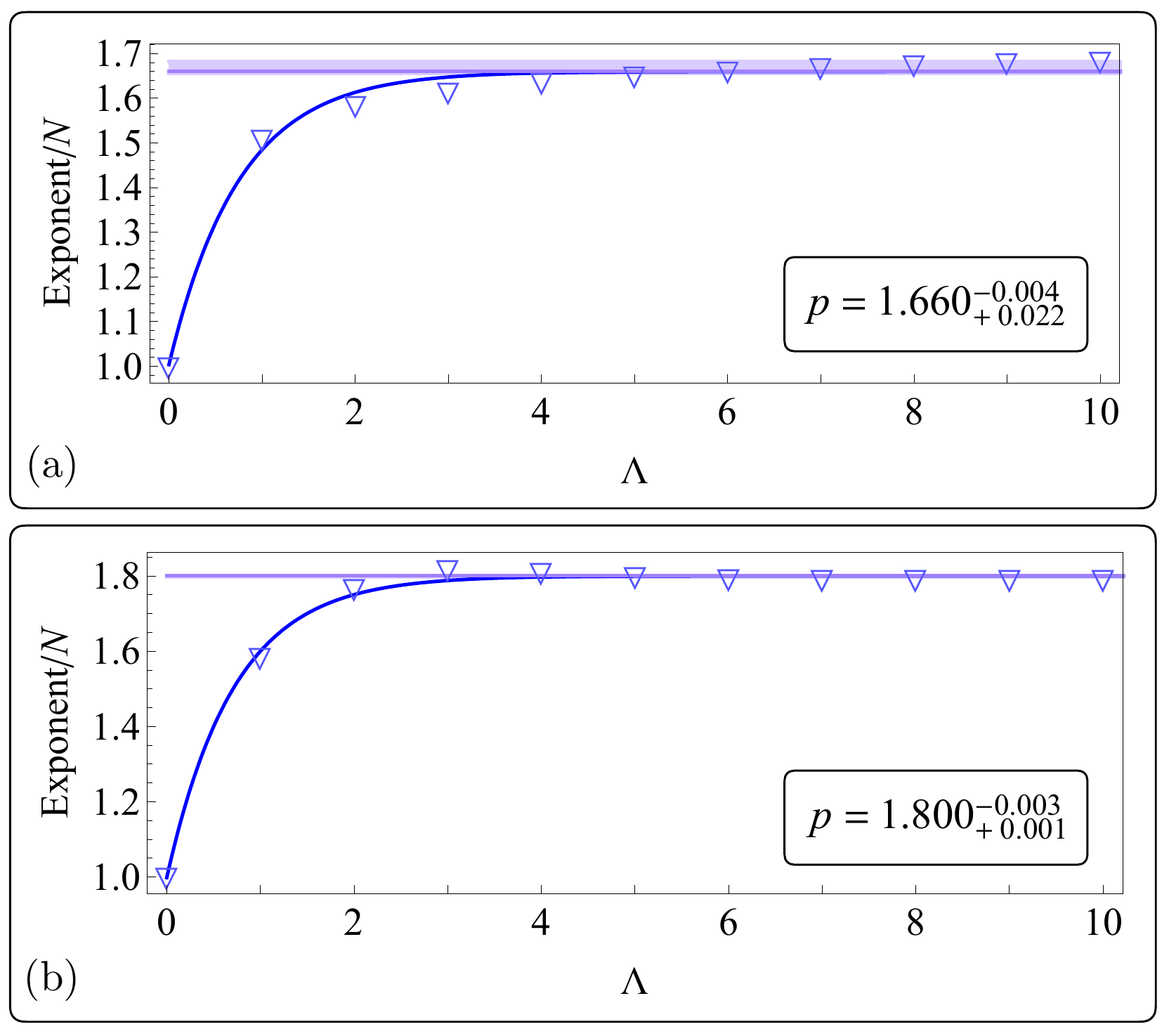}
\caption{
The dimension of the physical Hilbert space, $M$, within the KS (and LSH) formulation with PBC (a) and OBC (b) is approximated by $e^{p N}$, and the coefficient of the lattice size, $N$, in the exponent is obtained from fits to the $N$ dependence of $M$ for several values of $\Lambda$. The exponents approach, with an exponential form, a fixed value, and the empirical fit to this $\Lambda$ dependence obtains the asymptotic value of $p$ denoted by the horizontal lines in the plots and shown in the inset boxes. The uncertainty on these values is estimated by variations in the fit values when each data point is removed from the set, one at a time, and the remaining points are refit. The numerical values associated with these plots are listed in Supplemental Material.
}
\label{fig:kspbcobcexp}
\end{figure}

Despite significant state reduction after imposing physical constraints on the full basis states, without a finite cutoff on the electric-field quantum numbers, the physical Hilbert space will still be infinite-dimensional. In the remainder of this paper, we impose: $J_L,J_R \leq J_{\rm max}$, and denote the cutoff as $\Lambda=2J_{\rm max}$. Examining the dependence of the dimension of the Hilbert space, as well as that of observables, on this cutoff is one of the objectives of this work. After the imposition of this cutoff, the states in the physical Hilbert space can be obtained following the procedure outlined in Sec.~\ref{sec:KS} given the boundary conditions specified. For PBC, the $J_R$ value at site $x=0$ and the $J_L$ value at site $N-1$ are set equal. For OBC, the $J_R$ value at site $x=0$ is set to a constant $\epsilon_0$ smaller than the cutoff $J_{\max}$, while $J_L$ value at site $N-1$ is left free as long as it does not exceed the cutoff and that the Gauss's law at site $N-1$ is satisfied. In the rest of the paper, $\epsilon_0$ is set to zero, but the conclusions drawn can be extended to other values of this incoming `flux'.

The dimension of the physical Hilbert space, called $M$ throughout, for lattices of the size up to $N=10$ and cutoffs up to $\Lambda=10$ is provided in Tables \ref{tab:countsPBC} and \ref{tab:countsOBC} of Appendix~\ref{app:details} for PBC and OBC, respectively. There are a few interesting features to observe:
\begin{itemize}
\item[$\triangleright$]{For PBC, asymptotically the dimension of the physical Hilbert space grows linearly as a function of the cutoff for all $x$. This feature is evident from the plot of $M$ as a function of $\Lambda$ for various $N$ as shown in Fig.~\ref{fig:kspbcobclognloglambda}(a). This is a consequence of the observation that as $\Lambda$ increases, the number of new allowed states quickly saturates, i.e., introducing an additional possibility for the $J_{R,L}$ quantum numbers amounts to only adding ${2N}\choose{N}$ possible states. For $N=2,4,6,8,10$, the cutoff $\Lambda$ at which the growth of states become linear afterwards is $0,1,2,3,4$, respectively. The best empirical fits to this linear dependence are shown in the plot. Second, as expected, the dimension of the physical Hilbert space grows exponentially with the system's size at a fixed cutoff, as plotted in Fig.~\ref{fig:kspbcobclognn}(a). The growth, up to constant factors and higher-order terms in the exponent, can be approximated by $M \sim e^{pN}$. The coefficient of $N$ in the exponent approaches a constant value as a function of cutoff, as shown in Fig.~\ref{fig:kspbcobcexp}(a). This value can be obtained from a fit to points shown in the plot, as depicted in the figure. For moderate $N$ values such that the higher-order terms in the exponent are negligible, this $p$ value can be used to approximate the dimension of the physical Hilbert space with PBC as $\Lambda \to \infty$.}
\item[$\triangleright$]{For OBC, the dimension of the physical Hilbert space grows as a function of $\Lambda$ until it becomes a constant for $\Lambda \geq N$ ($\Lambda \geq N+2\epsilon_0$ for an arbitrary $\epsilon_0$), as depicted in Fig.~\ref{fig:kspbcobclognloglambda}(b). The reason for this behavior is that the $J$ quantum number only changes (by $\frac{1}{2}$) from the left to the right side of site $x$ if the site's total fermionic occupation number is equal to one. If the $J_R$ value at site $x=0$ is set to $\epsilon_0$, it can become at most $J_L=\epsilon_0+N/2$ at the last site. Increasing the cutoff beyond this value will not change the states present in the physical Hilbert space. This growth of the dimension of the physical Hilbert space to this saturation value at a fixed $N$ can be approximated by an exponential form, $M \sim e^{q \Lambda}$. The coefficient of $\Lambda$ in the exponent for various values of $N$ is plotted in Fig.~\ref{fig:ksobcexpdivlambda}, and is seen to asymptote to a constant value at large $N$. The fit to this asymptotic value is shown in the plot. This value can be used to approximate the dimension of the physical Hilbert space for an arbitrary large $N$ and any $\Lambda$. Similarly, the dependence of the dimension of the physical Hilbert space on the lattice size can be approximated by an exponential form, $M \sim e^{pN}$, for a fixed cutoff, and up to constant factors and higher order terms in the exponent. The coefficient of $N$ in the exponent asymptotes to a constant value at large $\Lambda$, as shown in Fig.~\ref{fig:kspbcobcexp}(b).}
\begin{figure}[!t]
\includegraphics[width=0.480\textwidth]{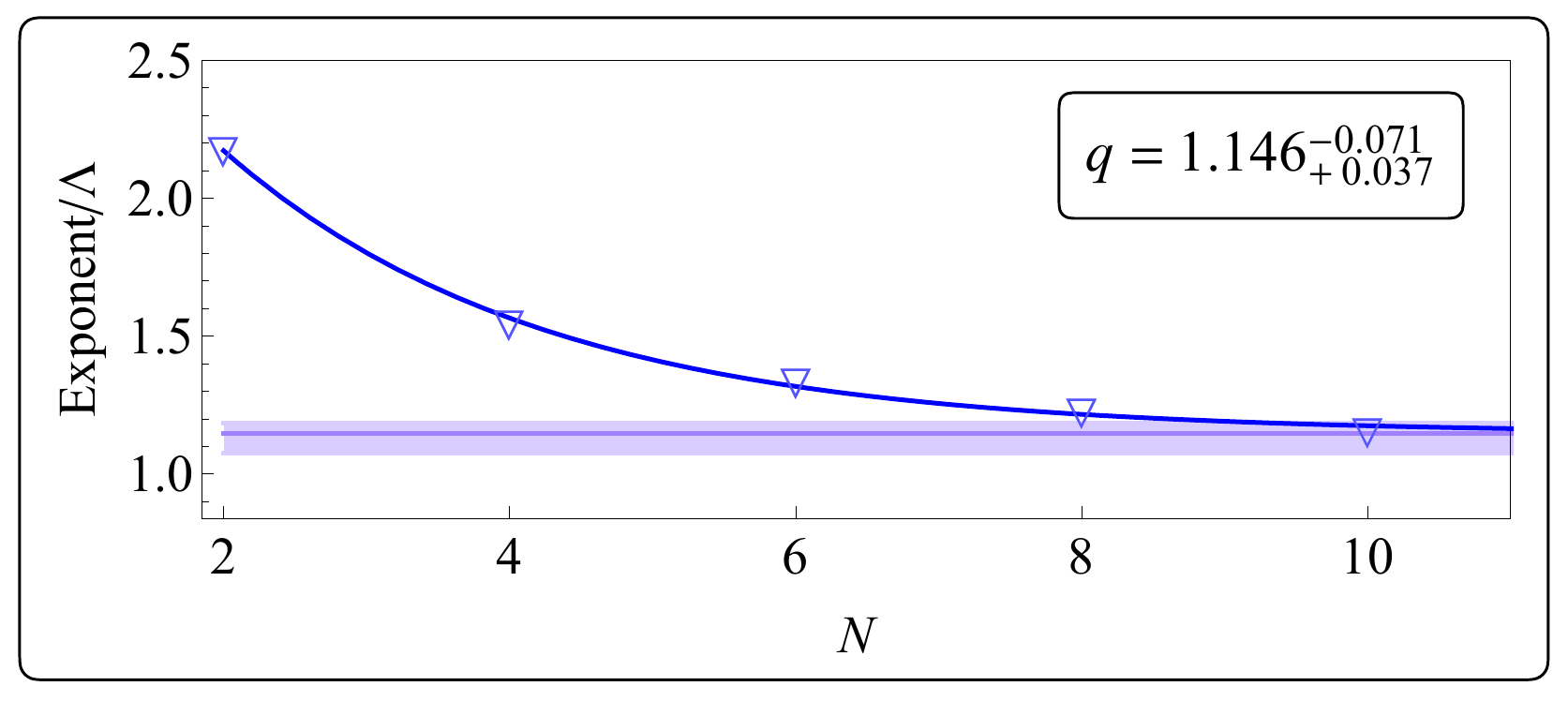}
\caption{
The dimension of the physical Hilbert space, $M$, within the KS (and LSH) formulation with OBC is approximated by $e^{q \Lambda}$, and the coefficient of the cutoff on the electric-field excitations, $\Lambda(=2J_{\rm max})$, in the exponent is obtained from fits to the $\Lambda$ dependence of $M$ for several values of $N$. The exponents approach, with an exponential form, a fixed value, and the empirical fit to this $N$ dependence obtains the asymptotic value of $q$ denoted by the horizontal line in the plot and shown in the inset box. The uncertainty on this value is estimated by variations in the fit values when each data point is removed from the set, one at a time, and the remaining points are refit. The numerical values associated with these plots are listed in Supplemental Material.
}
\label{fig:ksobcexpdivlambda}
\end{figure}

\item[$\triangleright$]{The size of the full Hilbert space before implementing physical constraints can be approximated by\footnote{Considering the Abelian Gauss's law that allows assigning only one $J$ quantum number to each link.} 
\begin{eqnarray}
M^{({\rm full})}(N,\Lambda)=\left[4\times\sum_j\left(2j+1\right)^2\right]^{N},
\label{eq:Nstatesfull}
\end{eqnarray}
with PBC, where $j=\{0,\frac{1}{2},1,\cdots,\frac{\Lambda}{2}\}$. To compare this with the dimension of the physical Hilbert space with PBC, one can again write the lattice-size dependence of the $M$ as $e^{p N}$. The coefficient of $N$ in this exponent as a function of $\Lambda$ can be plotted for both the full and physical Hilbert space, as is shown in Fig.~\ref{fig:kspbcexpfullvsphys}. As is evident, even for small values of the cutoff, the full Hilbert space grows much faster with the system's size than the physical Hilbert space. For example, with $\Lambda = 5$, the $p$ values differ by $\approx 7$. This means that for a lattice size $N=10$, for example, the dimension of the full Hilbert space is $\approx 30$ orders of magnitude larger than that of the physical Hilbert space. As a result, it is not plausible to perform a classical Hamiltonian simulation with a manageable cost if the physical constraints are not imposed \emph{a priori}. Implementing the physical constraints, nonetheless, introduces further complexity at the onset of the calculation and amounts to an additional preprocessing cost. We will come back to this point when comparing the simulation cost between the KS and LSH formulations in Sec.~\ref{sec:cost}.
}
\begin{figure}[!t]
\includegraphics[width=0.480\textwidth]{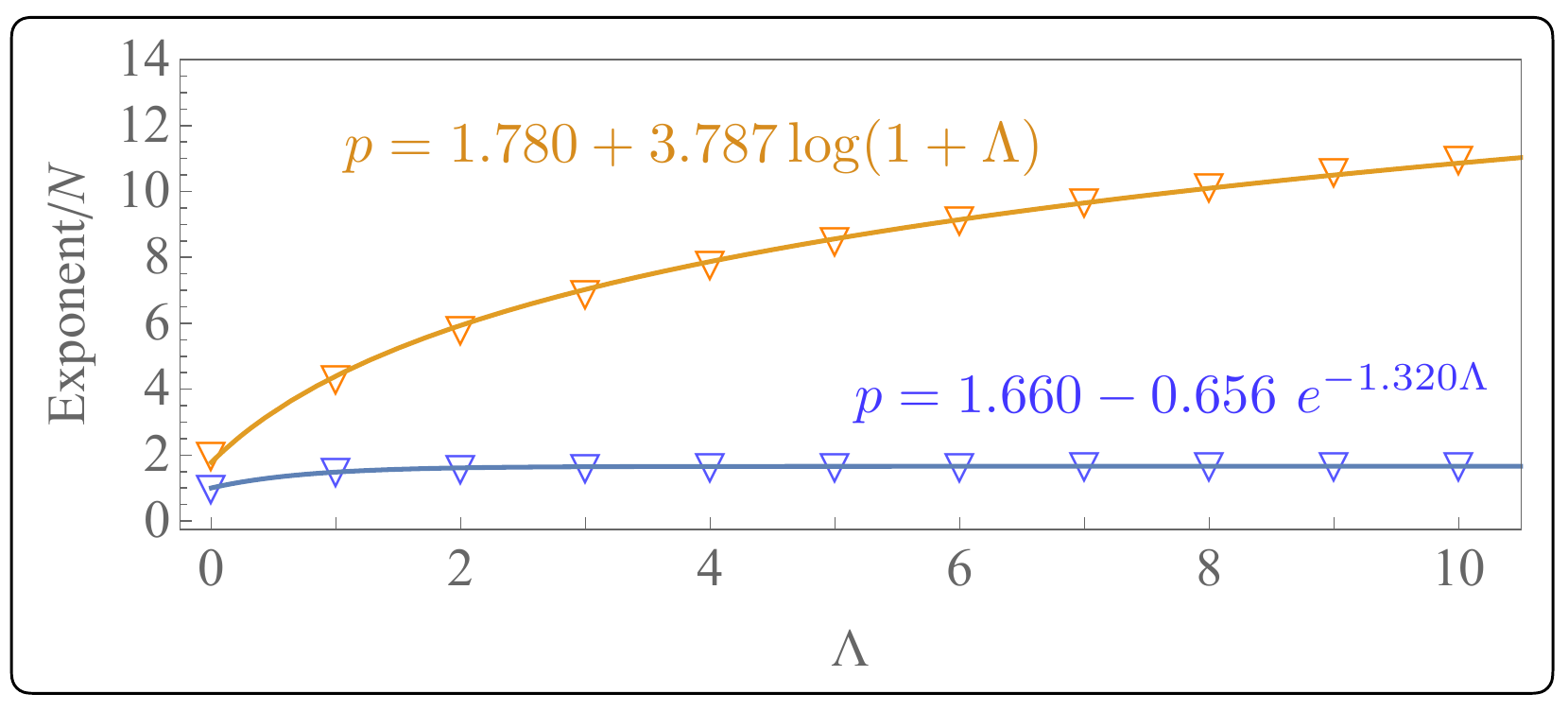}
\caption{
Shown in blue is the same as in Fig.~\ref{fig:kspbcobcexp}-a, i.e., the coefficient of the lattice size, $N$, in the exponent of $M \sim e^{pN}$ for several values of $\Lambda$, where $M$ denotes the dimension of the physical Hilbert space within the KS (and LSH) formulation with PBC. The same quantity can be plotted for the dimension of the full Hilbert space, as shown in orange, along with an empirical functional form obtained from a fit to the points. The numerical values associated with these plots are listed in Supplemental Material.
}
\label{fig:kspbcexpfullvsphys}
\end{figure}
\item[$\triangleright$]{A physical basis state is generally a superposition of the original angular-momentum basis states, see e.g., the example in Eqs.~(\ref{eq:KSN2nu2}) below. The number of terms in each superposition can become exponentially large in system's size. This creates significant complexity when generating the Hamiltonian matrix, due to the need to keep track of the Hamiltonian action on each constituent basis state. The maximum number of terms in a physical state is plotted in Fig.~\ref{fig:kspbcnumterms} as a function of $\Lambda$ for PBC, demonstrating this exponential growth. We will come back to this feature in Sec.~\ref{sec:cost} when analyzing the computational cost of the Hamiltonian simulation.}
\end{itemize}
%

\subsection{Purely fermionic formulation
\label{sec:HilbertF}}
\begin{figure}[!t]
\includegraphics[width=0.480\textwidth]{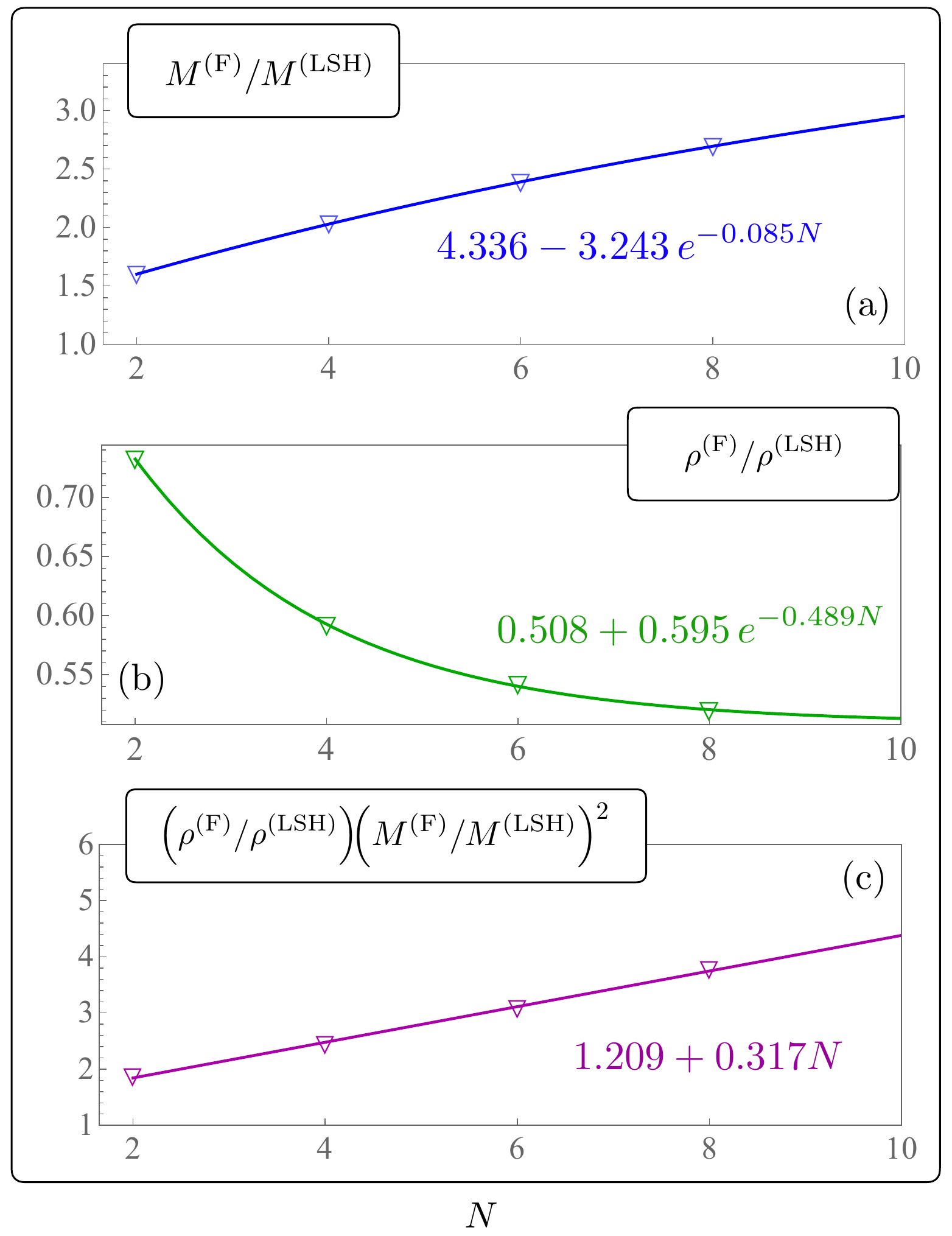}
\caption{
The upper panel depicts the ratio of the dimension of the Hilbert space in the fully fermionic formulation with OBC to the dimension of the physical Hilbert space within the KS (and LSH) formulation without removing the gauge DOF (but with a sufficiently large cutoff such that the dimension of the Hilbert space saturates to a fixed value), for several values of the lattice size, $N$. The middle panel depicts the density of the Hamiltonian matrix within the physical Hilbert space for each theory. The lower panel is the ratio of the Hamiltonian matrix densities multiplied by the square of the ratio of the size of the Hilbert spaces in each theory. This latter quantity enters the analysis of the computational complexity of matrix manipulation in Sec.~\ref{sec:cost}. The numerical values associated with these plots are listed in Supplemental Material.
}
\label{fig:allksvsfobc}
\end{figure}

As discussed in Sec.~\ref{sec:F}, the basis states that represent the Hilbert space of the purely fermionic representation of the KS formulation with OBC consist of the direct product of on-site fermionic states, see Eq.~(\ref{eq:Fbasis}), giving rise to $M=4^N$ basis states, where $N$ denotes the size of the lattice in 1+1 D as before. The dimension of the Hilbert space of the fermionic theory is larger than the dimension of the physical Hilbert space of the KS formulation in the angular momentum (and LSH) basis for cutoff values that allow the full physical Hilbert space to be constructed with OBC (i.e., $\Lambda \geq N+2\epsilon_0$). The ratio of the former to the latter is shown in Fig.~\ref{fig:allksvsfobc} for various $N$, along with an empirical fit form to the ratio as a function of the lattice size. This form shows that the ratio of the dimensions of the two Hilbert spaces asymptotes slowly to a fixed number.

To understand this mismatch between the number of (physical) states in both formulations, despite the fact the fermionic formulation is constructed to fully represent the physical Hilbert space, inspecting the following example will be illuminating. Consider the $N=2$ theory in the $\nu=1$ sector, where $\nu$ denotes the normalized fermion occupation number on the lattice defined in Eq.~(\ref{eq:nu}) below. In the KS formulation in the angular-momentum basis, the only four physical basis states are:\footnote{Such states are constructed efficiently in Ref.~\cite{Banuls:2017ena} by acting by the interacting Hamiltonian on the strong-coupling vacuum, i.e., state 1) shown, but they differ in relative signs with the states presented here. Nonetheless, only the signs denoted here give rise to gauge-invariant states as can be checked by acting by the Gauss's law operators in Eq.~(\ref{eq:Ga}) on the states shown.}
\begin{figure}[t!]
\includegraphics[width=0.480\textwidth]{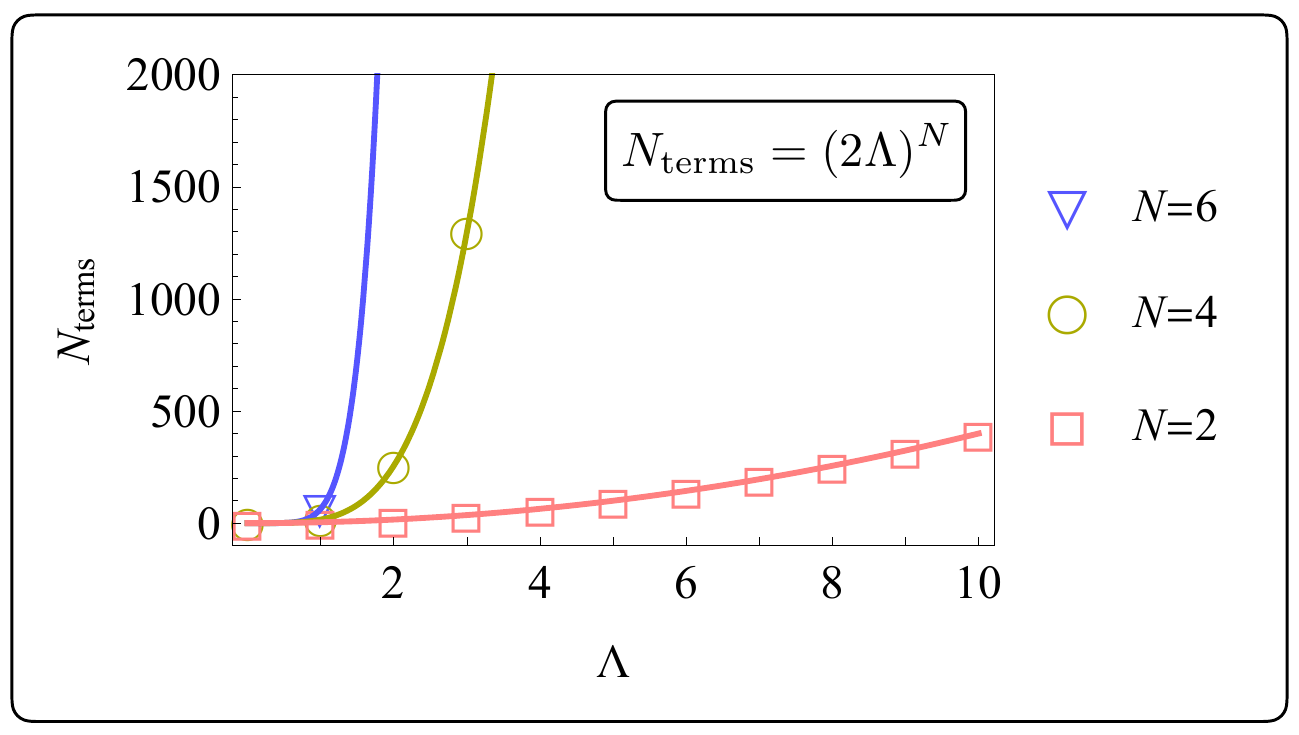}
\caption{Each physical state in the KS formulation in the angular-momentum basis is, in general, a linear combination of multiple basis states in Eq.~(\ref{eq:KSbasis}). Shown is the number of terms in the longest linear combination formed to represent a physical state in the KS theory with PBC, as a function of the cutoff, $\Lambda$, for various number of lattice sizes $N$. This number grows polynomially with $\Lambda$ for a fixed $N$, while it grows exponentially with $N$ for a fixed $\Lambda$, with the form shown.
}
\label{fig:kspbcnumterms}
\end{figure}
\begin{flalign}
&1)~ \left[\ket{0,0}  \ket{0,0} \ket{0,0}\right]^{(0)} \otimes \left[\ket{0,0}  \ket{1,1} \ket{0,0}\right]^{(1)},
\nonumber
\end{flalign}
\begin{align}
&2)~
\frac{1}{2} \left[\ket{0,0}  \ket{1,0} \ket{\frac{1}{2},-\frac{1}{2}}\right]^{(0)} \otimes \left[\ket{\frac{1}{2},\frac{1}{2}}  \ket{0,1} \ket{0,0}\right]^{(1)}
\nonumber\\
&~ -\frac{1}{2}\left[\ket{0,0}  \ket{1,0} \ket{\frac{1}{2},-\frac{1}{2}}\right]^{(0)} \otimes \left[\ket{\frac{1}{2},-\frac{1}{2}}  \ket{1,0} \ket{0,0}\right]^{(1)}
\nonumber\\
&~ -\frac{1}{2} \left[\ket{0,0}  \ket{0,1} \ket{\frac{1}{2},\frac{1}{2}}\right]^{(0)} \otimes \left[\ket{\frac{1}{2},\frac{1}{2}}  \ket{0,1} \ket{0,0}\right]^{(1)}
\nonumber\\
&~ +\frac{1}{2} 
\left[\ket{0,0}  \ket{0,1} \ket{\frac{1}{2},\frac{1}{2}}\right]^{(0)} \otimes \left[\ket{\frac{1}{2},-\frac{1}{2}}  \ket{1,0} \ket{0,0}\right]^{(1)},
\nonumber
\end{align}
\begin{flalign}
&3)~ \frac{1}{\sqrt{6}}\left[ \ket{0, 0} \ket{1, 0} \ket{\frac{1}{2}, -\frac{1}{2}} \right]^{(0)} \otimes \left[\ket{\frac{1}{2}, \frac{1}{2}}\ket{1, 0} \ket{1, -1}\right]^{(1)}
\nonumber\\
&~ -\frac{1}{2 \sqrt{3}}\left[\ket{0, 0} \ket{1, 0} \ket{\frac{1}{2}, -\frac{1}{2}}\right]^{(0)} \otimes \left[ \ket{\frac{1}{2}, \frac{1}{2}} \ket{0,1} \ket{1, 0} \right]^{(1)} 
\nonumber\\
&~ -\frac{1}{2 \sqrt{3}}\left[\ket{0, 0} \ket{1, 0} \ket{\frac{1}{2}, -\frac{1}{2}}\right]^{(0)} \otimes \left[ \ket{\frac{1}{2}, -\frac{1}{2}} \ket{1, 0} \ket{1, 0} \right]^{(1)}
\nonumber\\
&~ +\frac{1}{\sqrt{6}} \left[ \ket{0, 0} \ket{1, 0} \ket{\frac{1}{2}, -\frac{1}{2}}\right]^{(0)} \otimes \left[\ket{\frac{1}{2}, -\frac{1}{2}} \ket{0, 1} \ket{1, 1}\right]^{(1)}
\nonumber\\
%
%
&~ -\frac{1}{\sqrt{6}} \left[\ket{0, 0} \ket{0, 1} \ket{\frac{1}{2},\frac{1}{2}} \right]^{(0)} \otimes \left[\ket{\frac{1}{2}, \frac{1}{2}} \ket{1, 0} \ket{1, -1} \right]^{(1)}
\nonumber\\
&~ +\frac{1}{2 \sqrt{3}} \left[\ket{0, 0} \ket{0, 1} \ket{\frac{1}{2}, \frac{1}{2}} \right]^{(0)} \otimes \left[ \ket{\frac{1}{2}, \frac{1}{2}} \ket{0, 1} \ket{1,0} \right]^{(1)}
\nonumber\\
&~ +\frac{1}{2 \sqrt{3}} \left[ \ket{0, 0} \ket{0, 1} \ket{\frac{1}{2}, \frac{1}{2}}\right]^{(0)} \otimes \left[\ket{\frac{1}{2}, -\frac{1}{2}} \ket{1,0} \ket{1, 0} \right]^{(1)}
\nonumber\\
&~ -\frac{1}{\sqrt{6}} \left[\ket{0, 0} \ket{0, 1} \ket{\frac{1}{2}, \frac{1}{2}} \right]^{(0)} \otimes \left[ \ket{\frac{1}{2}, -\frac{1}{2}} \ket{0,1} \ket{1, 1} \right]^{(1)},
\nonumber
\end{flalign}
\begin{flalign}
&4)~ \left[\ket{0,0}  \ket{1,1} \ket{0,0}\right]^{(0)} \otimes \left[\ket{0,0}  \ket{0,0} \ket{0,0}\right]^{(1)},
\label{eq:KSN2nu2}
\end{flalign}
where each triplet in the square brackets denotes $[\ket{J_R,m_R} \otimes \ket{f_1,f_2} \otimes \ket{J_L,m_L}]^{(x)}$ at the corresponding site $x$, and the direct product symbol is suppressed in such triplets for brevity. On the other hand, in the purely fermionic representation of the same theory, the six basis states are
\begin{eqnarray}
1)~\ket{0,0}^{(0)} \otimes \ket{1,1}^{(1)},
\nonumber\\
2)~\ket{0,1}^{(0)} \otimes \ket{0,1}^{(1)},
\nonumber\\
3)~\ket{0,1}^{(0)} \otimes \ket{1,0}^{(1)},
\nonumber\\
4)~\ket{1,0}^{(0)} \otimes \ket{0,1}^{(1)},
\nonumber\\
5)~\ket{1,0}^{(0)} \otimes \ket{1,0}^{(1)},
\nonumber\\
6)~\ket{1,1}^{(0)} \otimes \ket{0,0}^{(1)}.
\label{eq:exampleF}
\end{eqnarray}
As is seen, while all the six possible fermionic configurations in the $\nu=1$ sector are present in the physical basis states of the KS formulation in the angular-momentum basis, only two proper linear combinations of states 2)-5) in the fermionic formulation appear in the KS formulation in the angular-momentum basis. Further inspection of the two representations reveals that the spectrum of both theories matches exactly for all values of the couplings, but with degeneracies present in the fermionic case. To conclude, the fermionic representation of the SU(2) LGT in 1+1~D with OBC has redundancies in the representation compared with the KS formulation in the angular-momentum basis, however it avoids complex linear combinations of basis states that arise in the latter due to the imposition of the Gauss's laws. As will be discussed in Sec.~\ref{sec:HilbertLSH}, the LSH formulation of the SU(2) LGT is free from the redundancies of the fermionic formulation, while at the same time it does not involve a cumbersome physical Hilbert-space construction.

\subsection{Purely bosonic formulation
\label{sec:HilbertB}}
The physical basis states of the bosonized SU(2) theory with OBC are, at the first sight, the direct product of the physical basis states of the KS theory discussed in Sec.~\ref{sec:HilbertKS} and the electric-field basis states satisfying the extra U(1) Gauss's law. Recall that the U(1) symmetry was introduced in the bosonized form to allow the elimination of fermionic DOF in favor of bosonic DOE in the SU(2) theory. The statement above is only true if the cutoff on the U(1) electric field is set sufficiently high such that all fermionic configurations allowed in the physical Hilbert space of the SU(2) theory can be realized. To make this statement more explicit, consider the example studied in Sec.~\ref{sec:HilbertF}, where $N=2$ and $\nu=1$ in the KS theory with OBC, and the incoming fluxes of the SU(2) and U(1) electric fields are set to zero. As was shown, while there are six allowed fermionic configurations in the purely fermionic representation, these reduce to four linear combinations of basis states for $\Lambda \geq 2$ in the physical Hilbert space of the KS formulation in the angular-momentum/fermionic basis (still encompassing all six possible fermionic configurations with $\nu=1$). Note that with $\Lambda \geq 2$, the physical Hilbert space of the SU(2) theory is complete. Now consider the purely bosonic formulation, with $\Lambda_0$ denoting the cutoff on the U(1) electric-field excitations. Obviously for $\Lambda_0 = 0$, the only state allowed is the strong-coupling vacuum states, i.e., state 1) in Eq.~(\ref{eq:KSN2nu2}), and the physical Hilbert space of the bosonized U(2) theory has dimension one in the specified sector. For $\Lambda_0 = 1$, there are three states contributing, corresponding to states 1), 2), and 3) in Eq.~(\ref{eq:KSN2nu2}), times the U(1) electric-field states $\ket{E}^{(0)} \otimes \ket{E}^{(1)}=\ket{0}^{(0)} \otimes \ket{0}^{(1)}$ for state 1) and $\ket{1}^{(0)} \otimes \ket{0}^{(1)}$ for states 2) and 3). Finally, for $\Lambda_0 \geq 2$, all four states in Eq.~(\ref{eq:KSN2nu2}) are allowed, and the corresponding U(1) electric-field states are those given above for states 1), 2), and 3), and $\ket{2}^{(0)} \otimes \ket{0}^{(1)}$ for state 4).

\begin{figure}[t!]
\includegraphics[width=0.480\textwidth]{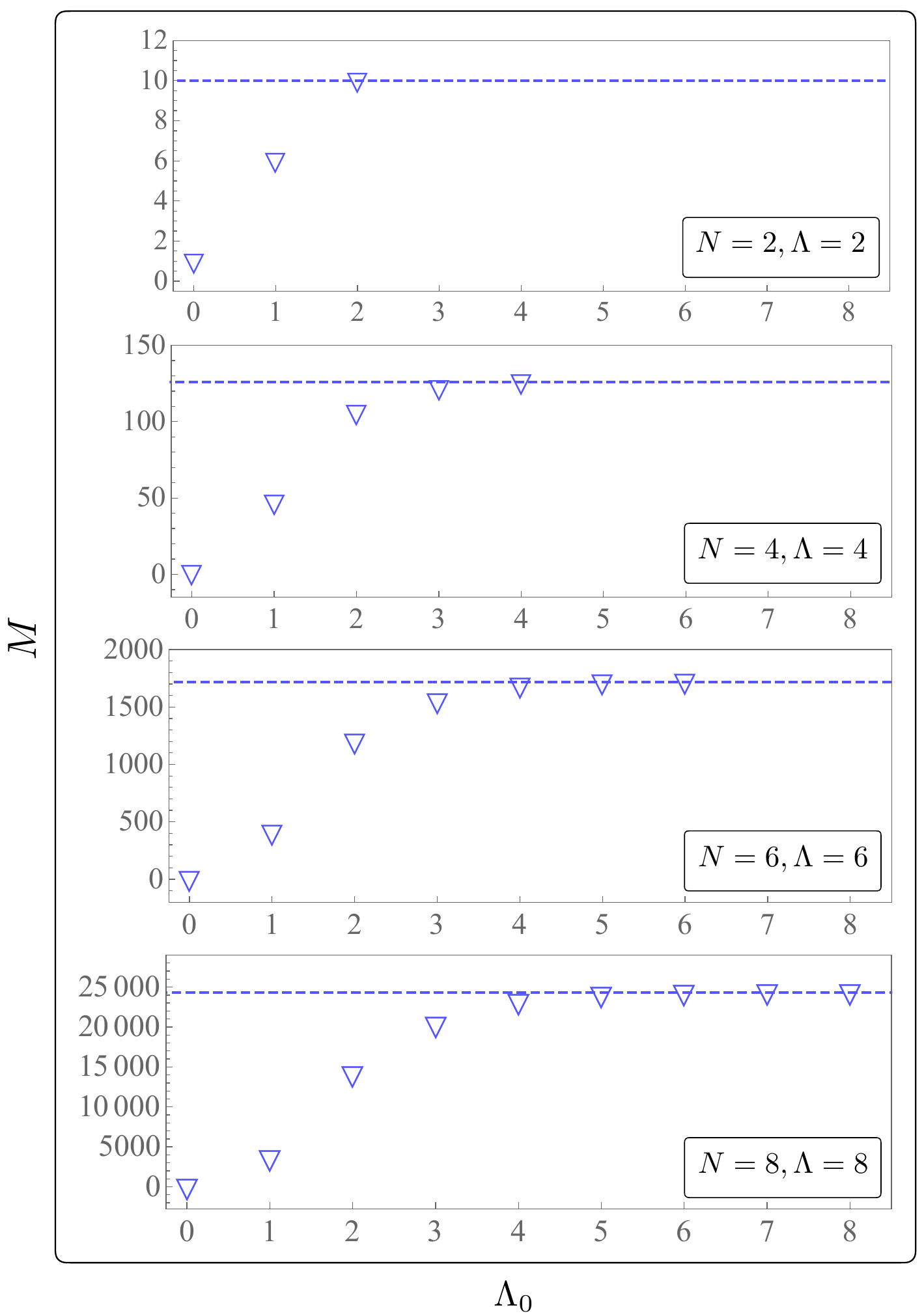}
\caption{The dimension of the physical Hilbert space, $M$, of the bosonized formulation of the KS theory as a function of $\Lambda_0$, the cutoff on the electric field excitations of the U(1) subgroup of the extended U(2) theory. The value of $\Lambda$, the cutoff on the electric field excitations of the SU(2) subgroup, is fixed to the smallest value at which the physical Hilbert space of the theory saturates to its full size. The dimension of the full physical Hilbert space is denoted by the dashed line in each plot. The numerical values associated with these plots are listed in Supplemental Material.
}
\label{fig:bosonized}
\end{figure}
In general, the dimension of the physical Hilbert space of the extended U(2) theory approaches that of the original SU(2) theory with $\Lambda \geq N$, and reaches a saturation value at $\Lambda_0 = N$. This trend has been depicted in Fig.~\ref{fig:bosonized} for $N=2,4,6,8$-site theories. Such an extra cutoff dependence can, in particular, be important in encoding the purely bosonic Hamiltonian onto qubits in a Hamiltonian simulation, in which one trades the few-dimensional Hilbert space of the fermions with the Hilbert space of U(1) bosons that grows with the lattice size when OBC are imposed. Nonetheless, in higher-dimensional gauge theories coupled to fermions, the bosonization may present some benefit by avoiding the non-local fermionic encodings, although the need for a sufficiently large U(1) cutoff may be more significant in higher dimensions. The pros and cons of such a reformulation of the original gauge theory in the context of quantum simulation requires further investigation.

\subsection{Loop-String-Hadron formulation
\label{sec:HilbertLSH}}
As described in Sec.~\ref{sec:LSH}, the Hilbert space of the LSH formulation is spanned by basis states 
\begin{eqnarray}
|n_l,n_i,n_o\rangle^{(x)},
\end{eqnarray}
for $x=0,1,2,...,N-1$, subject to the Abelian Gauss's law constraint $N_L(x)=N_R(x)$ along each link connecting sites $x$ and $x+1$. $N_L(x)$ and $N_R(x)$ quantum numbers are expressed in terms of the LSH quantum numbers according to Eqs.~(\ref{NL}) and (\ref{eq:NR}). In the following, an efficient procedure for generating the physical Hilbert space of the LSH formulation will be presented for both OBC and PBC. It can be shown that this Hilbert space is in one-to-one correspondence with the physical Hilbert space of the KS theory in the angular-momentum representation, and that LSH formulation is a more economical encoding of such a Hilbert space given its reliance on fully gauge-invariant DOF.
\begin{itemize}
\item[$\triangleright$] {OBC fixes the incoming electric flux into the lattice. In the language of LSH quantum numbers, this condition reads: $N_R(-1)=\epsilon_0$. Given this, and using Eqs.~(\ref{eq:AGL}), (\ref{NL}), and (\ref{eq:NR}) consecutively, it is straightforward to show that 
\begin{eqnarray}
n_l(x)&=& \epsilon_0+ \sum_{y=0}^{x-1}\left(n_o(y)-n_i(y)\right)\nonumber \\
&&~~~~~~-n_i(x)\left(1-n_o(x)\right).
\label{eq:nlOBC}
\end{eqnarray}
\begin{figure}[t!]
\includegraphics[width=0.315\textwidth]{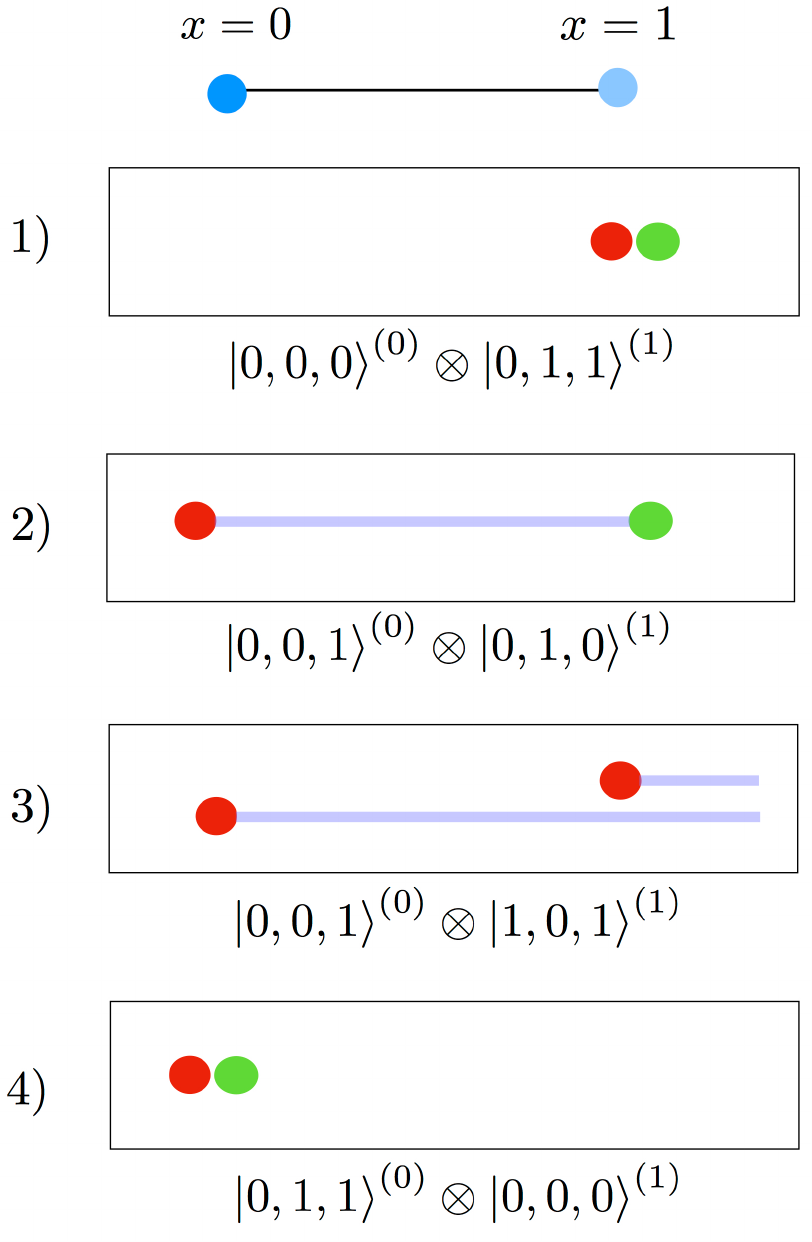}
\caption{
The LSH Hilbert space with OBC on a lattice of size $N=2$ and for the case of $\nu=1$ and $\Lambda=1$. The states are denoted as $\prod_{i=0}^{N-1}\ket{n_l,n_i,n_o}^{(i)}$. The four basis states shown are in one-to-one correspondence with the four physical basis states in the angular-momentum basis presented in Eq.~(\ref{eq:KSN2nu2}).
}
\label{fig:n2q1obc}
\end{figure}
Eq.~(\ref{eq:nlOBC}) implies that for the one-dimensional lattice with OBC, $n_l$ quantum number at any site is completely fixed by the boundary condition and string quantum numbers $n_i$ and $n_o$ at all the sites to its left. In other words, once the string quantum numbers are specified throughout the lattice,  all LSH quantum numbers are known, and hence a particular gauge-invariant state is specified. Note that for an $N$-site lattice, fixing the fermionic quantum numbers give rise to $4^N$ basis states, that is the same as the number of basis states one obtains with the purely fermionic formulation of the KS theory as described in Sec.~\ref{sec:HilbertF}. However, at this point, it should be noted that there exist certain string configurations (for a fixed value of $\epsilon_0$) that make the right-hand side of Eq.~(\ref{eq:nlOBC}) negative on one or more sites on the lattice. Such spurious string configurations must be discarded, and thus the Hilbert space of the LSH formulation is smaller in size than the purely fermionic formulation. In fact, the dimension of the LSH Hilbert space comes out to be of exactly the same as that of the physical Hilbert space of the KS theory in the angular-momentum basis. However, the cost of generating the physical Hilbert space is much less than that of the KS formulation in the angular momentum basis, as the states no longer need to satisfy SU(2) Gauss's laws at each site (since this has already been taken care of by the LSH construction), and the only remaining Gauss's law that is Abelian in nature is solved analytically. Moreover, while working with a cutoff $\Lambda$, the LSH Hilbert space is constrained to only contain those string configurations that yield $0\le N_{L/R}(x)\le \Lambda$ for all $x$. An example of the LSH Hilbert space with OBC on a lattice of size $N=2$ and with $\Lambda=2$ and $\nu=1$ is given in Fig.~\ref{fig:n2q1obc}. These basis states are in one-to-one correspondence with the physical basis states in the angular-momentum basis, i.e., the states enumerated in Eq.~(\ref{eq:KSN2nu2}).}
\item[$\triangleright$] {PBC implies that
\begin{eqnarray}
0 \leq N_R(-1)=N_L(N-1) \leq \Lambda,
\end{eqnarray}
yielding $(\Lambda+1)4^N$ states to start with. Identifying $\epsilon_0\equiv N_R(-1)$ and following the same prescription outlined above for OBC, all possible states in the LSH Hilbert space can be constructed subject to cutoff $\Lambda$. Note that with PBC, one obtains many copies of the same $\{n_i,n_o\}$ configurations corresponding to different winding numbers,  i.e., the number of closed loops that go around the lattice. A detailed account of the global symmetries of the KS theory will be presented in the next section.
}
\end{itemize}

The LSH Hilbert space, both for OBC and PBC, is identical to the physical Hilbert space of the KS theory, in the sense that each state in the LSH basis corresponds to one and only one state in the KS physical Hilbert space and vice versa. Therefore, all discussions regarding the scaling of the physical Hilbert space presented in Sec.~\ref{sec:HilbertKS} are valid for the LSH formulation as well. The only distinction is that the LSH Hilbert space and the associated Hamiltonian can be generated with far less computational complexity, as will be further discussed in Sec.~\ref{sec:cost}.

\section{Realization of Global Symmetries
\label{sec:symmetry}}
\noindent
The physical Hilbert space, projected out by imposing the Gauss's laws as studied in the previous sections, can be further characterized by global symmetry sectors as well as topological configurations. Identification of these symmetries can further simplify the Hamiltonian simulation, as the Hilbert space that needs to be studied can be further divided to smaller sectors. These symmetries are manifested differently in the case of  PBC and OBC, and are discussed in the following. 

The total fermion occupation number is conserved, as the operator
\begin{eqnarray}
\hat{Q} \equiv \sum_{x=0}^{N-1} \psi^\dagger(x)\psi(x),
\end{eqnarray}
commutes with the Hamiltonian with both boundary conditions. Note that in the LSH framework, this quantum number is simply $Q=\sum_x\left[n_i(x)+n_o(x)\right]$. With OBC, $Q$ can be any integer in the interval $[0,2N]$. With PBC, $Q$ can only be an even integer in the same interval, as an odd total fermionic occupation number creates an imbalance between the net flux of electric field into site $x=0$ and out of site $x=N-1$, which contradicts PBC.  For convenience, the global charge associated with the total number of fermions can be normalized by the lattice size as
\begin{table}[t!]
\label{tab:symmOBCN2}
\begin{tabular}{c|c|cccccccccc}
\multicolumn{1}{c}{} & \multicolumn{1}{c}{} &  & \multicolumn{9}{c}{$Q$}\tabularnewline
\cline{4-12} 
\multicolumn{1}{c}{} & \multicolumn{1}{c}{} &  & 0 &  & 1 &  & 2 &  & 3 &  & 4\tabularnewline
\cline{4-12} 
\multicolumn{1}{c}{} & \multicolumn{1}{c}{} &  &  &  &  &  &  &  &  &  & \tabularnewline
\multirow{3}{*}{$q$} & 0 &  & 1 &  &  &  & 3 &  &  &  & 1\tabularnewline
 & 1 &  &  &  & 2 &  &  &  & 2 &  & \tabularnewline
 & 2 &  &  &  &  &  & 1 &  &  &  & \tabularnewline
\end{tabular}
\caption{Breakdown of the physical Hilbert space of dimension $10$ with OBC and for $N=2$ and $\Lambda \geq 2$.}
\end{table}
\begin{eqnarray}
\nu \equiv \frac{Q}{N},
\label{eq:nu}
\end{eqnarray}
with $\nu \in [0,2]$. The strong-coupling vacuum lies in the $\nu=1$ sector of the Hilbert space.

Besides the conservation of the total number of fermions, there is an additional quantum number that divides the Hilbert space of each $Q$ sector to multiple disjoint sectors in general. This quantum number in the LSH language can be written as
\begin{eqnarray}
q \equiv \sum_{x=1}^{N-1}\left[n_o(x)-n_i(x)\right],
\end{eqnarray}
which using the identities in Eq.~(\ref{eq:NR}) can be written as $q=N_L(N-1)-N_R(-1)$, with a direct translation in the original KS formulation in the angular-momentum basis: $q=2\left[J_L(N-1)-J_R(-1)\right]$. In other words, this conserved quantum number distinguishes sectors with a different net flux of the outgoing electric field compared with the incoming electric field. For example, it is easy to verify that state 3) with $q=2$ in Eqs.~(\ref{eq:KSN2nu2}) and Fig.~\ref{fig:n2q1obc} does not evolve to states 1), 2), and 4) with $q=0$. This global charge is conserved with OBC since $J_R(-1)$ is fixed and there is no operator in the Hamiltonian to affect the $J_L(N-1)$ quantum number.  With OBC and $J_R(-1)=0$ , $q$ can be any integer in the interval $[0,\min(N,\Lambda)]$. With PBC, only the $q=0$ sector exists as the net electric field fluxes into and out of the one-dimensional lattice are equal, as mentioned above. An example of the breakdown of the physical Hilbert space with OBC into the $Q$ and $q$ sectors is given in Table~\ref{tab:symmOBCN2} for $N=2$ and $\Lambda \geq N$. A larger Hilbert space corresponding to $N=10$ is analyzed in Appendix~\ref{app:details}.
\begin{figure}[t!]
\includegraphics[width=0.490\textwidth]{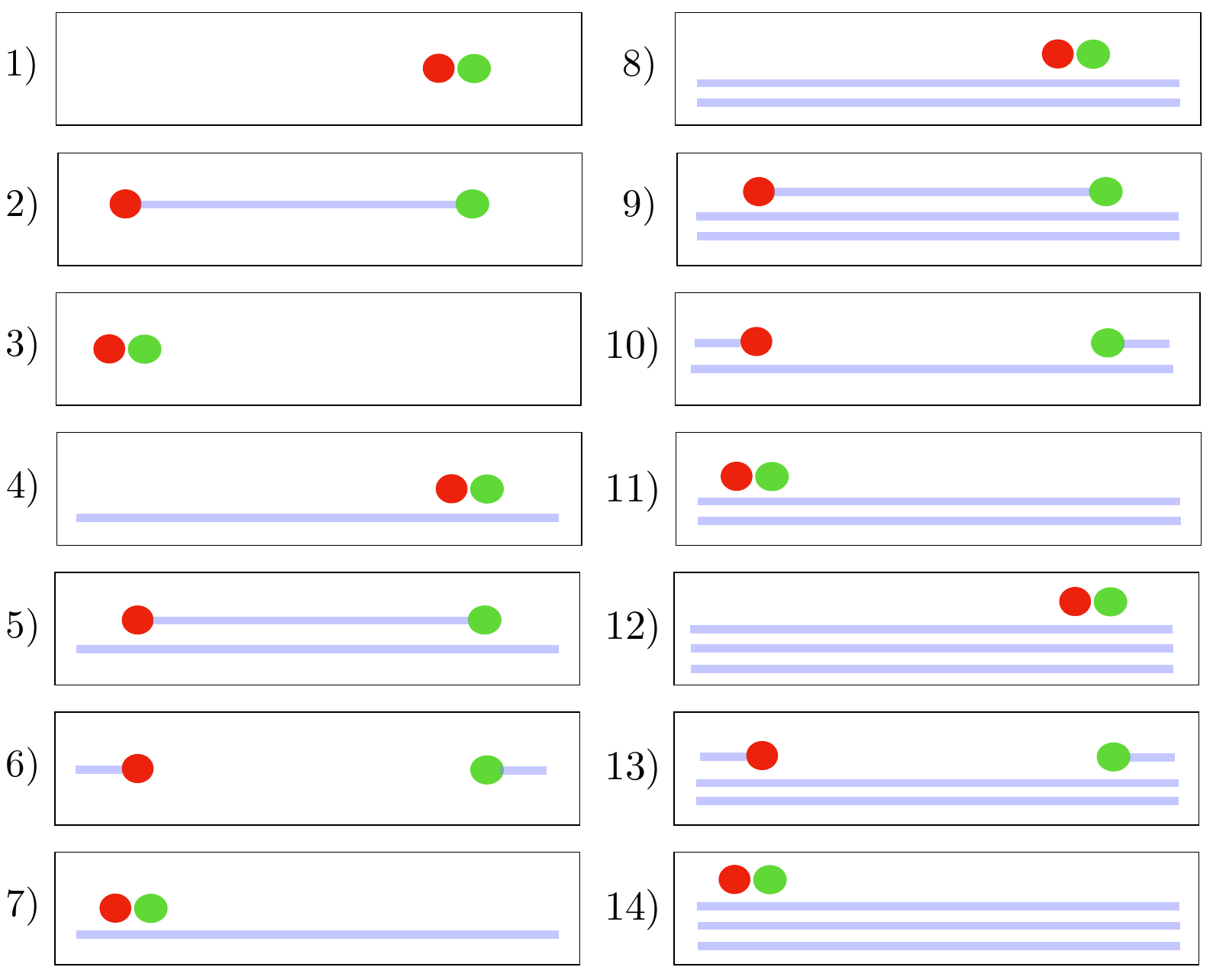}
\caption{
The physical Hilbert space within the LSH formulation with PBC with $N=2$, $\Lambda=3$, and $Q=2$.
}
\label{fig:PBC_LSH}
\end{figure}

In addition to the total fermionic number and the net flux, the Hamiltonian and the associated Hilbert space are symmetric under charge conjugation. While there is no $U(1)$ charge associated with the fields in the SU(2) LGT, the system is invariant, up to $\mu \to -\mu$, if the presence of fermions on the lattice is exchanged with their absence, i.,e, corresponding to a particle-hole exchange symmetry. One can verify that the associated charge conjugation operator, $\hat{C}$, commutes with the Hamiltonian of both PBC and OBC,
\begin{eqnarray}
[\hat C,\hat H]=0,
\end{eqnarray}
and that 
\begin{eqnarray}
\hat C^2=1,~ \{\hat{\mathcal{Q}},\hat C\}=0,
\end{eqnarray}
where $\hat{\mathcal{Q}} \equiv \hat{Q}-N \, \hat{\mathbb{I}}$, with $\hat{\mathbb{I}}$ being an identity operator. Therefore, for any given state in the Hilbert space with charge $Q$:
\begin{eqnarray}
\hat{Q}\hat C|\psi\rangle &=&-\hat C \hat{\mathcal{Q}} |\psi\rangle+N \hat{C} |\psi\rangle \nonumber \\
&=& (2N-Q)\hat C |\psi\rangle.
\end{eqnarray}
This implies that the charge-conjugated state exhibits a charge $2N-Q$. For an even $Q$-charge sector, the resulting spectrum is invariant under $\mu \to -\mu$, and hence the pair of charge-conjugated Hilbert spaces for $\{Q,2N-Q\}$ are physically identical. For odd $Q$-charge sectors, the charge-conjugated Hilbert space is only physically equivalent to the original Hilbert space once $\mu$ is replaced with $-\mu$.
\begin{table}
\label{tab:symmPBCN2}
\begin{tabular}{c|c|ccccccc}
\multicolumn{1}{c}{} & \multicolumn{1}{c}{} &  &  & \multicolumn{5}{c}{$Q$}\tabularnewline
\cline{5-9} 
\multicolumn{1}{c}{} & \multicolumn{1}{c}{} &  &  & 0 &  & 2 &  & 4\tabularnewline
\cline{5-9} 
\multicolumn{1}{c}{} & \multicolumn{1}{c}{} &  &  &  &  &  &  & \tabularnewline
\multirow{4}{*}{$l$} & 0 &  &  & 1 &  & 3 &  & 1\tabularnewline
 & 1 &  &  & 1 &  & 4 &  & 1\tabularnewline
 & 2 &  &  & 1 &  & 4 &  & 1\tabularnewline
 & 3 &  &  & 1 &  & 3 &  & 1\tabularnewline
\end{tabular}
\caption{Breakdown of the physical Hilbert space of dimension $22$ with PBC and for $N=2$ and $\Lambda = 3$ in terms of the $Q$ quantum number (all states satisfy $q=0$). The winding quantum number $l$, while not a conserved quantity, is also specified.
}
\end{table}

Finally, for PBC it is useful to characterize the states by a winding number variable $l$, such that for any state  $\ket{\psi}$ in the Hilbert space with a given cutoff $\Lambda$, $\left[{\rm Tr}(U(0)U(1)U(2)\ldots U(N-1))\right]^{l}\ket{\psi}$ is also a valid state of the physical Hilbert space with cutoffs up to $\Lambda+l$, where $0 \leq l<\infty$. Since the operator  $\left [{\rm Tr}(U(0)U(1)U(2)\ldots U(N-1))\right]^{l}$ does not commute with the Hamiltonian, it is not a symmetry of the theory, nonetheless, it provides a useful characterization of the states. As was explicitly realized in the previous section, the dimension of the Hilbert space for OBC is finite for arbitrary $\Lambda$. However with PBC, the Hilbert-space dimension grows linearly with the cutoff (once a saturating value of the cutoff is reached to accommodate all the fermionic configurations). In the linear region, the slope is obtained from the number of all fermionic configurations with charge $Q=0$, which is $\binom{2N}{N}$. As a result, the complete physical Hilbert space with PBC contains copies of the particular gauge-invariant Hilbert space with winding numbers varying from $0$ to $\Lambda$. Such a winding-number characterization of PBC Hilbert space is evident in the example shown in Fig.~\ref{fig:PBC_LSH}, where basis states $1)$, $2)$, $3)$, and $6)$ are repeated for different values of the winding numbers, $l=0,1,2,3$.  The breakdown of the PBC Hilbert space in terms of the fermionic occupation quantum number $Q$ and the winding number $l$ is worked out for a related example in Table~\ref{tab:symmPBCN2}. Finally, it should be noted that with PBC, the theory exhibits a discrete translational symmetry, and the eigenstates of a discrete momentum operator can be formed as well, see e.g., Ref.~\cite{Klco:2018kyo} for such a classification of momentum eigenstates in the case of the lattice Schwinger model.

\section{Comparative (classical) cost analysis
\label{sec:cost}}
\noindent
A classical algorithm for Hamiltonian simulation, in general, involves three steps: \emph{I}) Hilbert-space construction, \emph{II}) Hamiltonian generation, and \emph{III}) observable computation. The Hilbert space can be constructed by identifying the theory's DOF, symmetries, and a convenient basis to express the states. In the case of LGTs, where a major portion of the Hilbert space is irrelevant, to reduce the computational cost, one needs to project to the physical Hilbert space. This entails one of the following. One may impose the (non-trivial) Gauss's laws by reformulating DOF as is the case in the LHS formulation. Alternatively, the Gauss's law constraints can be imposed \emph{a posteriori} on convenient basis states. Another option is to generate the physical states by the consecutive action of the Hamiltonian on a trivial physical state such as the strong-coupling vacuum~\cite{hamer19822, Banuls:2017ena}. The associated computational cost of this step, therefore, depends largely on the Hamiltonian formulation used, as well as the algorithm itself. In the following, we analyze the first two approaches, noting the third approach is of comparable cost as it requires the Hamiltonian matrix to be acted on states by a number of times that grows exponentially with the system's size. After the basis states in the (physical) Hilbert space are identified, the next step of the simulation is to generate the Hamiltonian matrix. This step, obviously, depends on the formulation used as well. For example, some formulations may provide simpler operator structures, which could affect the sparsity of the matrix generated. Finally, the Hamiltonian matrix can be used to compute observables, such as spectrum, and static or dynamical expectation values of operators. This step often entails matrix diagonalization and matrix exponentiation, which can be sped up by efficient sparse-matrix algorithms especially when acted on a sparse state vector.

Having introduced various formulations of the SU(2) LGT in 1+1~D and analyzed their physical Hilbert-space dimensionality with regard to the system's size, electric-field cutoff, and boundary conditions, one can now analyze the classical-simulation cost within each formulation. For this purpose, we focus on the KS formulation in the physical Hilbert space, the LSH formulation, and the purely fermionic formulation, all with OBC, and will briefly comment on the case of PBC and the purely bosonic formulation in the end.

\subsection{Physical Hilbert-space construction}

\begin{center}
\emph{Purely fermionic formulation}
\end{center}
Within the purely fermionic formulation, redundant gauge symmetries are removed algebraically by an appropriate gauge transformation and after applying the Gauss's law repeatedly, as explained in Sec.~\ref{sec:F}. As such, the projection to the physical Hilbert space is essentially free, and the time complexity is:\footnote{Throughout this paper, computational cost, number of operations, and time complexity are used interchangeably, and are all meant to convey the same meaning.}
\begin{eqnarray}
\mathbb{T}_I^{({\rm F})} \sim \mathcal{O}(1).
\end{eqnarray}
Here, in principle, there is an additional cost associated with generating $4^N$ fermionic configurations $\prod_{i=1}^N \ket{f_1,f_2}^{(i)}$ with  $f_{1,2} \in \{0,1\}$. Nonetheless, with an efficient Kronecker-product algorithm introduced in the next subsection, the Hamiltonian can be generated without the need to generate and store these fermionic configurations.

\begin{center}
\emph{Loop-String-Hadron formulation}
\end{center}
An efficient algorithm and its associated cost for generating physical Hilbert space of the LSH formulation with OBC goes as follows. One first generates $4^N$ string configurations. The cost of generating each configuration can be realized as converting an integer label $k$ associated with each configuration to a binary number, which goes as $\log(k)$. This step can therefore be conducted with the time complexity $\mathcal{O}(N)$ for a lattice of $N$ sites. The binary digits in each generated configuration are then labeled by $n_i$ and $n_o$ string numbers, by e.g. assigning them to the even and odd digits, respectively. This step is essentially free. Next comes the generation of the $n_l$ quantum numbers. As mentioned before, with OBC, the Gauss's law is used to fix this number for any given string configuration. Since $n_l$ must be fixed at all links, $\mathcal{O}(N)$ number of operations is needed. There are additional $\mathcal{O}(N)$ operations required to pick each generated $n_l$ and check it against the requirement of not exceeding the cutoff as well as being a non-negative integer, as explained in Sec.~\ref{sec:HilbertLSH}. However, this step can be simultaneously performed as generating $n_l$ quantum numbers consecutively, to reduce the cost. The total cost of generating the physical Hilbert space with the LSH formulation is therefore:
\begin{eqnarray}
\mathbb{T}_I^{({\rm LSH})} \sim \mathcal{O}(N 2^{2N}).
\end{eqnarray}

Note that in order to reduce the dimensionality of the Hilbert space, one could additionally restrict the states to a given global-symmetry sector. For example, if only interested in the charge $\nu=1$ sector, there are $\mathcal{O}(N)$ operations involved to check the $\nu$-number of each string configuration, reducing the number of configurations needed for generation of $n_l$ quantum numbers from $4^N$ to $\bigl( \begin{smallmatrix}2N\\ N\end{smallmatrix}\bigr) \approx \frac{4^N}{\sqrt{N}}$. Since this is not an exponential speed up as a function of the size of the system, such finer decompositions of the physical Hilbert space will not be considered in the rough estimate of the computational cost in the remainder of this section. Such symmetry considerations, however, will be advantageous in practical implementations.

\begin{center}
\emph{Angular-momentum representation}
\end{center}
An efficient algorithm for the generation of the physical Hilbert space of the KS in the angular-momentum basis with OBC starts by making  a gauge-invariant state at site $x$, i.e., one that satisfies the non-Abelian Gauss's laws in Eq.~(\ref{eq:GLawJ}). If there is one and only one fermion at site $x$, then a gauge-invariant state is obtained from the relation 
\begin{eqnarray}
&& \ket{(J_R,\tfrac{1}{2})J_{Rf}J_L;00}^{(x)}=\sum_{m_R,m_f,m_{Rf},m_L}
\nonumber\\
&&\braket{J_R,m_R;\tfrac{1}{2},m_f | J_{Rf},m_{Rf}} \braket{J_{Rf},m_{Rf};J_L,m_L | 0,0}
\nonumber\\
&&  \qquad~~~ \ket{J_R,m_R}^{(x-1)}\otimes {\ket{\tfrac{1}{2},m_f}}^{(x)} \otimes \ket{J_L,m_L}^{(x)},
\label{eq:KSPhysState}
\end{eqnarray}
with the notation defined in Sec.~\ref{sec:KS}. While $m_f$ only takes values $\pm \tfrac{1}{2}$ in these sums, each sum over $m_R$, $m_{Rf}$, and $m_L$ involves of the order of $J_R$ operations. This is because the value of $J_R$ fixes the value of $J_{Rf}$ to be equal to $J_R\pm\tfrac{1}{2}$, and that in order for the final total angular momentum to be zero, the value of $J_L$ needs to be equal to $J_R\pm\tfrac{1}{2}$. Now to generate a set of all possible physical states, such construction at the site should be repeated for all possible values of $J_R$, i.e., $0 \leq J_R \leq \Lambda$. As a result, the number of operations required to generate a complete set of physical states at a given site is $\mathcal{O}(\Lambda^4)$. If, however, there is either no fermion or there are two fermions at site $x$, the above relation is modified by setting $\tfrac{1}{2} \to 0$, the expression simplifies to only two summations, and the final number of operations required to generate a gauge-invariant set of states is $\mathcal{O}(\Lambda^3)$, which is subdominant compared with the first case and can be ignored in the limit $\Lambda \gg 1$. Note that in both cases, there is an additional cost involved amounting to checking and removing the generated $J_L$ values that violate $0 \leq J_L \leq \Lambda$, but this step can be checked simultaneously in the sum above, and the total asymptotic cost in the limit $\Lambda \gg 1$ remains the same. The next step involves an $N$-fold Kronecker product of states in each set to connect states at adjacent sites throughout the lattice. This adds a cost with the time complexity $\mathcal{O}\left(\Lambda^{4N}\right)$. Finally, the boundary condition on $J_R$ at site $x=-1$ must be imposed, along with the Abelian Gauss's law to ensure that $J_R$ and $J_L$ belonging to the same link are set equal. This can be achieved by a search and elimination algorithm, and involves an additional cost that scales as $\mathcal{O}\left(\Lambda^{4N}\right)$, which is the conservative scaling of the number of basis states formed in the previous step. As a result, the total time complexity of generating the physical Hilbert space is: $\mathbb{T}_I^{({\rm J})} \sim \mathcal{O}\left(\Lambda^4+\Lambda^{4N}+\Lambda^{4N}\right) \approx \mathcal{O}\left(\Lambda^{4N}\right)$, where $\approx$ sign in this section is meant the approximate scaling in the limit: $N \gg 1$. The time complexity if $\Lambda$ is fixed to a constant much smaller than $N$ is: $\mathbb{T}_I^{({\rm J})} \sim \mathcal{O}\left(\Lambda^{4N}+\Lambda^{3N}\right)$.

An alternative algorithm can be realized by first generating $4^N$ fermionic configurations throughout the lattice. Each configuration generation involves a time complexity that scales as $\mathcal{O}(N)$. Next, given the value of $J_R$ at the boundary $x=-1$, all $J_L$ and $J_R$ values can be produced throughout the lattice, given the known fermionic occupation at each site and the Abelian Gauss's law. This involves a maximum number of operations that goes as $\mathcal{O}(N+N 2^N)$, since at each site and given a $J_R$ value, there may be two possibilities for the $J_L$ value as the fermion occupation may be equal to one. Now given the set of configurations for fermions and angular momenta generated, the non-Abelian Gauss's law can be implemented following the relation in Eq.~(\ref{eq:KSPhysState}), introducing an additional cost $\mathcal{O}(\Lambda^3)$. The Kronecker-product cost is the same as before but there will be no need to impose the boundary condition and Abelian Gauss's law anymore as these are already implemented. In summary, this algorithm involves a time complexity that scales as
\begin{eqnarray}
\mathbb{T}_I^{({\rm J})} \sim \mathcal{O}\left(N2^{3N}+(2\Lambda)^{3N} \right).
\end{eqnarray}
For $\Lambda \sim N$ and $N \gg 1$, this later algorithm therefore is asymptotically faster than what was described earlier. The cost, however, remains super-exponential in $N$ this limit.

\subsection{Hamiltonian generation}

\begin{center}
\emph{Purely fermionic formulation}
\end{center}
In the fermionic formulation of the KS theory with OBC, the Hamiltonian becomes non-local with $\mathcal{O}(N^2)$ terms in the Hamiltonian, see e.g., the electric Hamiltonian in Eq.~(\ref{eq:HEF}). However, a considerable advantage is that the operator structures encountered are only of the type $\psi^\dagger(x)\psi(y)$, which make the Hamiltonian generation amenable to Kronecker-product algorithms, eliminating the need to generate and store the fermionic Hilbert space \emph{a priori}. Explicitly, each $\psi^\dagger(x)\psi(y)$ operator can be written as at the Kronecker product of $\mathbb{I}_{4 \times 4}$ at all sites but $x$ and $y$, $A_{4 \times 4}$ at site $x$ and $B_{4 \times 4}$ at site $y$, in an ordered manner, where $\mathbb{I}$ is the identity matrix, and $A$ and $B$ are matrices formed by the action of $\psi^\dagger(x)$ and $\psi(y)$, respectively, on the four allowed fermionic configurations at the respective sites. Since the on-site matrices are sparse and involve $\mathcal{O}(d_i)$ elements (with $d_i=4$ being the dimensionality of the matrices), performing the Kronecker product along a lattice of length $N$ comes with the time complexity $\mathcal{O}(4^N)$. The total time complexity of this algorithm for generating the Hamiltonian is therefore:
\begin{eqnarray}
\mathbb{T}_{II}^{(\rm F)} \sim \mathcal{O}\left(N^2 2^{2N}\right).
\end{eqnarray}
Note that there are additional $\mathcal{O}(N)$ operations involved in finding the position of $x$ and $y$ along the chain, but that is a subdominant cost compared with the subsequent Kronecker-product operation. 

\begin{center}
\emph{Loop-String-Hadron formulation}
\end{center}
In order to generate the LSH Hamiltonian, first note that the dimensionality of the physical Hilbert space can be approximated by $4^{N-1}$ at large $N$, estimated by comparing the asymptotic ratio of the physical Hilbert space dimension in the fermionic formulation to that of the LSH formulation, as shown in Fig.~\ref{fig:allksvsfobc}. On the other hand, in the LSH formulation, the Hamiltonian remains local, with the total number of operators scaling as $\mathcal{O}(N)$. Generating the Hamiltonian matrix elements amounts to picking one state out of the Hilbert space at a time, acting by the Hamiltonian operator on the state to arrive at another state (which requires $\mathcal{O}(N)$ operations), and find the position of the state's assigned index in a previously produced look-up table of states indices, with a time complexity that scales as $\mathcal{O}(4^{N-1})$, and is therefore dominant compared with Hamiltonian operation cost. This step identifies the position of the element in the Hamiltonian matrix and its value. The total time complexity of Hamiltonian generation for the LSH formulation, therefore, scales as
\begin{eqnarray}
\mathbb{T}_{II}^{(\rm LSH)} \sim \mathcal{O}\left(N 2^{4N}\right).
\end{eqnarray}
Note that the more efficient Kronecker-product algorithm that was applied in the fermionic case could not be taken advantage of in the LSH formulation as the one-to-one mapping between the Kronecker product of on-site physical states and the global physical state is lost given the imposition of the boundary condition and the Abelian Gauss's laws. Such a convenient feature is lost in the angular momentum representation of the physical states as well.

\begin{center}
\emph{Angular-momentum representation}
\end{center}
Similar to the LSH formulation, the dimension of the physical Hilbert space in the angular-momentum basis scales as $\mathcal{O}(4^{N-1})$, and there are $\mathcal{O}(N)$ operators in the Hamiltonian. The Hamiltonian matrix elements can be generated by picking one physical state at a time and find the resulting state after the operation of each term in the Hamiltonian on the chosen state. Despite the LSH states though, the physical states are linear combinations of a multitude of basis states in general. While in principle, there are of the order of $\mathcal{O}(\Lambda^{3N})$ terms for each physical state, as discussed before, many of these states have a vanishing contribution due to the corresponding vanishing Clebsch-Gordan coefficients. Empirically, the maximum number of terms obtained in a physical state with OBC is seen to scale as $\mathcal{O}\left((2\Lambda)^{N/2-1}\right)$.\footnote{Compare this with the scaling of the maximum number of terms in a physical state for PBC shown in Fig.~\ref{fig:kspbcnumterms}.} Furthermore, the action of the interaction Hamiltonian on each basis state involves $\mathcal{O}(\Lambda)$ operations as is evident from Eq.~(\ref{eq:UonState}). Finally, the obtained state itself is a linear combination of other physical states, and requires $\mathcal{O}(4^{N-1}\Lambda(2\Lambda)^{N/2-1})$ to find its overlap to other states in the physical Hilbert space. As a result, the total cost of generating the Hamiltonian matrix in the angular-momentum representation scales as
\begin{eqnarray}
\mathbb{T}_{II}^{(\rm J)} \sim \mathcal{O}\left(N 2^{5N}\Lambda^{N}\right).
\end{eqnarray}
In the limit $\Lambda \sim N$, this step introduces another super-exponential cost to the Hamiltonian simulation in this basis, in addition to the cost of generating the physical Hilbert space as derived in the previous subsection.

\subsection{Observable computation}
\begin{center}
\emph{General scaling relations}
\end{center}
The cost of Hamiltonian matrix manipulation required to evaluate observables depends upon the dimensionality of the Hamiltonian, its sparsity, and the sparsity of the state vector in computing expectation values. For a square matrix with dimensions $M \times M$, the matrix density is defined as the ratio of the number of non-zero elements in the matrix to $M^2$. Computation of spectrum in the Hamiltonian formulation amounts to evaluating the (first $m$) eigenvalues, while time-dependent expectation values require the action of the exponential of a matrix on a state vector. A rigorous analysis of the computational cost of these matrix manipulations given a desired accuracy in the estimated outcome is beyond the scope of this work. Given that the goal of this section is to make a comparison among different Hamiltonian formulations of the SU(2) LGT, with a focus on select algorithms, we point out a generic time complexity measure for the operations without attempting to find rigorous theoretical bounds on the algorithms considered.

The extremal eigenvalues of a matrix can be obtained using an iterative algorithm, which generates convergence to the desired eigenvalues. To demonstrate the time complexity of such algorithms, consider a Hermitian matrix with the following eigenvalue equation: $A\ket{v_a} = \lambda_a \ket{v_a}$. The power series algorithm for extremal-eigenvalue estimation requires choosing a random initial unit vector $\ket{b}=\sum_a c_a \ket{v_a}$, $|c_a|^2=1$. One repeats applying $A$ to $\ket{b}$, $A^k\ket{b}=\sum_n c_n \lambda_a^k \ket{v_a}$, and normalizes the output state after each iteration. This procedure suppresses the contribution of smaller eigenvalues and quickly converges to the maximal eigenvalue. The Arnoldi algorithm~\cite{arnoldi1951principle} enhances this procedure by generating a Krylov subspace with the set of output vectors $\left \{ \ket{b},A\ket{b},A^2\ket{b} ,..., A^{m-1}\ket{b} \right\}$ that are orthogonalized against each other. To generate the lowest eigenvalues, $A$ is replaced with $A^{-1}$ in the procedure outlined. The limiting feature of this approach is the cost of matrix multiplication. As the procedure is repeated, it becomes more costly to expand the Krylov subspace, due to the growing density of the output vectors. Applying the Lanczos algorithm~\cite{lanczos1950iteration, ojalvo1970vibration, paige1971computation, paige1972computational}, a generalization of the Arnoldi algorithm to Hermitian matrices, requires $\mathcal{O}(\xi \rho_H M^2)$ operation for an $M \times M$ Hamiltonian with matrix density $\rho_H$. Empirical observations have shown that $\xi \sim \frac{3}{2}m$, where $m$ is the number of desired extremal eigenvalues~\cite{ojalvo1970vibration}.

Among methods to compute matrix exponentiation is the truncated Taylor series method~\cite{moler2003nineteen}. For the unitary time-evolution operator, the $k$-th order truncated series is defined as: $\left . e^{-iHt}\right|_k=\sum_{l=0}^{k-1}(-iHt)^l/l!$. The number of terms, $k$, needed in the Taylor expansion to reach the desired accuracy depends on  time, Hamiltonian's dimension, and on some form of a Hamiltonian norm, such as the size of its largest eigenvalue. With identical eigenvalues and similar Hilbert space sizes in all formulations considered, the number of terms to be computed will be of the same order in all cases and will not be constrained any further. The time complexity of multiplying an $M \times M$ Hamiltonian with density $\rho_H$ and an $M$-dimensional vector with density $\rho_v$ goes as $\mathcal{O}(\rho_H\rho_vM^2)$~\cite{bell2009implementing}, hence a $k$-th order truncation of $e^{-iHt}$ on state vector $\ket{v}$ scales as: $\mathcal{O}(k\rho_H\rho_vM^2)$. Although the initial state can be chosen to be sparse, the time evolution will eventually saturate the sparsity bound determined by the action of the $k$-th order Taylor series on the state, which results in mixing the initial state with states in the same symmetry sector.  

To simplify the cost analysis, in the following the time complexity 
\begin{eqnarray}
\mathbb{T}_{III} \sim \mathcal{O}(\rho_H M^2)
\end{eqnarray}
will be used as a general measure of the spectral and dynamical observables and the companying factors mentioned above will be dropped. Once again, it should be noted that due to the increase in $\rho_v$ as the system evolves, the time-evolution cost will ultimately approach the time complexity $\mathcal{O}(M^2)$.

\begin{center}
\emph{Computing cost in different formulations}
\end{center}
Since the Hamiltonian in the angular-momentum basis in the physical Hilbert space is the same as that in the LSH basis, the cost analysis of the observable computation is the same in both formulations. What makes a difference in computational costs of one versus the other is clearly the Hilbert-space and Hamiltonian-generation cost, as will be studied more closely in the next subsection. As a result, only a cost comparison is presented here for the fermionic and LSH formulations. For the LGT formulations of SU(2) theory in 1+1~D considered in this work, the Hamiltonians are sparse and their density goes roughly as $1/M^p$ with $p>0$, given the locality of the interactions in the original KS formulation. Even in the fermionic formulation, in which the gauge DOF are traded with the locality of interactions, the number of operators in the Hamiltonian remains small compared with the size of the Hilbert space, rendering the Hamiltonian matrix extremely sparse. A comparison of both the dimension of the Hilbert space and the density of the Hamiltonian matrix between the fermionic and LSH  formulations of SU(2) LGT in 1+1~D with OBC is shown in Fig.~\ref{fig:allksvsfobc}. As the plots demonstrate, while the dimension of the LSH Hilbert space remains smaller than that of the fermionic representation for all values of $N$, the density of the Hamiltonian matrix in the fermionic formulation decreases compared with that of the LSH formulation as $N$ increases. The determining factor in the time complexity of the eigenvalue computation and matrix exponentiation is, nonetheless, $\rho_H M^2$, and the ratio of this quantity between the two formulations is plotted in the lower panel of Fig.~\ref{fig:allksvsfobc}. As is observed, the ratio grows slowly as $\mathcal{O}(N)$ as $N \to \infty$, demonstrating the slightly higher cost of observable computation in the fermionic formulation. In the comparative analysis of the next subsection, the asymptotic cost of observable computation for the LSH formulation (hence the angular-momentum formulation in the physical sector) will be estimated simply as $\mathcal{O}(\rho_HM^2)$ with $M = 4^N$ and $\rho_H = 2^{-N}$, where the density is approximated from an empirical fit of the density of the LSH Hamiltonian for small lattice sizes. For the fermionic Hamiltonian, the cost of the LSH Hamiltonian should be multiplied by $N$.

\subsection{Total cost and comparisons}
\begin{figure}[!t]
\includegraphics[width=0.490\textwidth]{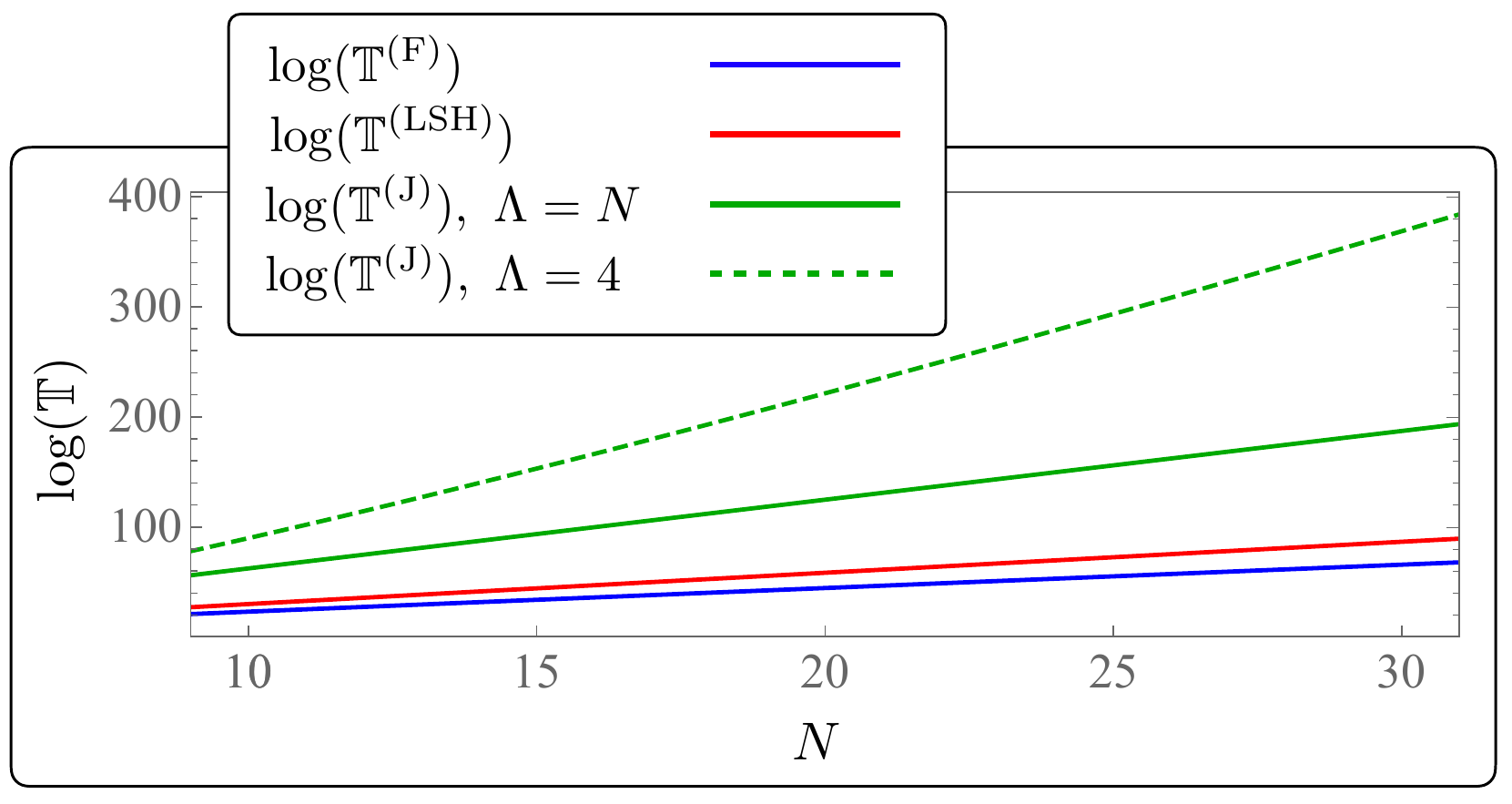}
\caption{The asymptotic cumulative cost of the three steps of given classical numerical algorithms for Hamiltonian simulation of the KS SU(2) LGT in 1+1~D with the fermionic formulation (F), LSH formulation, and the angular-momentum formulation in the physical sector (J), as a function of lattice size $N$ for large $N$.
}
\label{fig:costcomp}
\end{figure}
\begin{figure}[h]
\includegraphics[width=0.490\textwidth]{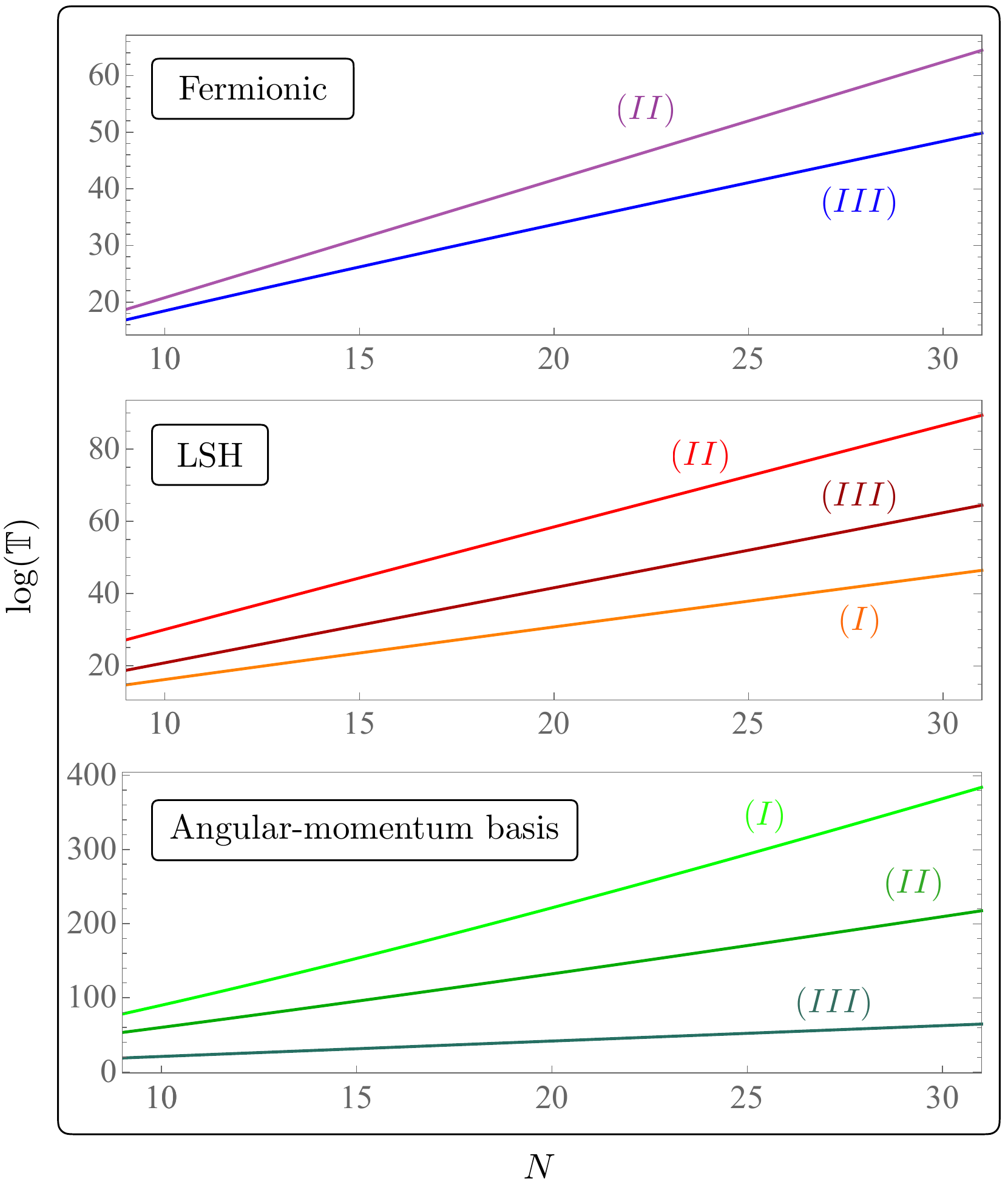}
\caption{The asymptotic cost of each step of given classical algorithms for Hamiltonian simulation of the KS SU(2) LGT in 1+1~D with the fermionic formulation (F), LSH formulation, and the angular-momentum basis in the physical sector (J), as a function of the lattice size $N$ for large $N$. Step $(I)$ refers to Hilbert-space construction, step $(II)$ refers to Hamiltonian generation, and step $(III)$ denotes observable computation assuming a generic scaling for sparse matrix manipulations, see the text. The step $(I)$ for the fermionic formulation is of $\mathcal{O}(1)$ with the chosen algorithm and is not shown. 
}
\label{fig:costsepall}
\end{figure}
Given the time complexity of various steps of a Hamiltonian simulation outlined above for the three formulations of SU(2) LGT in 1+1~D studied, the cumulative cost of the simulation can be estimated by adding $\mathbb{T}_{I}$, $\mathbb{T}_{II}$, and $\mathbb{T}_{III}$ for each formulation from the previous subsection. Figure~\ref{fig:costcomp} plots this asymptotic scaling cost, i.e., in the limit where $N \gg 1$. For the angular-momentum formulation, two scenarios are considered: The cutoff on the gauge excitations is set to a constant value, and the cutoff is set to $N$, corresponding to the saturating value of the cutoff with OBC. As already evident from the analytical scaling relations, the least time complexity is offered by the fermionic representation, while the angular-momentum formulation provides the worst scaling and is unfit for even small-scale simulations. As an explicit comparison, for $N=20$, angular-momentum formulation (with $\Lambda=N$) requires 160 orders of magnitude larger computing resources than the LSH formulation, while the fermionic formulation requires 20 orders of magnitude lesser resources than the LSH formulation.

It is also interesting to examine the cost of each step of the simulation, i.e., Hilbert-space constructions (I), Hamiltonian generation (II), and observable computation (III), for the algorithms outlined. While the conclusions are already evident from the analytic scaling forms provided in the previous subsection, they can be observed more crisply from the plots shown in Fig.~\ref{fig:costsepall}. For both the fermionic and LSH formulations, the least costly step is constructing the physical Hilbert space, while interestingly for the angular-momentum formulation this step is the most costly  step involved. Hamiltonian generation is the most costly step of the simulation for the fermionic and LSH formulations. Once the Hamiltonian is generated, as already explained, the observable computation costs roughly equal computing resources in all cases. In other words, in the angular-momentum formulation much of the computing effort is put into generating a Hamiltonian that is of much smaller dimensionality than the naive construction and is the same as that of LSH. That is the ultimate reason behind why the LSH formulation is developed and promoted, as it simplifies \emph{a priori} the construction of the Hamiltonian and its associated Hilbert space without requiring the implementation of these steps on the inefficient angular-momentum basis states. 
\begin{figure*}[!t]
\includegraphics[width=0.990\textwidth]{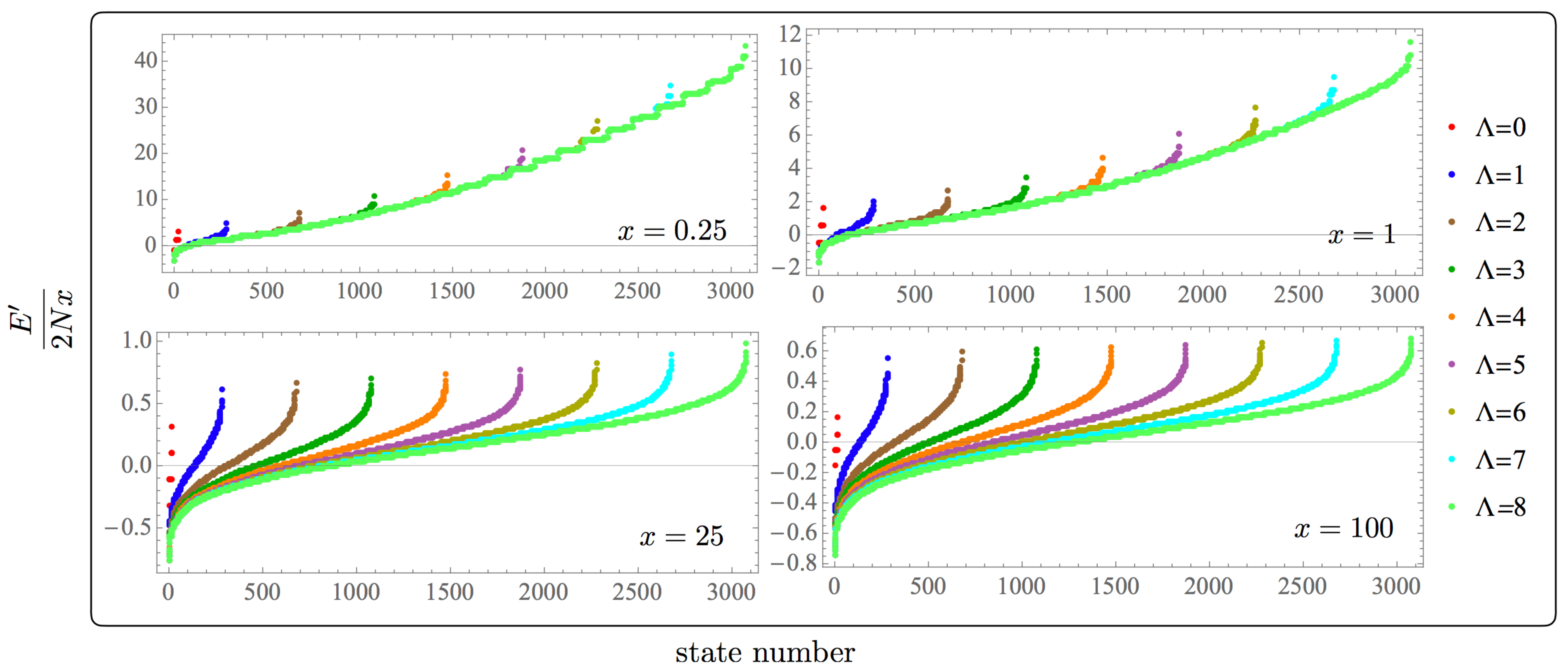}
\caption{
The spectra of the KS Hamiltonian in the physical Hilbert space with PBC for $N=6$, $\nu=1$, and various values of $x$ and $\Lambda$. More precisely, the quantity plotted is $\frac{E'}{2Nx}$, where $E'$ is the scaled energy corresponding to the scaled Hamiltonian in Eq.~(\ref{eq:HprimeKS}). The numerical values associated with these plots are provided in Supplemental Material.
}
\label{fig:energyvslambdapbc}
\end{figure*}

Before completing this section, a few comments are in order. First, it should be noted that the algorithms described can be made far more efficient by parallelizing operations when possible. The degree of parallelization will depend on the formulation considered, as well as the algorithm devised for each step of the computation. Second, the benefit of fermionic representation over the LSH formulation is weakened by the fact that only the LSH formulation can be generalized to other boundary conditions and to higher dimensions, hence its great benefit over the other formulations when the gauge DOF are present is a significant result. In particular, if the fully bosonic formulation of Sec.~\ref{sec:B} is considered within the angular-momentum basis, its computational cost will be greater than that of the angular-momentum representation with both bosons and fermions present, given the introduction of an extra U(1) cutoff in constructing the Hilbert space and the associated Hamiltonian, with a saturation value that scales as $N$. Last but not least, it must be now obvious that the extreme advantage offered by the LSH formulation over the angular-momentum formulation will persist with PBC as the complexity of Hilbert-space construction and the Hamiltonian generation is even more significant with PBC. While in the LSH formulation with PBC, the Hamiltonian operator still projects one basis state to another, in the angular-momentum formulation, not only the maximum number of basis states in a given physical state can scale as $\mathcal{O}(\Lambda^N)$, as shown in Fig.~\ref{fig:kspbcnumterms}, but also the Hamiltonian operator, in general, transforms a basis state to a linear combination of many distinct basis states.

\section{Spectrum and dynamics in truncated theory
\label{sec:specdyn}}
\noindent
The analysis of the previous section revealed that the most suitable formulation to construct the Hilbert space, generate the Hamiltonian, and analyze system's properties is the LSH formulation. It reproduces the same Hamiltonian as the KS case in the angular-momentum basis in the physical sector, and it does so by eliminating the need to diagonalize the Gauss's law operator in the angular-momentum basis at the level of states. As was discussed in Sec.~\ref{sec:HilbertKS}, this latter step generally leads to a proliferation of terms in a given physical state when expressed in the angular-momentum basis, see Fig.~\ref{fig:kspbcnumterms}, increasing the cost of the Hamiltonian-matrix generation. The fermionized form leads to the same Hilbert space with OBC as the KS formulation in the physical Hilbert space and  the LSH formulation but involves redundancies in the spectrum as mentioned above, making it less suitable for computation than the LSH framework. It further introduces non-local fermionic interactions which make it less efficient to simulate with certain classical and quantum-simulation algorithms. The bosonic formulation, once augmented by the U(1) gauge fields with sufficiently large truncation on their excitations, reproduces the KS theory exactly, as discussed in Sec.~\ref{sec:HilbertB}. As a result, it is less favored compared with the LSH framework when it comes to computational cost. Finally, the QLM does not correspond to the KS Hamiltonian unless certain limits are implemented. As a result, the comparison between the KS and QLM spectrum and dynamics will not be meaningful away from those limits, as is the case in this study. Given these considerations, in this section, the LSH formulation (or identically KS formulation in the physical Hilbert space) will be mostly considered for the analysis of the dependence of the spectrum and dynamics on the cutoff on the gauge-field excitations, but an example of the effect of the extra gauge DOF on the spectrum of the purely bosonic formulation will be analyzed as well.

\begin{figure}[!t]
\includegraphics[width=0.490\textwidth]{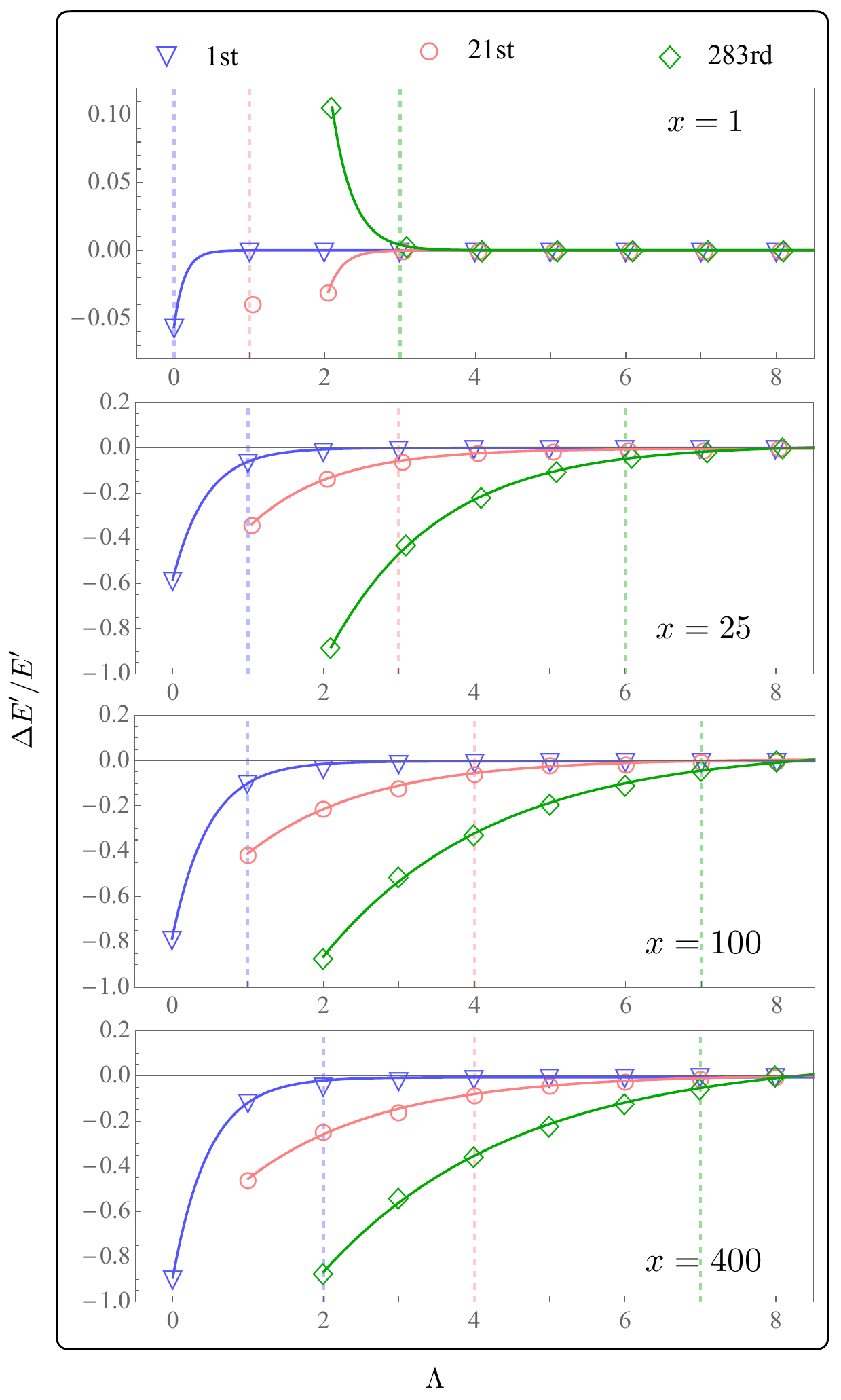}
\caption{
The quantity $\frac{\Delta E'}{E'} \equiv \frac{E'(\Lambda)-E'(\Lambda=8)}{E'(\Lambda)}$ as a function of $\Lambda$ for various values of $x$, and for the 1st, 21st, and 283rd lowest-lying states in the spectrum of the KS Hamiltonian in the physical Hilbert space with $N=6$ and $\nu=1$ with PBC. $E'(\Lambda)$ is the scaled energy corresponding to the scaled Hamiltonian in Eq.~(\ref{eq:HprimeKS}). The dashed lines denote the first $\Lambda$ values at which the corresponding scaled energies become equal or less than $10\%$ of their values at $\Lambda = 8$ (which are approximated as the $\Lambda \to \infty$ values). When needed for presentational clarity, the points are artificially displaced along the horizontal axis by a small amount. The numerical values associated with these plots are provided in Supplemental Material.}
\label{fig:deltaenergyvslambdapbc}
\end{figure}
\begin{figure}[!h]
\includegraphics[width=0.490\textwidth]{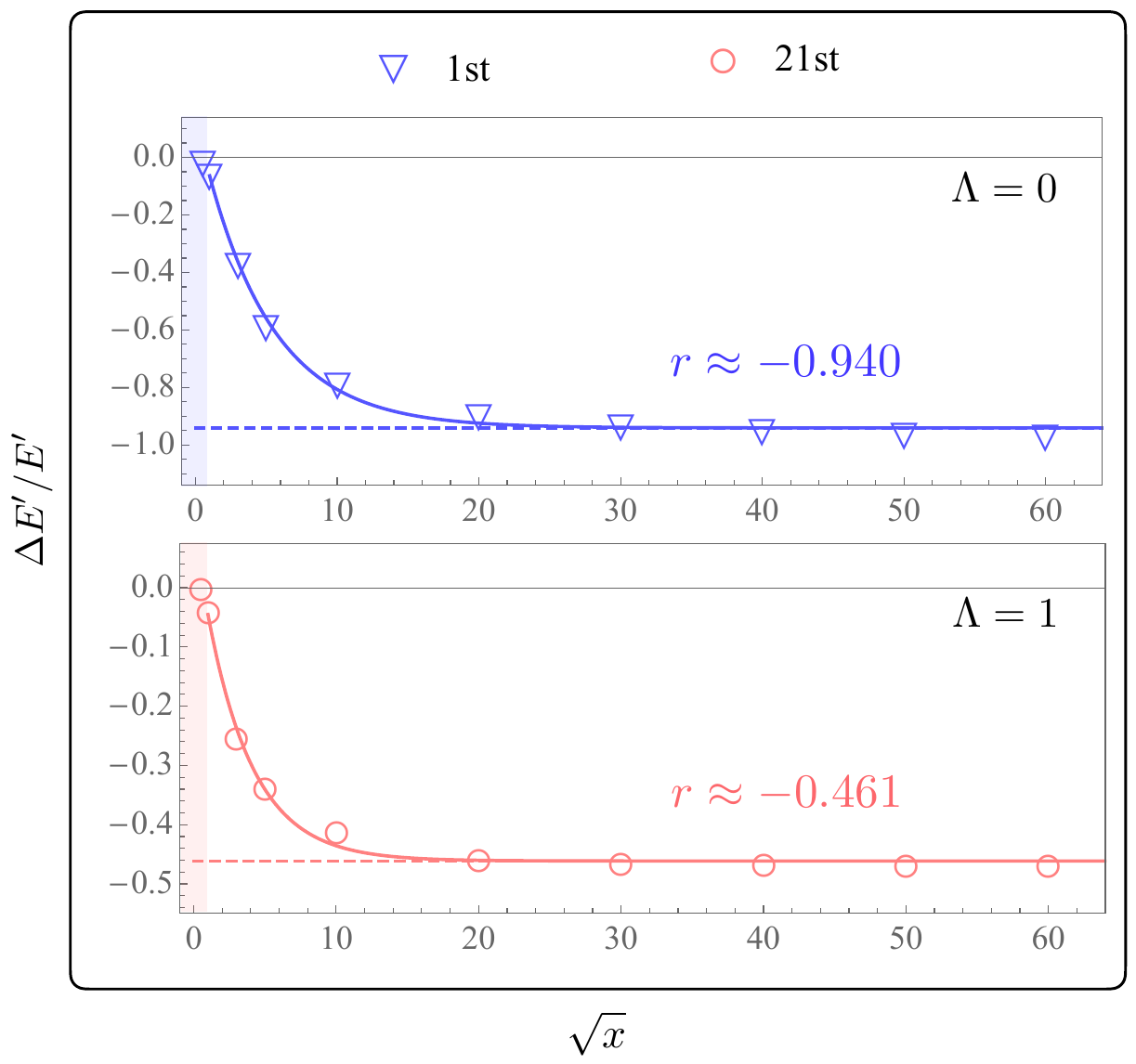}
\caption{
The quantity $\frac{\Delta E'}{E'} \equiv \frac{E'(\Lambda)-E'(\Lambda=8)}{E'(\Lambda)}$ as a function of $\sqrt{x}$ for given values of $\Lambda$ as denoted in the plots, and for the 1st and 21st states in the spectrum of the KS Hamiltonian in the physical Hilbert space with $N=6$ and $\nu=1$ with PBC. $E'(\Lambda)$ is the scaled energy corresponding to the scaled Hamiltonian in Eq.~(\ref{eq:HprimeKS}). The asymptotic ($x \to \infty$) values of the quantity, $r$, are obtained from the fits to data points in each case with an exponentially varying function of $\sqrt{x}$, and are denoted in the plots. The colored regions denote the $\sqrt{x}$ values excluded from the fits. The numerical values associated with these plots are provided in Supplemental Material.
}
\label{fig:deltaenergyvsxpbc}
\end{figure}

For convenience, the KS Hamiltonian in Eq.~(\ref{eq:HKS}) can be multiplied by $\frac{2}{a g^2}$ to yield a dimensionless scaled Hamiltonian $H'^{({\rm KS})}$:
\begin{eqnarray}
H'^{({\rm KS})} &\equiv& \frac{2}{a g^2}H^{({\rm KS})}
\nonumber\\
&=&x\sum_{x=0}^{N_1}\left[\psi^{\dagger}(x) U(x) \psi(x+1)+{\rm h.c.} \right]+
\nonumber\\
&& \sum_{x=0}^{N_2} \bm{E}^2(x)+
\mu \sum_{n=0}^{N_3}(-1)^n\psi^\dagger(x) \psi(x),
\label{eq:HprimeKS}
\end{eqnarray}
where $x = \frac{1}{a^2g^2}$, $\mu = \frac{2m}{g^2a}$, and $N_1$, $N_2$, and $N_3$ are defined in Sec.~\ref{sec:formalism}. The limit $x \to 0$ corresponds to the strong-coupling limit of the theory, while the limit $x \to \infty$ at a fixed $\frac{m}{g}$ provides a trajectory in parameter space along which the continuum limit can be taken. The matrix elements of this Hamiltonian can be formed using the KS angular-momentum or LSH bases, giving rise to identical results in the physical sector, which serves as a strong check of the newly-developed LSH formulation for the 1+1~D case. While efficient classical simulations such as those based on tensor networks have enabled studies of SU(2) lattice gauge theories with a large number (hundreds) of sites~\cite{Banuls:2019rao, Banuls:2017ena, Sala:2018dui, Silvi:2016cas}, enabling the continuum limit of the results to be taken systematically, such considerations are not the focus of this work. Instead, the aim is to arrive at qualitative conclusions regarding the cutoff dependence of several quantities of interest in small lattices where exact diagonalization is achievable on a typical desktop computer. A comprehensive study of SU(2) physics within the LSH formulation using state-of-the-art classical Hamiltonian algorithms is underway and will be presented elsewhere. Note that such methods still encounter the issues stemming from inefficient formulations of DOF and implementation of non-Abelian symmetries especially toward higher dimensions, and hence the analysis of this work is relevant in optimizing such studies.

\subsection{Spectrum analysis
\label{sec:spec}}
%
%
\begin{figure*}[!t]
\includegraphics[width=0.990\textwidth]{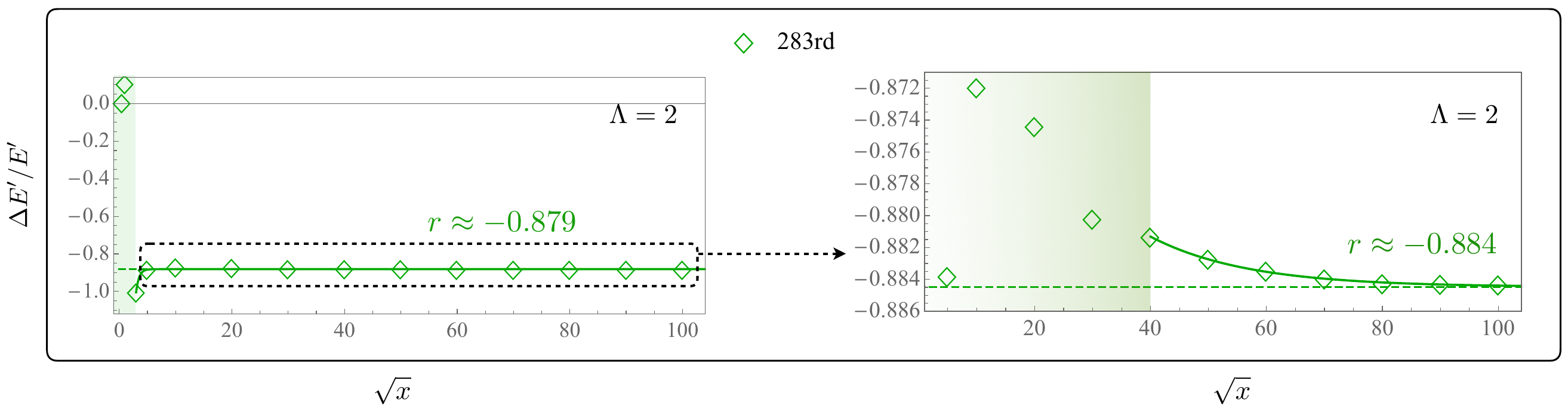}
\caption{
The same quantity as in Fig.~\ref{fig:deltaenergyvsxpbc} corresponding to the 283rd state in the spectrum for $\Lambda=2$. With coarser resolution, the asymptotic ($x \to \infty$) value of the function can be obtained from a fit to the exponential form shown in the left panel. By zooming into the large-$x$ region of the plot, a finer structure can be observed in the data as a function of $\sqrt{x}$, revealing an exponential asymptote to the continuum value that starts at a much larger value of $\sqrt{x}$ as shown in the right panel, with an asymptotic value that is within sub-percent of that obtained in the left plot. The numerical values associated with these plots are provided in Supplemental Material. 
}
\label{fig:deltaenergyvsxpbc}
\end{figure*}
\begin{figure}[!h]
\includegraphics[width=0.485\textwidth]{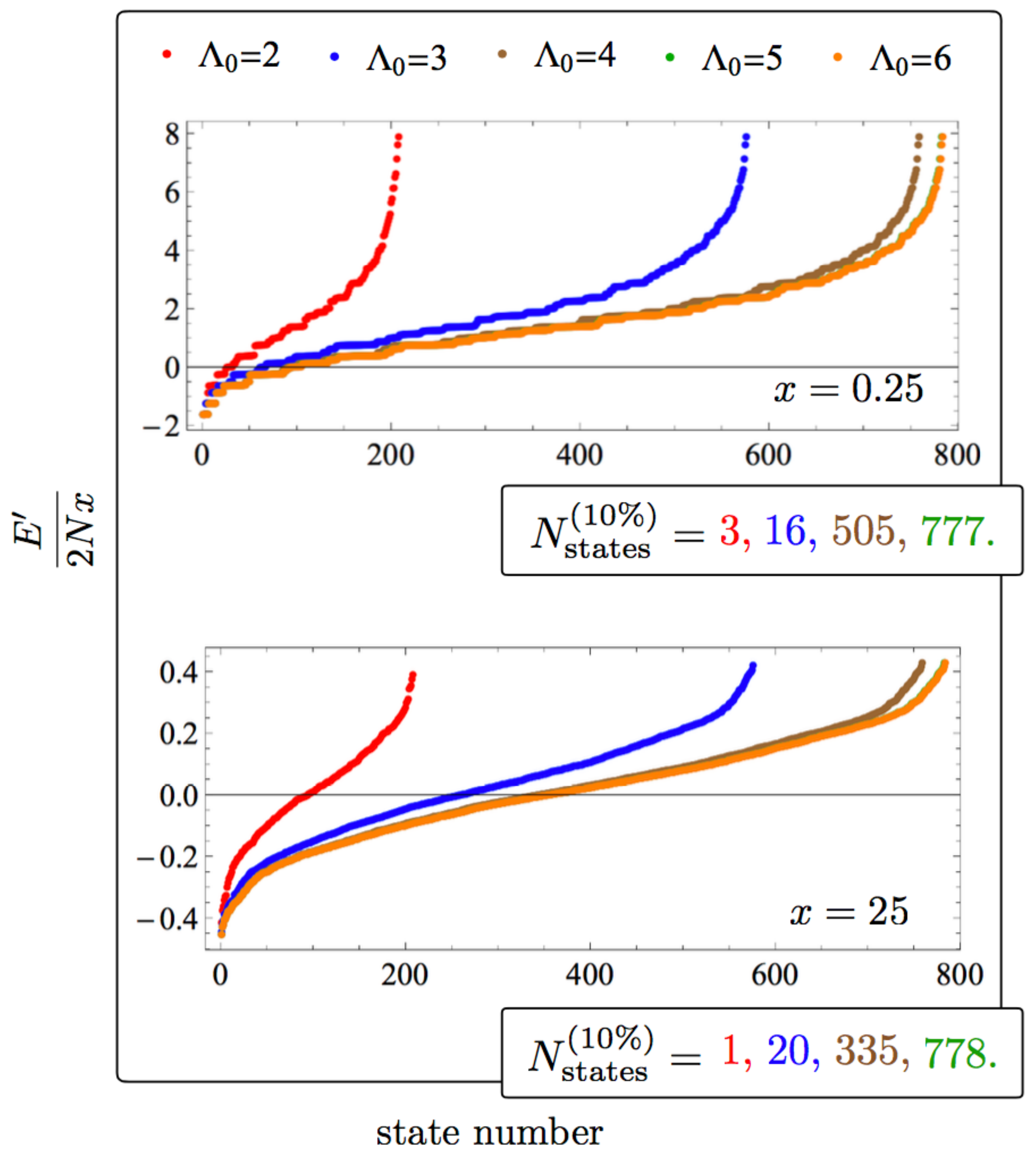}
\caption{
The spectra of the bosonized KS Hamiltonian in the physical Hilbert space with OBC for $N=8$ and $\nu=1$ at two values of $x$ and with $\Lambda=8$, and for several values of the cutoff on the U(1) electric field of the extended U(2) gauge theory, $\Lambda_0$. At $\Lambda_0=6$, the Hilbert space in this sector coincides with that of the SU(2) theory. The quantity plotted is $\frac{E'}{2Nx}$, where $E'$ is the scaled energy corresponding to the scaled Hamiltonian in Eq.~(\ref{eq:HprimeKS}). $N_{\rm states}^{(10\%)}$ denotes the total number of eigenstates in the physical Hilbert space of the bosonized theory with a given $\Lambda_0$ whose scaled energies in the units considered are $\leq 10\%$ of the exact energy (corresponding to the saturated value $\Lambda_0=6$). The numerical values associated with these plots are provided in Supplemental Material.}
\label{fig:evergyvslambda0bosonized}
\end{figure}
\begin{figure}[!t]
\includegraphics[width=0.490\textwidth]{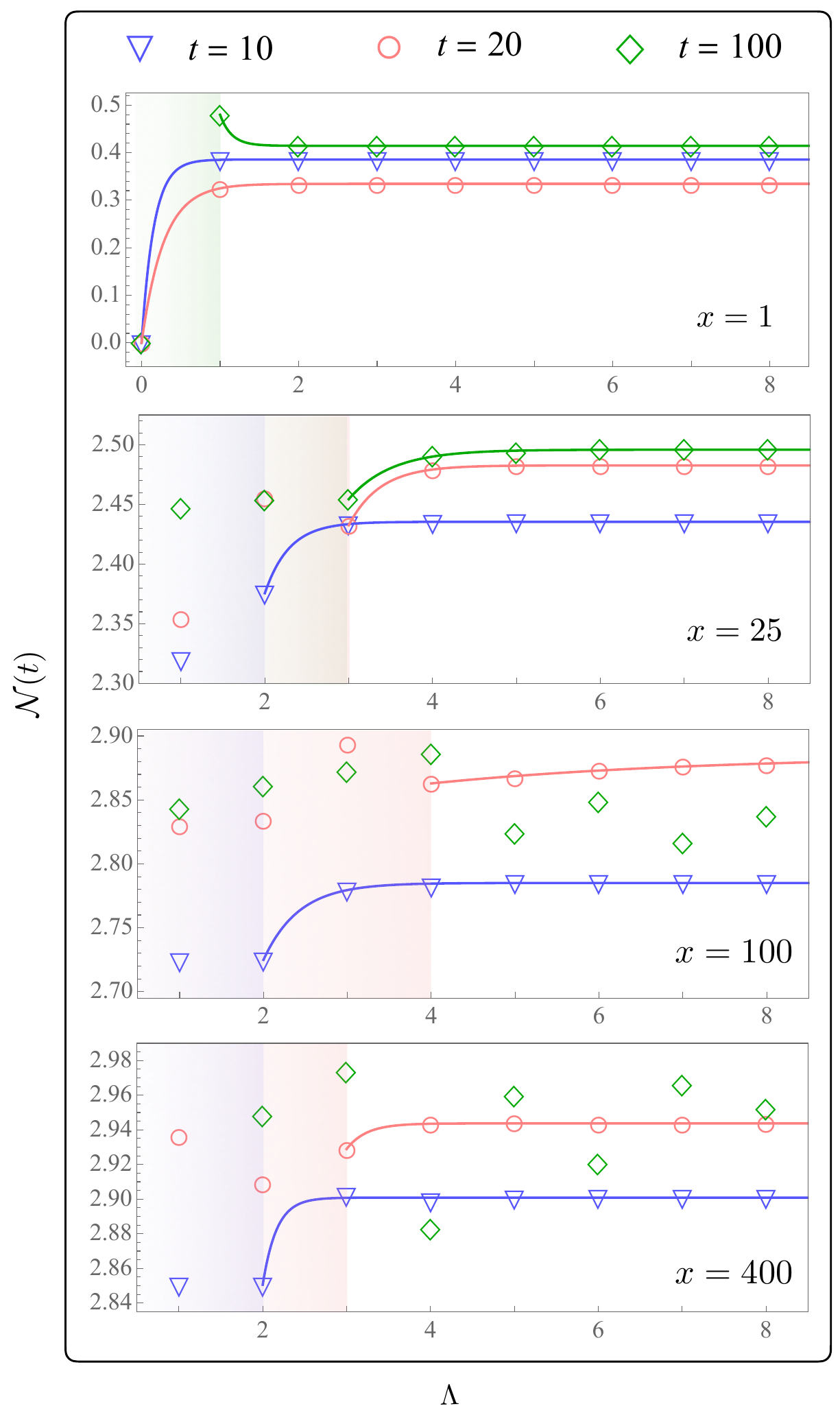}
\caption{The quantity $\mathcal{N}(t)$ defined in Eq.~(\ref{eq:Nt}) as a function of $\Lambda$ for various values of $x$ and $t$ in the KS Hamiltonian in the physical Hilbert space with $N=6$ and $\nu=1$ with PBC. $t$ is the absolute time as defined in the text in units of $a$. When possible, the points are fit to $\mathcal{N}=Ae^{-B\Lambda}+C$ and the colored regions associated with each $t$ are excluded from such fits. The numerical values associated with these plots are provided in Supplemental Material. 
}
\label{fig:timedepvslambda}
\end{figure}
\begin{figure*}[!t]
\includegraphics[width=0.990\textwidth]{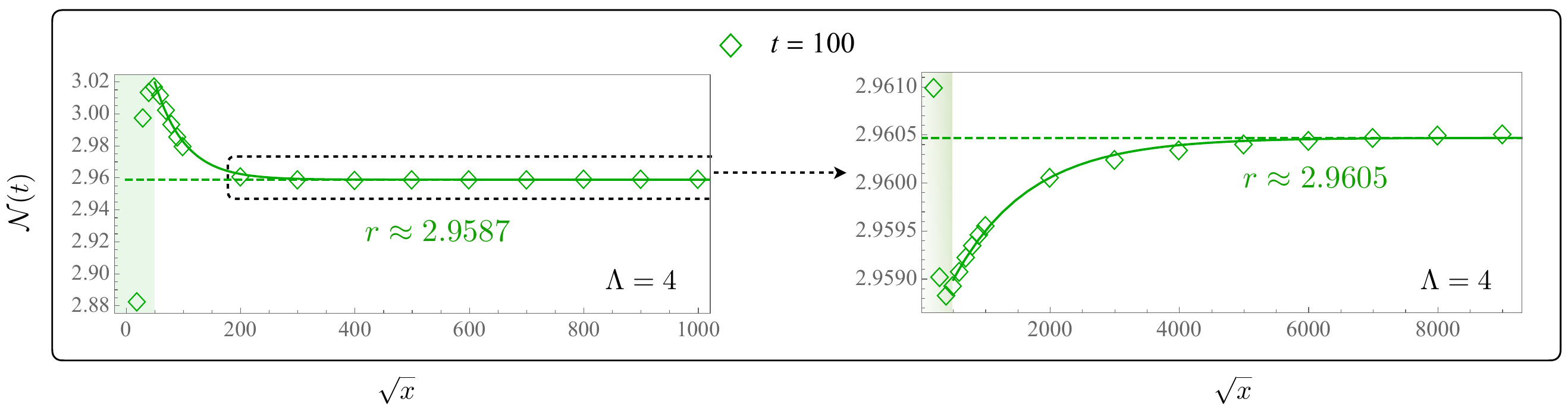}
\caption{The left  panel shows the quantity $\mathcal{N}(t)$ defined in Eq.~(\ref{eq:Nt}) as a function of $\sqrt{x}$ for $\Lambda=4$ and $t=100$ in the KS Hamiltonian in the physical Hilbert space with $N=6$ and $\nu=1$ with PBC. The points are fit to $\mathcal{N} \sim e^{-r\sqrt{x}}$ and the colored regions are excluded from such fits. By zooming into the large-$x$ region of the plot, a finer structure can be observed in the data as a function of $\sqrt{x}$, revealing an exponential asymptote to the continuum value that starts at a much larger value of $\sqrt{x}$ as shown in the right panel, with an asymptotic value that is within sub-percent of that obtained in the left plot. The numerical values associated with these plots are provided in Supplemental Material. 
}
\label{fig:ntvssqxt100}
\end{figure*}
\noindent
Since the gauge-matter interaction term, i.e., the term proportional to $x$ in Eq.~(\ref{eq:HprimeKS}), mixes the electric-field eigenstates, it is anticipated that for large $x$ (small $ga$), the truncation of the Hamiltonian matrix, imposed by the cutoff on the electric-field excitations, will introduce a more significant deviation from the exact spectrum compared with the small-$x$ values. The ratio of the scaled energy, $E'$, obtained from diagonalizing the scaled Hamiltonian in Eq.~(\ref{eq:HprimeKS}) to the corresponding $x$ value, nonetheless, will approach a constant value as $x \to \infty$, which represents the continuum limit once $N \to \infty$. The same is true for the asymptotic large-$x$ behavior of the ratios of energies at a given value of $x$. As a result, these ratios provide a more suitable means to analyze the scaling behavior of spectral quantities. Figure \ref{fig:energyvslambdapbc} plots the dimensionless quantity $\frac{E'}{2Nx}$ for the $\nu=1$ sector of the KS Hamiltonian with $N=6$ and with PBC, for several values of $x$ and $\Lambda$. As is visually evident, $\Lambda=2$ appears sufficient to obtain almost cutoff-independent energy eigenvalues for a large fraction of the spectrum associated with a given $\Lambda$ when $x < 1$. The higher-energy states exhibit a more significant cutoff dependence, as is expected, and such cutoff dependence becomes more prominent as $x$ increases. Nonetheless, as $x \to \infty$, the shown quantity asymptotes to fixed values for all states in the spectrum, hence stabilizing the effect of the cutoff.

To make these features more explicit, three eigenstates of this systems are selected, corresponding to the 1st, 21st, and 283rd lowest eigenenergies. The 1st eigenenergy exists for all $\Lambda$ values considered. The 21st eigenenergy is the first eigenenergy that does not appear in the spectrum of the $\Lambda=0$ theory, as the $\Lambda=0$ theory has a 20-dimensional physical Hilbert space given the quantum numbers specified. Finally, the 283rd eigenenergy does not exist in the spectrum of the $\Lambda=1$ theory, as the $\Lambda=1$ theory has a 282-dimensional physical Hilbert space given the quantum numbers considered. Figure~\ref{fig:deltaenergyvslambdapbc} plots the quantity $\frac{\Delta E'}{E'} \equiv \frac{E'(\Lambda)-E'(\Lambda=8)}{E'(\Lambda)}$ as a function of $\Lambda$ for various $x$ values and for the three states identified.\footnote{For $x \ll 1$ as noted above, the lowest-lying energy levels are minimally affected by the truncation cutoff and the scale dependence for $\Lambda > 1$ is almost vanishing, so this case has not been included in the figure.} As is seen for all $x$, as $\Lambda$ increases, the quantity plotted exponentially approaches the exact value, which is taken to be the value at $\Lambda =8$. This scaling behavior may only exhibit itself for sufficiently large $\Lambda$, as is seen clearly in the case of the 21st state at $x=1$. This indicates that a sufficiently large $\Lambda$ is required to be able to approximate the $\Lambda \to \infty$ values using these asymptotic relations. This is similar, in spirit, to the numerical procedure outlined in Ref.~\cite{Konik:2007cb}, where an iterative process is outlined to include higher-energy `shells' in the Hilbert space in order to drive the quantities to the scaling regime. In this region, analytic relations based on RG can then be applied to arrive at cutoff-independent observables. In the same plots, the values of $\Lambda$ at which the scaled energy deviates by $10\%$ or less from the $E'(\Lambda=8)$ value are denoted as dashed vertical lines. As is seen, the required $\Lambda$ values to reach this moderate accuracy increases with increasing $x$ but stabilizes at fixed values toward the continuum limit. For higher-energy states in the theory, as is evident from the behavior of the 283rd eigenenergy, the energy corresponding to the $\Lambda=8$ value is not a good approximation for the exact value and higher cutoffs must be included in the analysis. The scaling of the shown quantity, nonetheless, is expected to follow the same exponential form for sufficiently large $\Lambda$ adjusted to the large eigenenergy considered.

Finally, to demonstrate the convergence of the quantity $\frac{\Delta E'}{E'}$ to a constant value at large $x$, the same eigenenergies picked in Fig.~\ref{fig:deltaenergyvslambdapbc} are taken for the smallest $\Lambda$ values for which those eigenenergies exist, and are plotted as a function of $x$ in Fig.~\ref{fig:deltaenergyvsxpbc}. After a threshold $x$ value (that is larger for higher-energy states), the dependence of the ratio on $x$ can be approximated by $r+ae^{-ax}$, where $r,a,b$ are constants to be fit. The values of the constant $r$ obtained from the fits shown correspond to the asymptotic large-$x$ value of the ratios, and are provided in the plots for each of the eigenenergies selected.

For OBC, similar qualitative features are seen for the cutoff dependence of the spectrum, as are evident from Figs.~\ref{fig:energyvslambdaobc}-\ref{fig:deltaenergyvsxobc} in Appendix~\ref{app:OBC}. Such cutoff dependence, however, is less severe than the PBC case in general, consistent with the fact that the gauge fields are not truly dynamical DOF in gauge theories in 1+1~D with OBC. In particular, once the value $\Lambda = N$ is reached, the spectrum of the theory is the same as that corresponding to $\Lambda \to \infty$. Nonetheless, for large lattice sizes, it is not computationally viable to implement this saturation value, since the Hilbert space increases as $e^{q\Lambda}$ with the $q$ value obtained from a fit to results at small lattice sizes. As a result, the scaling behavior presented for select spectral quantities in Appendix~\ref{app:OBC} will still serve to guide approaching cutoff-independent results from computations performed at smaller values of $\Lambda$.

As discussed in Sec.~\ref{sec:HilbertB}, the bosonized KS theory exhibits a physical Hilbert space that is identical to that of the KS theory with fermions in the limit where the cutoff on the U(1) electric-field excitations of the extended U(2) theory is equal or larger than the saturation value $\Lambda_0=N$. To examine the effect of the extra U(1) cutoff, consider, as an example, the bosonized theory with OBC and with $N=8$ and $\nu=1$. In this sector, the full physical Hilbert space is reached once the cutoff on the SU(2) electric-field excitations is set to $\Lambda \geq 6$, giving rise to a total of $784$ states. However, such a value is only achieved if $\Lambda_0 \geq 6$ as well. In particular, the number of allowed states in the physical Hilbert space is $0,0,208,576,759,783$ for $\Lambda_0 = 0,1,2,3,4,5$, respectively. The values of the scaled energies obtained from $H'=\frac{2}{ag^2}H$ for a small and a large value of the coupling $x$ (divided by $2Nx$) are plotted in Fig.~\ref{fig:evergyvslambda0bosonized} for $\Lambda_0 = 2,3,4,5,6$, demonstrating a strong $\Lambda_0$ dependence in the spectrum. For comparison, the total number of states whose scaled energies in the shown unit are less than or equal to $10\%$ of the exact value (corresponding to $\Lambda_0=6$) are shown in the figure.

\subsection{Dynamics analysis
\label{sec:dyn}}
Dynamical quantities, i.e., those obtained from time-dependent expectation values, exhibit cutoff dependence as well. In fact, intuitively one expects that when the system evolves, the unitary evolution operator $e^{-iHt}$ ultimately mixes the initial state with all states within the Hilbert space with the same (conserved) quantum numbers~\cite{csahinouglu2020hamiltonian}, where $H$ denotes the Hamiltonian of the system and $t$ is time. In other words, various parts of the system entangle quickly as the system evolves. This feature is at the heart of the inefficiency of tensor-network methods in dynamical studies as the evolution time increases.  As a result, quantities evolved to large times compared with $1/||H||$ suffer from Hilbert-space truncation errors more significantly. Here, $||H||$ is some form of a Hamiltonian norm, such as the absolute value of the largest energy eigenvalue.

To demonstrate this point quantitatively in the case of the KS SU(2) LGT in 1+1~D, consider a select quantity obtained from
\begin{eqnarray}
\mathcal{N}(t)&=& \frac{1}{2\mu} \langle {\rm vac}(x=0)|e^{iH'^{({\rm KS})}t'} \times
\nonumber\\
&&(H'^{({\rm KS})}_M+N\mu)e^{-iH'^{({\rm KS})}t'}|{\rm vac}(x=0) \rangle,
\label{eq:Nt}
\end{eqnarray}
where $\ket{{\rm vac}(x=0)}$ denotes the strong-coupling vacuum, $t'$ is a dimensionless time: $t' \equiv \frac{t}{2x}$, and all other quantities are defined after Eq.~(\ref{eq:HprimeKS}). Here $t$ is time in units of $a$. The quantity in Eq.~(\ref{eq:Nt}), therefore, counts the occupation number of particles (as opposed to antiparticles) on the lattice as a function of time, and is identically zero for the strong-coupling vacuum. Figure~\ref{fig:timedepvslambda} plots $\mathcal{N}(t)$ as a function of $\Lambda$ for various values of $x$ and $t$ in the KS Hamiltonian within the physical Hilbert space with $N=6$ and $\nu=1$ with PBC. Note that $t$ is the absolute time, and hence $\mathcal{N}(t)$ at different $x$ values can be directly compared at a fixed $t$. At small $x$ values, the mixing among states expressed in the electric-field basis is small, and the dependence of the time-evolved quantity on the cutoff can be approximated by $Ae^{-B\Lambda}+C$ even for rather small $\Lambda$. As $x$ becomes larger, larger values of $\Lambda$ are needed to enter this asymptotic scaling region, at which point the $\Lambda \to \infty$ value of the quantity can be estimated based on the exponential scaling. This feature stabilizes as the function of $x$ for large $x$ to ensure convergence to a continuum limit (once $N \to \infty$ is taken). Not surprisingly though, at larger values of $t$, it takes a large value of $\Lambda$ to enter the asymptotic region. For example, at $t=100$ no convergence is seen in the quantity considered for $x=100$ and $x=400$ in the range of cutoffs considered. This example demonstrates that studying dynamics in this theory with high precision will be computationally demanding, given the extreme sensitivity to the cutoff considered at long evolution times.

Another feature of dynamical quantities is that their continuum limit seems to be achieved more slowly, requiring more computational resources compared with static quantities. Although such a limit should be considered as an ordered double limit in which $N \to \infty$ and $x \to \infty$, this point can be demonstrated at a fixed $N$ and for various $x$ values. Examples of this feature are shown in Fig.~\ref{fig:ntvssqxt100}, in which the quantity $\mathcal{N}(t)$ is plotted as a function of $x$ for the KS theory with PBC and $\Lambda=4$ at a select time, $t=100$. This observation is consistent with the slow convergence of high-lying energies in the spectrum as a function of $x$ (see Fig.~\ref{fig:deltaenergyvsxpbc}), considering that time evolution leads to contributions to observables from all states in the same symmetry sector. Nonetheless, if percent-level precision is needed in this quantity, one would not need computations to be performed at extreme values of $x$.

It is worth noting that the point at which the quantities can be described by exponentially converging functions as $\Lambda \to \infty$ or as $x \to \infty$, depends on the size of the system, the choice of $m/g$, and the quantity considered. To formally understand the underlying mechanism for such  an scaling and its breakdown, beyond the empirical observations of this work in small systems, is the subject of a future investigation and potentially has its root in the Nyquist-Shannon sampling theorem~\cite{Klco:2018zqz}. A more systematic understanding of Hamiltonian renormalization can shed light on this question, as we have alluded to in Sec.~\ref{sec:conclusions}.

\section{Conclusions and outlook
\label{sec:conclusions}}
\begin{figure*}[!t]
\includegraphics[width=0.80\textwidth]{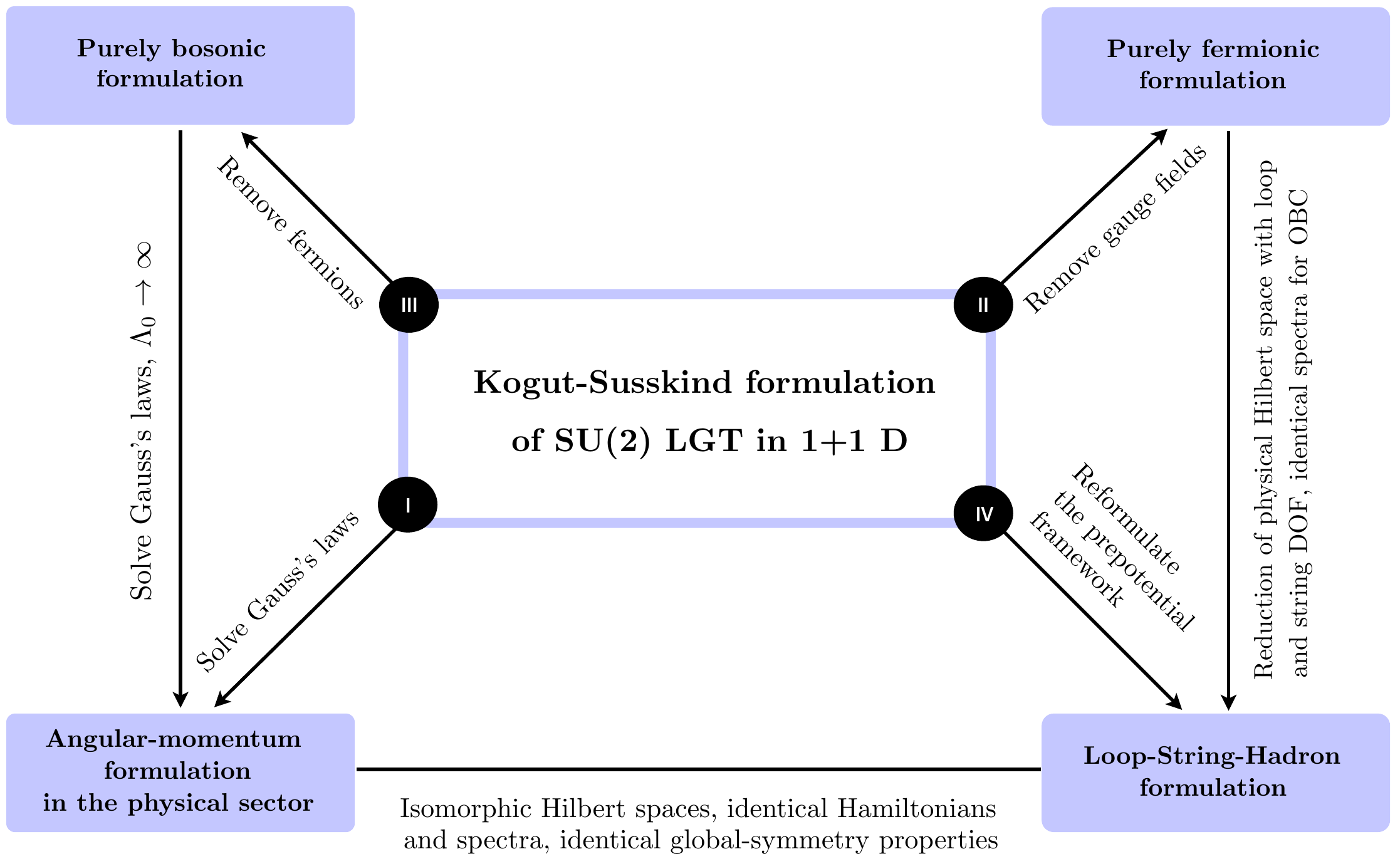}
\caption{Various formulations of the KS SU(2) LGT in 1+1~D studied in this work, and the connection among them.}
\label{fig:summary}
\end{figure*}
\noindent
The present paper provides a stepping stone to establishing the most efficient Hamiltonian formulation(s) for the case of a non-Abelian gauge field theories in 1+1~Dimensions, with implications for other non-Abelian gauge theories and for higher dimensions. The theory studied in this paper is the SU(2) lattice gauge theory coupled to matter. Its various forms, characterized by the retained degrees of freedom and the basis states employed, are thoroughly analyzed and scrutinized. These forms include the original Kogut-Susskind formulation within the angular-momentum basis, the purely fermionic and purely bosonic formulations (possible only with open boundary conditions), and the recently developed Loop-String-Hadron formulation, with both periodic and open boundary conditions. A schematic of these formulations and the relation among them are shown in Fig.~\ref{fig:summary}. Given the extent and the scope of the conclusions reached, a summary of the main findings of the work is presented here, along with a number of comments on the outlook of this work and future directions:
\begin{itemize}
\item[$\rhd$]{The dimension of the Hilbert space grows quickly with the size of the lattice, $N$, as $\sim e^{pN}$. In the angular-momentum basis, a naive counting of the full Hilbert space reveals that the exponent grows logarithmically with the cutoff on the electric-field excitations. With open boundary conditions, a saturation value of $\Lambda=N+2\epsilon_0$ ensures that the full untruncated theory is achieved, but at the cost of expediting the growth of the full Hilbert space by $\sim e^{N\log{N}}$. The physical Hilbert space in the same representation still grows exponentially with the system's size, but with a $p$ value that approaches a constant of $\mathcal{O}(1)$ as a function of the cutoff. The empirical values are obtained in Sec.~\ref{sec:Hilbert}. Additionally, as a function of the cutoff on the electric-field excitations, the dimension of the physical Hilbert space approaches a constant with OBC but grows linearly with PBC. The considerable reduction in the Hilbert-space dimension by restricting computations to the physical sector comes at a significant cost, since imposing the non-Abelian Gauss's laws locally amounts to a super-exponential cost in generating the Hilbert space. There is a similar cost associated with generating the Hamiltonian matrix, simply due to the fact that each physical state in the original angular-momentum basis is a linear combination of many basis states, with the number of terms in given physical states growing exponentially with the cutoff. Furthermore, the action of Hamiltonian on such states generates other populated linear combination of states. Given these features, the angular-momentum basis of the KS theory appears the least appealing framework for Hamiltonian simulation.}
\item[$\rhd$]{In 1+1~D, the gauge DOF are non-dynamical with OBC and can be fully fixed by the boundary value of the electric field. This means that restricting to the physical Hilbert space comes as no computational cost. Nonetheless, the fermionic basis states lead to redundancies, and while the spectra of both the original theory in the physical sector and the fermionic theory are identical, there are degeneracies in the spectrum in the latter case. The difference in the dimension of the physical Hilbert spaces in these formulations is not significant though, and the ratio remains constant as the system's size grows. The larger Hilbert space is compensated by a slightly faster decline in the Hamiltonian-matrix density of the fermionic formulation, and the cost of computing observables, i.e., eigenvalue evaluation or matrix exponentiation, is only $\mathcal{O}(N)$ higher in the fermionic formulation compared with the angular-momentum formulation in the physical sector. Nonetheless, the purely fermionic formulation is the least costly among all the formulations considered in the limit of large lattice sizes, as expected. The drawback is that this formulation has no analog in higher dimensions as the boundary conditions and Gauss's law constraints are not sufficient to fully remove the gauge DOF. As a result, other formulations with better generalizability perspective and similar computational-resource requirements are favorable.}
\item[$\rhd$]{Alternatively, the fermionic DOF can be removed at the expense of enlarging the gauge group to U(2) and introducing additional complexity due to the need for sufficiently large cutoffs on the additional U(1) electric field excitations, and for appropriate encoding of the fermionic statistics. This theory, once the saturating value of the U(1) cutoff is imposed with OBC, and the Gauss's laws associated with both the U(1) and SU(2) symmetries are solved, has an identical physical Hilbert space to that in the original KS theory. All the complexities arising from imposing the non-Abelian constraints in the angular-momentum basis remain relevant to this formulation, and hence no particular advantage is gained by the purely bosonic formulation with classical Hamiltonian-simulation algorithms. Such a formulation, nonetheless, is argued to be beneficial in digital quantum simulations as it avoids non-local fermion-to-hard-core-boson transformations.}
\item[$\rhd$]{The complexities with the angular-momentum representation of the KS theory and the increasing mixing of the local basis states in the physical Hilbert space are avoided altogether in the LSH formulation. Here, the building blocks of the Hamiltonian are operators that act directly on local gauge-invariant basis states, namely strings and loops, and only an Abelian Gauss's law on the link between the lattice sites is left to be imposed \emph{a posteriori}. Furthermore, the action of each operator in the Hamiltonian maps a given state to one and only one state, a feature that is absent in the angular-momentum basis in the physical Hilbert space. While the string  quantum numbers can only take values $0$ or $1$, the loop quantum numbers are non-negative integers, and their spectrum must be truncated in practical applications. With the same truncation cutoff, the Hamiltonian matrices in the physical Hilbert space with the angular-momentum and LSH  bases are identical, nonetheless the cost of generating such a matrix is exponentially suppressed in the LSH formulation. Similarly, while in the original angular-momentum basis, the construction of Hilbert space is the most costly step in the simulation algorithm, this step is the least  costly with the LSH formulation, and offers up to hundreds of orders of magnitude reduction in the cost of the simulation compared with the original formulation for lattices with tens of sites, as demonstrated in Sec.~\ref{sec:cost}. Although the purely fermionic formulation remains slightly less complex and computationally less costly to simulate (despite its all-to-all fermionic interactions), the LSH framework is already generalizable to higher dimensions, and is shown to retain the same crucial features as the 1+1~D case. Namely, a complete basis of local gauge-invariant states exists to eliminate the need for the imposition of non-Abelian Gauss's laws (that are more complicated to implement in higher dimensions) at the level of basis states, and that the Abelian Gauss's law is still implemented in only one dimension, thanks to a point-splitting procedure developed in Ref.~\cite{Anishetty:2018vod, Raychowdhury:2018tfj}. As a result, this framework appears to be the most appealing among all the formulations considered for the purpose of Hamiltonian simulation of the SU(2) LGT. The generalization of such an efficient formulation to the SU(3) LGT coupled to matter is, therefore, an important next step in approaching the ultimate goal of Hamiltonian simulation of QCD.}
\item[$\rhd$]{Hamiltonian-truncation methods, common in a variety of problems from condensed-matter physics to nuclear physics, have been well developed over decades and have a direct connection to the concept of Renormalization Group. In particular, Wegner~\cite{wegner1994flow} and Glazek and Wilson~\cite{glazek1993renormalization} independently studied the concept of renormalization within a Hamiltonian approach. They established powerful methods such as similarity renormalization group to continuously flow the transformed Hamiltonian to a banded-diagonal form such that the high-energy and low-energy modes in a given theory are maximally decoupled. This procedure leads to smaller truncation errors resulting from considering only a partial sector of the Hilbert space in generating and processing the Hamiltonian matrix. More concretely, such approaches put the system in the asymptotic scaling regime such that the limit of $\Lambda \to \infty$ can be taken using asymptotic formulae, see e.g., Ref.~\cite{Konik:2007cb}. In other words, a Hamiltonian truncation approach can be used to derive the RG equations for the parameters of the theory, promoting the truncated bare Hamiltonian to an effective renormalized Hamiltonian. A dedicated study of renormalized Hamiltonian gauge theories which extends upon previous work is a necessity as one moves toward using the Hamiltonian approach in the simulations of gauge theories. In this first study, however, we have only presented empirical observations for the scale dependence of the spectrum and dynamics in the SU(2) gauge theory in 1+1~D in small lattices to illuminate several seemingly general features. In particular, the exponential dependence of observables on the excitation of gauge fields is established for energies and given dynamical expectation values, but a sufficiently large cutoff is needed to enter this scaling regime for high-energy states, as well as the long-time-evolved states. Approaches in constructing a renormalized theory in the Hamiltonian formulation will be studied in future work to shed light on the interesting empirical features observed. In this context, the formal analysis of digitization effects in bose-fermi systems~\cite{somma2015quantum, Macridin:2018oli} and in scalar field theories~\cite{Klco:2018zqz} will potentially be relevant.}
\item[$\rhd$]{Despite the steady progress in advanced simulation algorithms such as tensor networks, the enormous cost of Hamiltonian simulation, even with the most efficient formulations identified, suggests that the quantum simulation of LGTs of interest may be the ultimate path to reaching accurate and large-scale dynamical simulations. It is, therefore, important to extend the analysis of this work to verify the efficiency of the LSH formulation compared to other formulations in the context of quantum simulation, and to find the most efficient quantum algorithms for simulating it, similar to studies conducted to date for various quantum field theories, such as scalar field theories~\cite{Jordan:2011ci, Klco:2018zqz} and the lattice Schwinger model~\cite{Shaw:2020udc}. The insights obtained from the thorough comparative study of this work are the key to embarking on such investigations in a quantum-computing setting. Analog quantum-simulation protocols within the LSH formulation are being developed, see e.g., Ref.~\cite{Dasgupta:2020}, and similar developments are planned for digital protocols. In particular, explicitly solving Gauss's laws in a digital quantum computation and projecting out the physical Hilbert space of the non-Abelian SU(2) LGTs in any dimension has only been achieved using the LSH formalism~\cite{Raychowdhury:2018osk}. Presumably, a mixture of various formulations, such as purely bosonic formulations or partially fermionic formulation with the LSH basis, could provide more computational benefits in this context, but further investigations need to be conducted to reach accurate conclusions.}
\end{itemize}

\section*{Acknowledgments}
\noindent
We are grateful to Rudranil Basu for valuable discussions and to Jesse Stryker for valuable comments on the manuscript. ZD and AS are supported by the US Department of Energy's Office of Science Early Career Award DE-SC0020271. ZD is further supported by the Maryland Center for Fundamental Physics, University of Maryland, College Park. IR is supported by the U.S. Department of Energy's Office of Science, Office of Advanced Scientific Computing Research, Quantum Computing Application Teams program, under fieldwork proposal number ERKJ347.

\bibliography{bibi.bib}

\appendix
\section{Quantum Link Model
\label{app:QLM}}
\begin{figure*}[t!]
\includegraphics[width=0.735\textwidth]{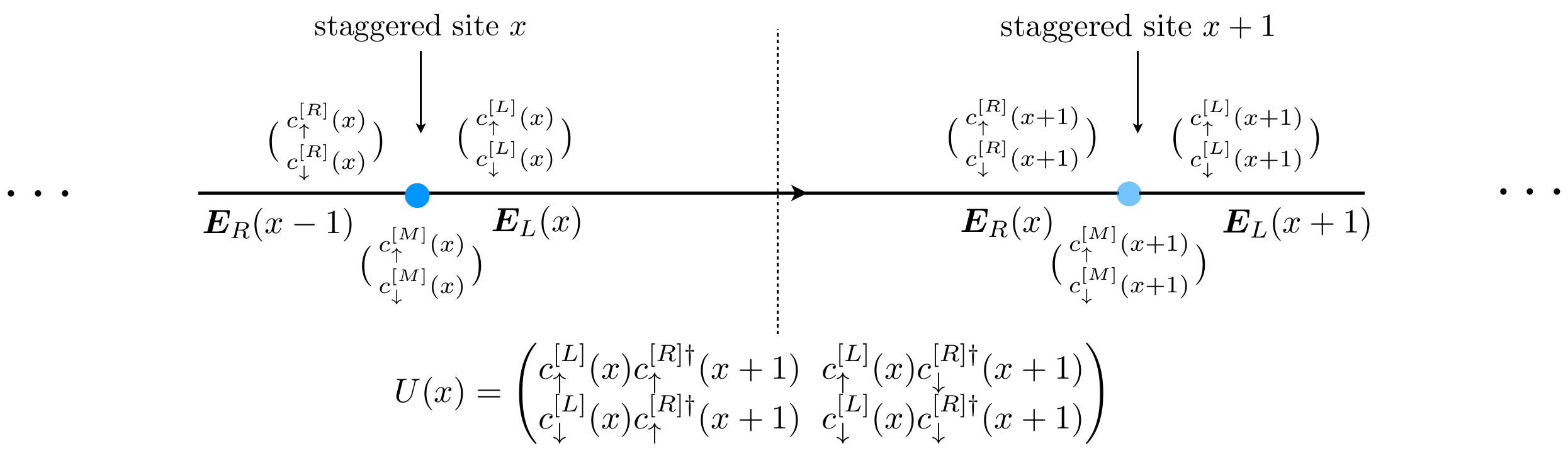}
\caption{Rishon constituents at two adjacent staggered sites $x$ and $x+1$ in a one-dimensional lattice within the QLM formulation of SU(2) LGT.
}
\label{fig:QLM_link}
\end{figure*}
\noindent
An alternate formulation of LGTs within the Hamiltonian framework, known as QLM has been developed more than two decades ago~\cite{Chandrasekharan:1996ih, Brower:1997ha}, and has been particularly popular in the context of proposals for quantum simulations of Abelian and non-Abelian LGTs~\cite{Banerjee:2012pg,Banerjee:2012xg,Stannigel:2013zka,Kasper:2015cca,Mil:2019pbt}. In this section, we briefly describe the QLM for the SU(2) LGT using the `rishon' representation in the same setting as discussed in earlier sections, i.e., with staggered fermionic fields on a one-dimensional spatial lattice, following the presentation of Ref.~\cite{Silvi:2016cas}. Like in all other formulations, the fermionic matter field is in the fundamental representation of SU(2) and is defined at each lattice site. Gauge fields, defined in terms of link variables, couple to the matter fields and constitute the interaction part of the Hamiltonian. One important difference at this point is that the link operator is constructed as a composite operator made up of `rishons', which are fermions. The rishon and matter constituents on two staggered sites on the lattice are shown in Fig.~\ref{fig:QLM_link}. The mapping between the variables is:
\begin{eqnarray}
\{\hat{\bm E}_{L/R}(x),\hat{U}(x),\psi (x)\} \rightarrow \{ \hat{c}^{[\tau]}_{s}(x) \},
\end{eqnarray}
where $s=\uparrow,\downarrow$ is the SU(2) component and $\tau=M,L,R$, where $M$ refers to matter, and $L(R)$ refers to rishons on the left (right) side of the link. The canonical conjugate variables, as well as the link variable and the matter field, are constructed as
 \begin{align}
 &\hat{E}^a_L(x)= -\hat{c}^{[L]\dagger}_{s}(x)T^a_{s,s'}\hat{c}^{[L]}_{s'}(x),\\
 &\hat{E}^a_R(x)= \hat{c}^{[R]\dagger}_{s}(x+1)T^a_{s,s'}\hat{c}^{[R]}_{s'}(x+1),\\
 &\hat{U}_{s,s'}(x) = \hat{c}^{[L]}_{s}(x)\hat{c}^{[R]\dagger}_{s'}(x+1)\label{rU},\\
&\psi^\dagger_s(x)= \hat{c}^{[M]\dagger}_{s}(x),
 \end{align}
where $T^a=\frac{1}{2} \tau^a$, and $\tau^a$ is the $a^{\rm th}$ Pauli matrix, with $a=1,2,3$. These definitions yield the same commutation relations between the electric fields and link variable as in Eqs.~(\ref{eq:EUcomm}). However, unlike the KS theory, $[U_{s,s'},U^\dagger_{t,t'}]$ can be non-vanishing in the QLM formulation, and the link operator has necessarily a finite-dimensional representation.

This construction re-expresses the different parts of the SU(2) LGT Hamiltonian in terms of rishon operators as
\begin{eqnarray}
H_I^{(\rm QLM)}&=& t\sum_{x,s,s'} \left[\hat{c}^{[M]\dagger}_{s}(x)\hat{U}_{s,s'}(x)\hat{c}^{[M]}_{s'}(x+1)+{\rm h.c.} \right]\nonumber \\
&=& t\sum_{x,s,s'} \Big[\hat{c}^{[M]\dagger}_{s}(x)\hat{c}^{[L]}_{s}(x)\hat{c}^{[R]\dagger}_{s'}(x+1)\hat{c}^{[M]}_{s'}(x+1)
\nonumber \\
&&
\hspace{5.01 cm} +\,{\rm h.c.} \Big],
\label{QLM_HI}
\\
H_E^{(\rm QLM)}&=& \frac{g_0^2}{2} \sum_{x}\left[\hat{\bm E}^2_R(x-1)+ \hat{\bm E}^2_L(x)\right] \nonumber \\
&\equiv& \frac{3g_0^2}{8} \sum_{x}\Bigg[\Big(\hat{n}^{[R]}_{\uparrow}(x-1)+\hat{n}^{[R]}_{\downarrow}(x-1)\nonumber \\ &&-2\hat{n}^{[R]}_{\uparrow}(x-1)\hat{n}^{[R]}_{\downarrow}(x-1)\Big)\label{QLM_HE} \nonumber\\
&&+\left(\hat{n}^{[L]}_{\uparrow}(x)+\hat{n}^{[L]}_{\downarrow}(x)-2\hat{n}^{[L]}_{\uparrow}(x)\hat{n}^{[L]}_{\downarrow}(x)\right)\Bigg], \nonumber\\
\\
H_M^{(\rm QLM)}&=& m\sum_{x} (-1)^x \left[\hat{n}^{[M]}_{\uparrow}(x)+ \hat{n}^{[M]}_{\downarrow}(x)\right]\label{QLM_HM},
\end{eqnarray}
where $\hat{n}^{[\tau]}_{s}(x) \equiv {\hat{c}^{[\tau]^\dagger}_s}(x) \, \hat{c}^{[\tau]}_s(x)$ is the fermionic occupation number at site $x$ for the $s=\uparrow,\downarrow$ components of the $\tau=L,R,M$ type fermion.
The Hamiltonian, being the sum of the three parts given in Eqs.~(\ref{QLM_HE}), (\ref{QLM_HM}), and (\ref{QLM_HI}), conserves the total number of fermions at each site, i.e.,
\begin{eqnarray}
\sum_{s=\uparrow,\downarrow} \left[n^{[M]}_{s}(x)+n^{[L]}_{s}(x)+n^{[R]}_{s}(x)\right]={\rm const.},
\end{eqnarray}
implying that the local symmetry is instead $SU(2)\otimes U(1)$. Hence, in order to recover the SU(2) gauge symmetry of interest, one needs to add the following U(1) symmetry-breaking term to the Hamiltonian,
\begin{eqnarray}
H_{\rm break}^{(\rm QLM)}&=& \frac{\epsilon}{2}\sum_{x}\left[\det \hat{U}(x,x+1)+{\rm h.c.}
\right] \nonumber \\
&=& \epsilon \sum_{x} \Big[\hat{c}^{[L]\dagger}_{\uparrow}(x)\hat{c}^{[L]\dagger}_{\downarrow}(x)\hat{c}^{[R]}_{\downarrow}(x+1)\hat{c}^{[R]}_{\uparrow}(x+1) \nonumber \\
&&\hspace{4.5 cm}+ {\rm h.c.}
\Big].
\label{eq:QLM_Hbreak}
\end{eqnarray}
Hence, the total Hamiltonian of the SU(2) QLM in 1+1~D is given by
\begin{equation}
H^{(\rm QLM)}=H_I^{(\rm QLM)}+H_M^{(\rm QLM)}+H_E^{(\rm QLM)}+H_{\rm break}^{(\rm QLM)},
\label{eq:HQLMtottot}
\end{equation}
with the Gauss's law operator\footnote{To compare with Ref.~\cite{Silvi:2016cas}, note that $\bm{E}_L \equiv -\bm{J}_L$ and $\bm{E}_R \equiv \bm{J}_R$.}
\begin{eqnarray}
\hat{G}^a(x)=-\hat{E}^a_L(x)+\hat{E}^a_R(x-1)+c^{[M]\dagger}_{s}(x)T^a_{s,s'}c^{[M]}_{s'}(x).
\nonumber\\
\label{QLM_GL}
\end{eqnarray}

It should be noted that the U(1) symmetry-breaking term in Eq.~(\ref{eq:QLM_Hbreak}) is only non-vanishing if the total number of fermionic rishons on each link is equal to two~\cite{Banerjee:2012xg}:
\begin{equation}
n^{[L]}_{\uparrow}(x)+n^{[L]}_{\downarrow}(x)+n^{[R]}_{\uparrow}(x+1)+n^{[R]}_{\downarrow}(x+1) =2.
\end{equation}
Since the Hamiltonian in Eq.~(\ref{eq:HQLMtottot}) conserves the rishon number on the link, this constraint is preserved.

\begin{figure}[t!]
\includegraphics[width=0.480\textwidth]{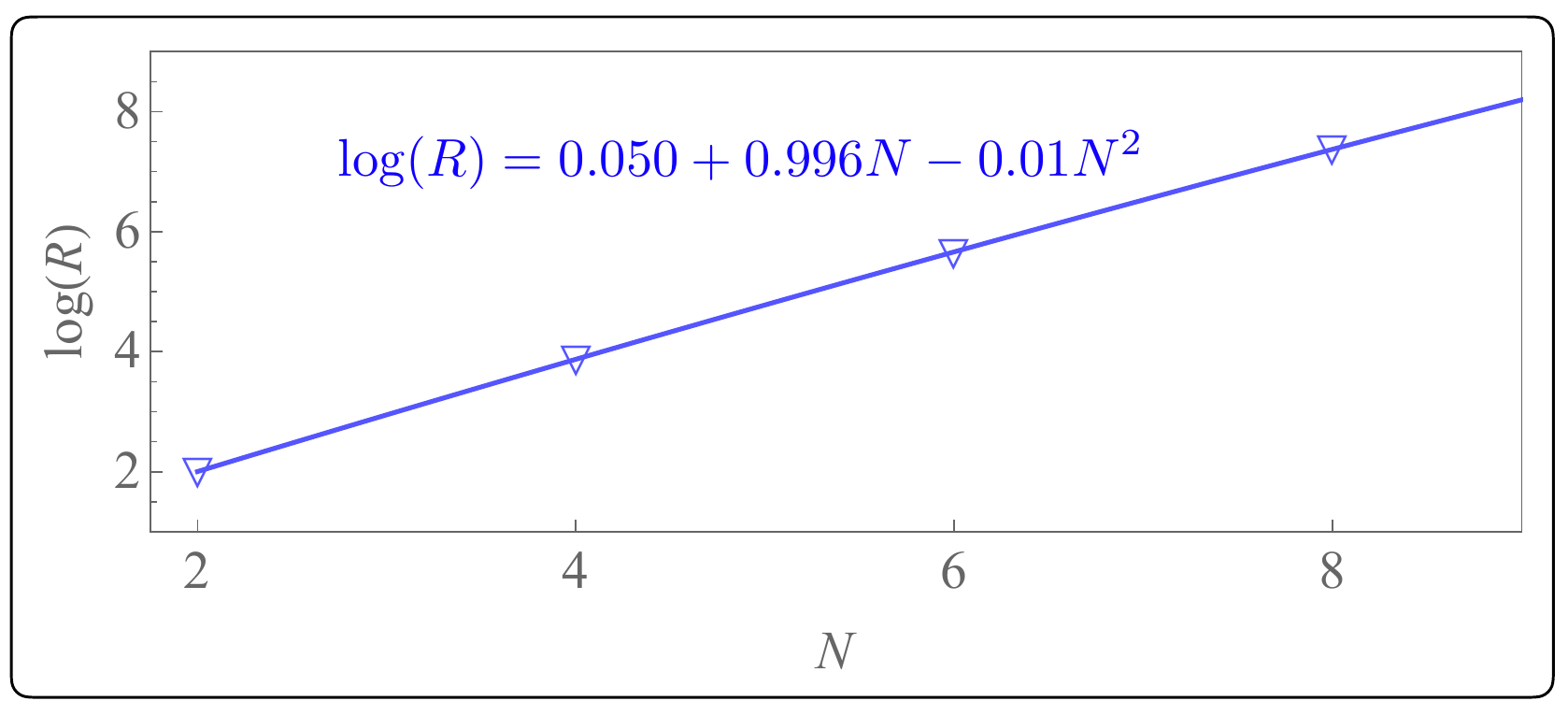}
\caption{
The (logarithm of the) ratio of the dimension of the physical Hilbert space within the QLM formulation to that within the KS (and LSH) formulation (the latter with a sufficiently large cutoff such that the number of basis states saturates to a fixed value), both with OBC, for several values of the lattice size, $N$. The empirical fit to the $N$ dependence of this quantity is shown in the figure. The numerical values associated with this plot are listed in Supplemental Material.
}
\label{fig:ksvsqlmobc}
\end{figure}

The QLM is only equivalent to the KS SU(2) gauge theory in the continuum limit and through a dimensional reduction~\cite{Chandrasekharan:1996ih, Brower:1997ha}. As a result, the spectrum and dynamics of the theory outlined here are not the same as those of the KS LGT obtained in this paper, even when the same lattice size and couplings are considered, a feature that can be verified by numerical computations. Nonetheless, it is interesting to ask whether the QLM is computationally efficient in the context of Hamiltonian simulation. While a thorough analysis of the computational cost of constructing the physical Hilbert space, generating the Hamiltonian, and computing observables with the QLM are not analyzed in this paper, one important observation can be made regarding the dimensionality of the physical Hilbert space of the QLM. Figure~\ref{fig:ksvsqlmobc} plots the ratio of the dimension of the physical Hilbert space in the QLM to that in the KS formulation (or equivalently the LSH formulation) when the cutoff is set to its saturating value with OBC. As is seen from the empirical fit to the values for the first lowest $N$ values, this ratio grows exponentially with $N$, and so from a computational standpoint, such a finite-dimensional representation of the SU(2) LGT is still costly. Note that here, one should also account for a comparable computational cost to the LSH formulation in generating the Hilbert space: there are six types of fermions present locally, requiring $2^6$ configurations to be generated at each site. Additionally, the Gauss's law constraints and the fixed-rishon number per link must be imposed when constructing the Hilbert space.

It is worth noting that while in the LSH formulation, the infinite-dimensional bosons are present (and are cut off at some finite value), with OBC their value can be fixed given the string configurations. In higher dimensions, there may still be an advantage in working with the finite-dimensional QLM since with other formulations, the Hilbert space grows with the cutoff on the link quantum numbers as these can no longer be fully fixed with boundary conditions. Nonetheless, the LSH formulation appears to be a competitive formulation of the SU(2) LGT, that not only is equivalent to the original KS theory, but also its economical resource requirements can bring it to the same footing as the QLM when it comes to quantum-simulation proposals on analog and digital simulators, an avenue that will be explored in the upcoming studies.
%

\section{Physical Hilbert-space dimensionality 
\label{app:details}}
\noindent
In this appendix, the numerical values of the dimension of the physical Hilbert space in the KS SU(2) LGT as a function of lattice sites $N$ are listed in Tables~\ref{tab:countsPBC} and \ref{tab:countsOBC} for PBC and OBC, respectively.

\begin{table*}[t!]
\begin{tabular}{ccccccccccc}
\multicolumn{11}{c}{$N=2$}\tabularnewline
\hline 
0 & 1 & 2 & 3 & 4 & 5 & 6 & 7 & 8 & 9 & 10\tabularnewline
\hline 
4 & 10 & 16 & 22 & 28 & 34 & 40 & 46 & 52 & 58 & 64\tabularnewline
\hline 
\multicolumn{11}{c}{}\tabularnewline
\multicolumn{11}{c}{$N=4$}\tabularnewline
\hline 
0 & 1 & 2 & 3 & 4 & 5 & 6 & 7 & 8 & 9 & 10\tabularnewline
\hline 
16 & 82 & 152 & 222 & 292 & 362 & 432 & 502 & 572 & 642 & 712\tabularnewline
\hline 
\multicolumn{11}{c}{}\tabularnewline
\multicolumn{11}{c}{$N=6$}\tabularnewline
\hline 
0 & 1 & 2 & 3 & 4 & 5 & 6 & 7 & 8 & 9 & 10\tabularnewline
\hline 
64 & 730 & 1,648 & 2,572 & 3,496 & 4,420 & 5,344 & 6,268 & 7,192 & 8,116 & 9,040\tabularnewline
\hline 
\multicolumn{11}{c}{}\tabularnewline
\multicolumn{11}{c}{$N=8$}\tabularnewline
\hline 
0 & 1 & 2 & 3 & 4 & 5 & 6 & 7 & 8 & 9 & 10\tabularnewline
\hline 
256 & 6,562 & 18,720 & 31,582 & 44,452 & 57,322 & 70,192 & 83,062 & 95,932 & 108,802 & 121,672\tabularnewline
\hline 
\multicolumn{11}{c}{}\tabularnewline
\multicolumn{11}{c}{$N=10$}\tabularnewline
\hline 
0 & 1 & 2 & 3 & 4 & 5 & 6 & 7 & 8 & 9 & 10\tabularnewline
\hline 
1,024 & 59,050 & 216,256 & 399,502 & 584,248 & 769,004 & 953,760 & 1,138,516 & 1,323,272 & 1,508,028 & 1,692,784\tabularnewline
\hline 
\end{tabular}
\caption{
The dimension of the physical Hilbert space of the KS (and LSH) formulation with PBC, for $N=2,4,\cdots,10$ and $\Lambda(=2J_{\rm max})=0,1,\cdots,10$.
}
\label{tab:countsPBC}
\end{table*}
\begin{table*}
\begin{tabular}{ccccccccccc}
\multicolumn{3}{c}{$N=2$} & & & & & & & & \tabularnewline
\cline{1-3} 
0 & 1 & 2 & & & & & & & & \tabularnewline
\cline{1-3} 
4 & 9 & 10 & & & & & & & & \tabularnewline
\cline{1-3} 
 & & & & & & & & & & \tabularnewline
\multicolumn{5}{c}{$N=4$} & & & & & & \tabularnewline
\cline{1-5} 
0 & 1 & 2 & 3 & 4 & & & & & & \tabularnewline
\cline{1-5} 
16 & 81 & 116 & 125 & 126 & & & & & & \tabularnewline
\cline{1-5} 
 & & & & & & & & & & \tabularnewline
\multicolumn{7}{c}{$N=6$} & & & & \tabularnewline
\cline{1-7} 
0 & 1 & 2 & 3 & 4 & 5 & 6 & & & & \tabularnewline
\cline{1-7} 
64 & 729 & 1,352 & 1,625 & 1,702 & 1,715 & 1,716 & & & & \tabularnewline
\cline{1-7} 
 & & & & & & & & & & \tabularnewline
\multicolumn{9}{c}{$N=8$} & & \tabularnewline
\cline{1-9} 
0 & 1 & 2 & 3 & 4 & 5 & 6 & 7 & 8 & & \tabularnewline
\cline{1-9} 
256 & 6,561 & 15,760 & 21,250 & 23,494 & 24,157 & 24,292 & 24,309 & 24,310 & & \tabularnewline
\cline{1-9} 
 & & & & & & & & & & \tabularnewline
\multicolumn{11}{c}{$N=10$}\tabularnewline
\hline 
0 & 1 & 2 & 3 & 4 & 5 & 6 & 7 & 8 & 9 & 10\tabularnewline
\hline 
1,024 & 59,049 & 183,712 & 278,125 & 326,382 & 345,401 & 351,176 & 352,485 & 352,694 & 352,715 & 352,716\tabularnewline
\hline 
\end{tabular}
\caption{
The dimension of the physical Hilbert space of the KS (and LSH) formulation with OBC, for $N=2,4,\cdots,10$ and $\Lambda(=2J_{\rm max})=0,1,\cdots,10$. For $\Lambda > N$, the dimension of the physical Hilbert space saturates to the value at $\Lambda = N$, and is hence not shown.
}
\label{tab:countsOBC}
\end{table*}
\begin{table*}
\label{tab:symm}
\begin{tabular}{c|c|cccccccccccccccccccccc}
\multicolumn{1}{c}{} & \multicolumn{1}{c}{} &  & \multicolumn{21}{c}{$Q$}\tabularnewline
\cline{4-24} 
\multicolumn{1}{c}{} & \multicolumn{1}{c}{} &  & 0 &  & 1 &  & 2 &  & 3 &  & 4 &  & 5 &  & 6 &  & 7 &  & 8 &  & 9 &  & 10\tabularnewline
\cline{4-24} 
\multicolumn{1}{c}{} & \multicolumn{1}{c}{} &  &  &  &  &  &  &  &  &  &  &  &  &  &  &  &  &  &  &  &  &  & \tabularnewline
\multirow{11}{*}{$q$} & 0 &  & 1 &  &  &  & 55 &  &  &  & 825 &  &  &  & 4,950 &  &  &  & 13,860 &  &  &  & 19,404\tabularnewline
 & 1 &  &  &  & 10 &  &  &  & 330 &  &  &  & 3,300 &  &  &  & 13,860 &  &  &  & 27,720 &  & \tabularnewline
 & 2 &  &  &  &  &  & 45 &  &  &  & 990 &  &  &  & 6,930 &  &  &  & 20,790 &  &  &  & 29,700\tabularnewline
 & 3 &  &  &  &  &  &  &  & 120 &  &  &  & 1,848 &  &  &  & 9,240 &  &  &  & 19,800 &  & \tabularnewline
 & 4 &  &  &  &  &  &  &  &  &  & 210 &  &  &  & 2,310 &  &  &  & 8,250 &  &  &  & 12,375\tabularnewline
 & 5 &  &  &  &  &  &  &  &  &  &  &  & 252 &  &  &  & 1,980 &  &  &  & 4,950 &  & \tabularnewline
 & 6 &  &  &  &  &  &  &  &  &  &  &  &  &  & 210 &  &  &  & 1,155 &  &  &  & 1,925\tabularnewline
 & 7 &  &  &  &  &  &  &  &  &  &  &  &  &  &  &  & 120 &  &  &  & 440 &  & \tabularnewline
 & 8 &  &  &  &  &  &  &  &  &  &  &  &  &  &  &  &  &  & 45 &  &  &  & 99\tabularnewline
 & 9 &  &  &  &  &  &  &  &  &  &  &  &  &  &  &  &  &  &  &  & 10 &  & \tabularnewline
 & 10 &  &  &  &  &  &  &  &  &  &  &  &  &  &  &  &  &  &  &  &  &  & 1\tabularnewline
\end{tabular}
\caption{Breakdown of the physical Hilbert space of dimension $352,716$ with OBC and for $N=10$ and $\Lambda \geq 10$. The dimensionalities of the sectors with charge $2N-Q$ are the same as those with charge $Q \in [0,10]$, and are not shown.}
\end{table*}
%

\section{Observables with open boundary conditions 
\label{app:OBC}}
\noindent
This appendix contains the same plots as in Sec.~\ref{sec:specdyn} for the spectrum and dynamics of the KS SU(2) LGT but with OBC. These plots were removed from the main text for brevity, and while the overall features and the general conclusions remain the same as those for the PBC, they are included in this appendix for completeness.
\begin{figure*}[]
\includegraphics[width=0.990\textwidth]{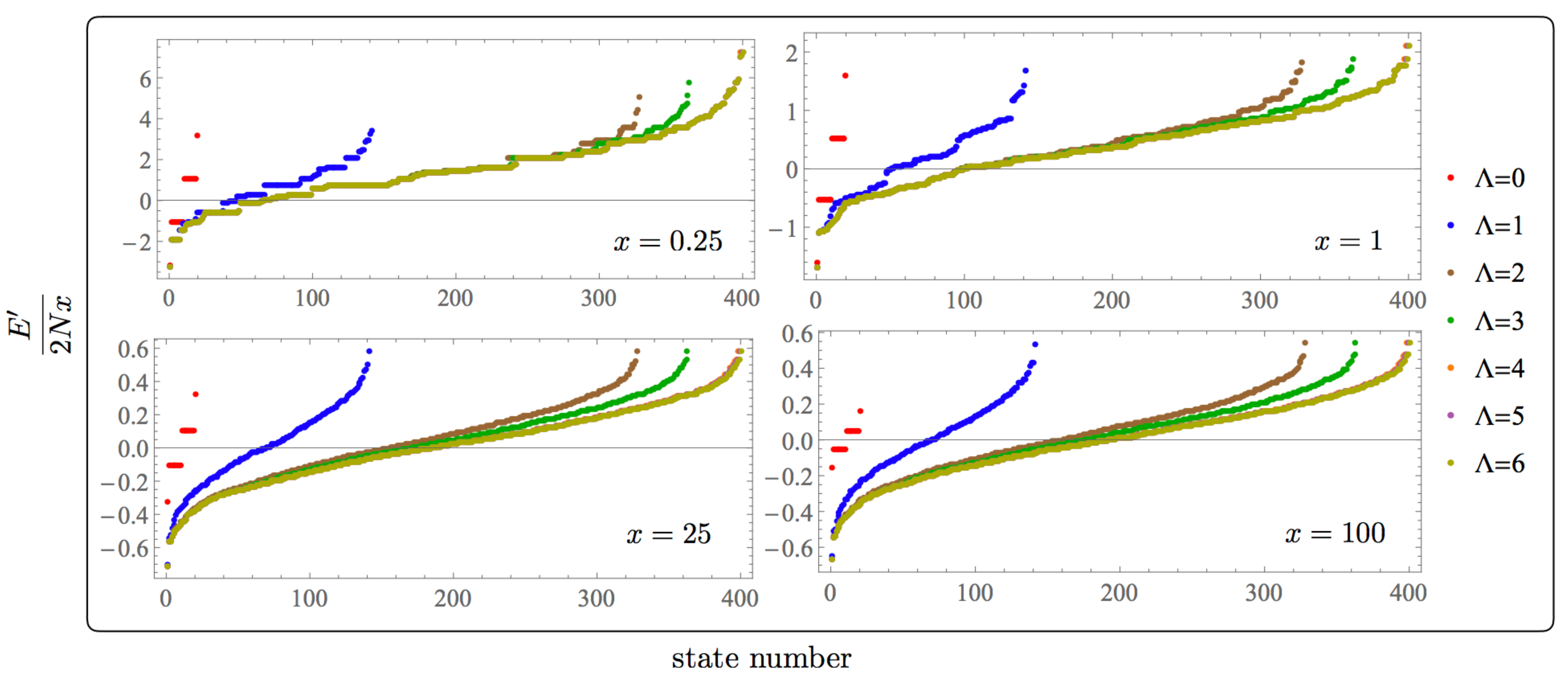}
\caption{
The spectra of the KS Hamiltonian in the physical Hilbert space with OBC for $N=6$, $\nu=1$, and various values of $x$ and $\Lambda$. More precisely, the quantity plotted is $\frac{E'}{2Nx}$, where $E'$ is the scaled energy corresponding to the scaled Hamiltonian in Eq.~(\ref{eq:HprimeKS}). The numerical values associated with these plots are provided in Supplemental Material.
}
\label{fig:energyvslambdaobc}
\end{figure*}
\begin{figure}[htp]
\includegraphics[width=0.490\textwidth]{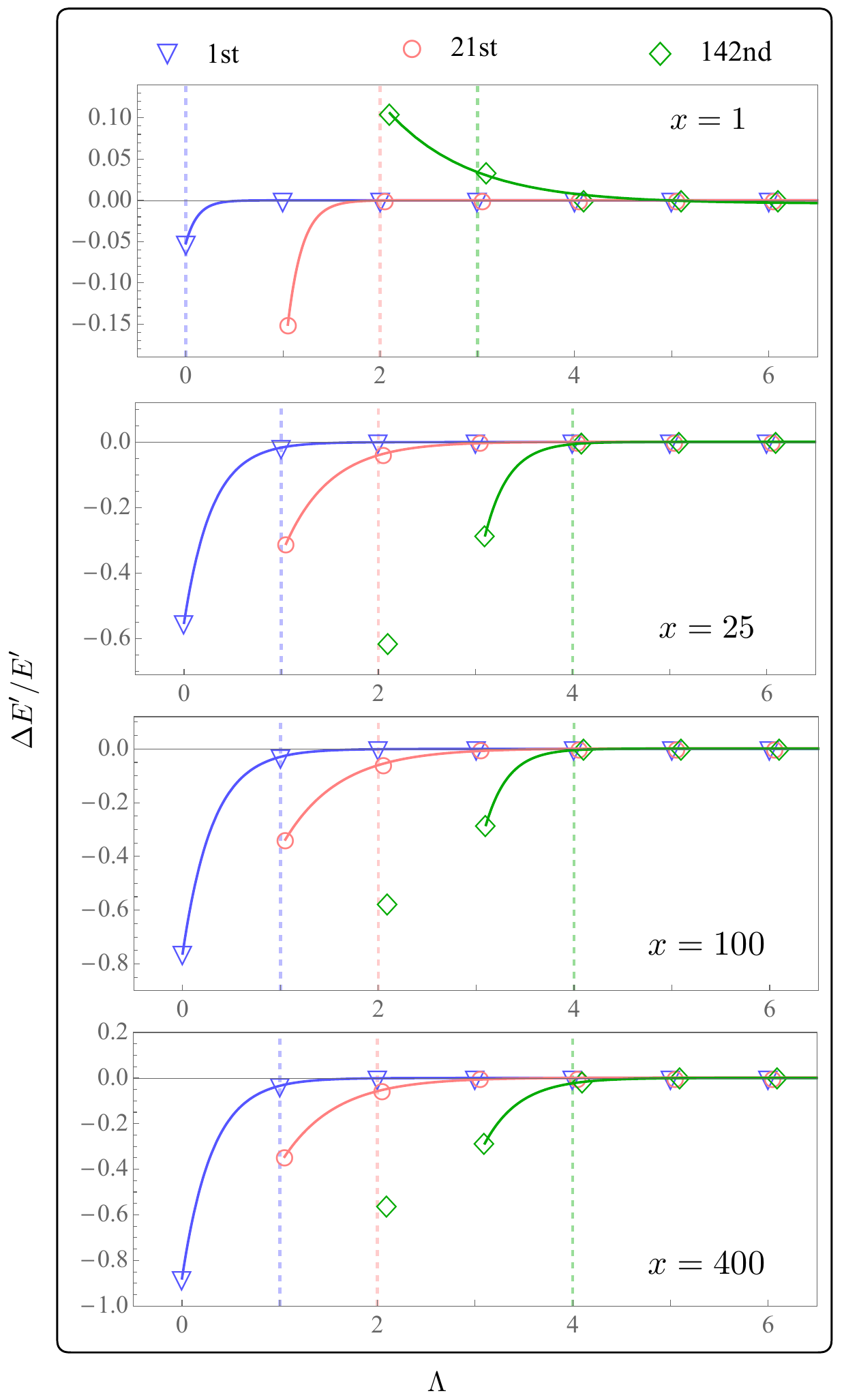}
\caption{
The quantity $\frac{\Delta E'}{E'} \equiv \frac{E'(\Lambda)-E'(\Lambda=8)}{E'(\Lambda)}$ as a function of $\Lambda$ for various values of $x$, and for the 1st, 21st, and 142nd states in the spectrum of the KS Hamiltonian in the physical Hilbert space with OBC and with $N=6$ and $\nu=1$. $E'(\Lambda)$ is the scaled energy corresponding to the scaled Hamiltonian in Eq.~(\ref{eq:HprimeKS}). The dashed lines denote the first $\Lambda$ values at which the corresponding scaled energies become equal or less than $10\%$ of their value at $\Lambda = 8$ (which are the $\Lambda \to \infty$ values). When needed for presentational clarity, the points are artificially displaced along the horizontal axes by a small amount. The numerical values associated with these plots are provided in Supplemental Material.
}
\label{fig:deltaenergyvslambdaobc}
\end{figure}
\begin{figure}[htp]
\includegraphics[width=0.490\textwidth]{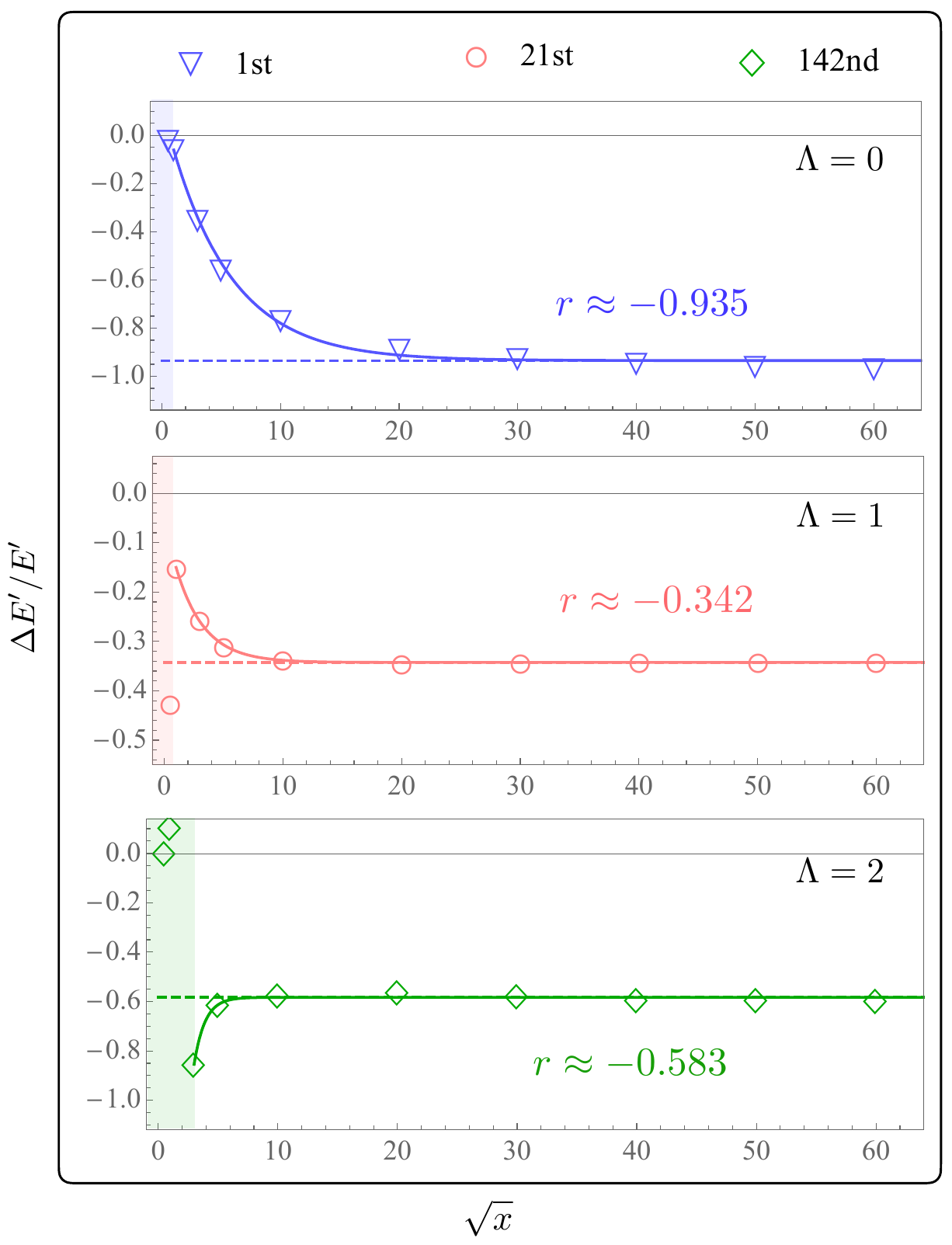}
\caption{
The quantity $\frac{\Delta E'}{E'} \equiv \frac{E'(\Lambda)-E'(\Lambda=8)}{E'(\Lambda)}$ as a function of $\sqrt{x}$ for given values of $\Lambda$ as denoted in the plots, and for the 1st, 21st, and 142nd states in the spectrum of the KS Hamiltonian in the physical Hilbert space with OBC and with $N=6$ and $\nu=1$. $E'(\Lambda)$ is the scaled energy corresponding to the scaled Hamiltonian in Eq.~(\ref{eq:HprimeKS}). The asymptotic ($x \to \infty$) values of the quantity are obtained from the fits to data points in each case with an exponentially varying function of $x$, and are denoted in the plot. The colored regions denote the $\sqrt{x}$ values excluded from the fits. The numerical values associated with these plots are provided in Supplemental Material.
}
\label{fig:deltaenergyvsxobc}
\end{figure}
\clearpage

\begin{figure}[htp]
\includegraphics[width=0.490\textwidth]{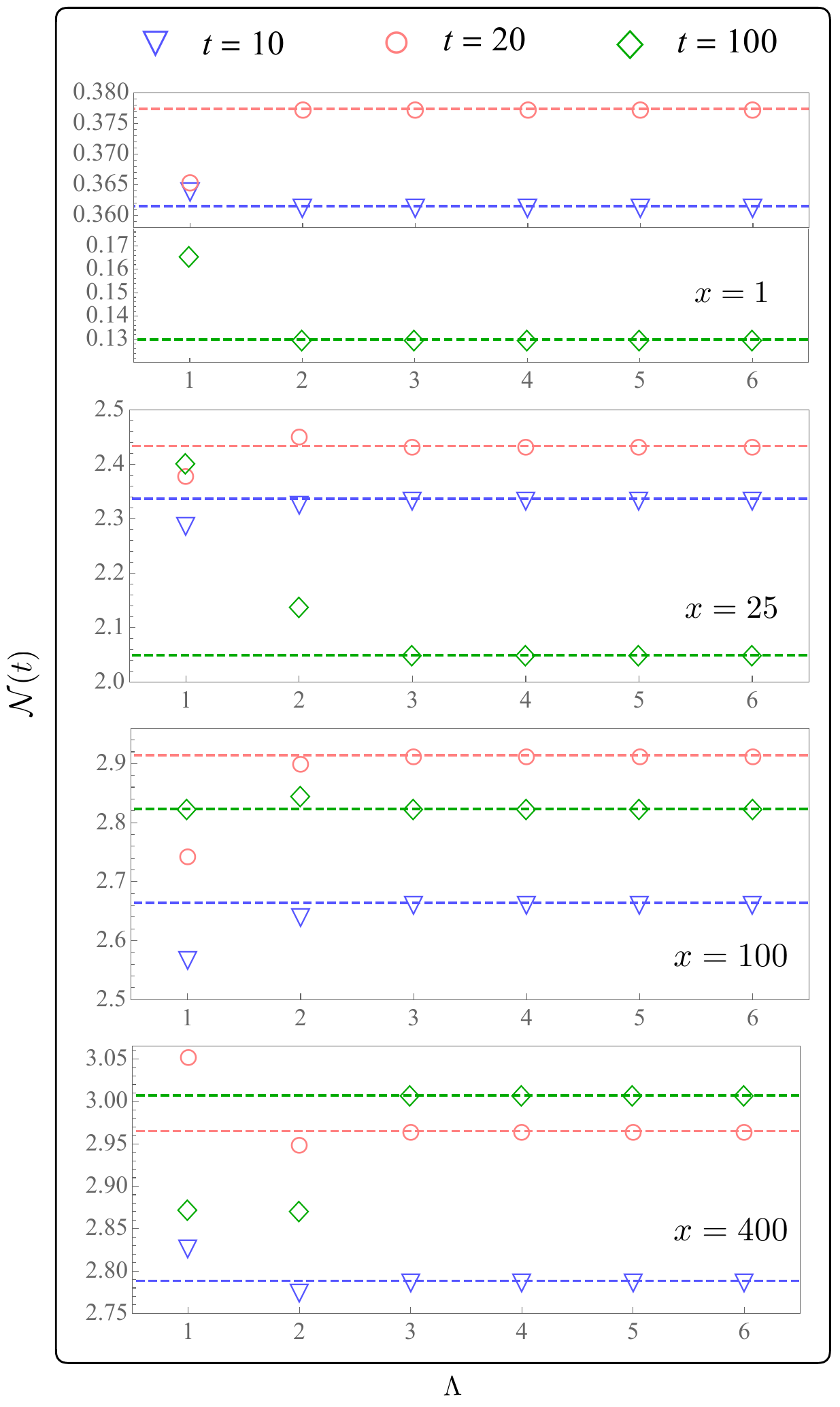}
\caption{The quantity $\mathcal{N}(t)$ defined in Eq.~(\ref{eq:Nt}) as a function of $\Lambda$ for various values of $x$ and $t$ in the KS Hamiltonian in the physical Hilbert space with $N=6$ and $\nu=1$ with OBC. $t$ is the absolute time as defined in the text in units of $a$. The exact values of the quantity are shown in dashed lines. Note that for $\Lambda=6$, the full Hilbert space with this quantum number is achieved, nonetheless the quantity shown reaches the exact values with $\Lambda < 6$ in all cases. As a result, no fit to the $\Lambda$ dependence of the data is performed in contrast to the PBC case. The numerical values associated with these plots are provided in Supplemental Material. 
}
\label{fig:timedepvslambdaobc}
\end{figure}
\begin{figure}[htp]
\includegraphics[width=0.490\textwidth]{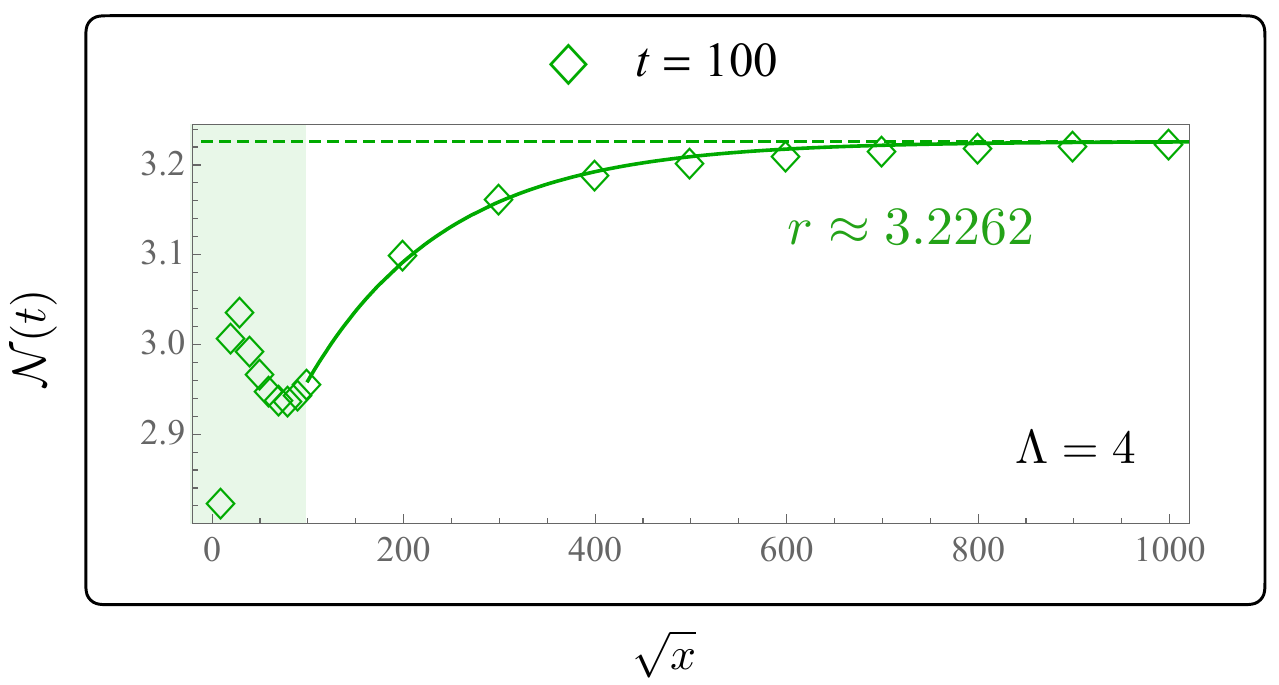}
\caption{The quantity $\mathcal{N}(t)$ defined in Eq.~(\ref{eq:Nt}) as a function of $\sqrt{x}$ for $\Lambda=4$ and $t=100$ in the KS Hamiltonian in the physical Hilbert space with $N=6$ and $\nu=1$ with OBC. The points are fit to $\mathcal{N} \sim e^{-r\sqrt{x}}$ and the colored regions are excluded from such fits. The numerical values associated with these plots are provided in Supplemental Material. 
}
\label{fig:ntvssqxt100obc}
\end{figure}

\clearpage

\end{document}